\documentclass[superscriptaddress,prc,twocolumn,showpacs,preprintnumbers,amsmath,amssymb,bibnotes,altaffilletter]{revtex4}

\usepackage{graphicx}
\usepackage{dcolumn}
\usepackage{bm}
\usepackage{psfrag}
\usepackage{txfonts}
\usepackage{float} 
\usepackage{xspace}

\newcommand{\asymmA}{A}
\newcommand{\Ckv}{Cherenkov\xspace}
\newcommand{\nacl}{NaCl\xspace}
\newcommand{\be}{$^8$B\xspace}
\newcommand{\lie}{$^8$Li\xspace}
\newcommand{\ns}{$^{16}$N\xspace}
\newcommand{\cf}{$^{252}$Cf\xspace}
\newcommand{\cl}{$^{35}$Cl\xspace}
\newcommand{\dto}{D$_2$O\xspace}
\newcommand{\hto}{H$_2$O\xspace}
\newcommand{\teff}{T_{\rm eff}}

\newcommand{\tltze}{$^{208}$Tl\xspace} 
\newcommand{\bitof}{$^{214}$Bi\xspace}
\newcommand{\utte}{$^{238}$U\xspace} 
\newcommand{\tttt}{$^{232}$Th\xspace}

\newcommand{\nhits}{$N_{\rm hits}$\xspace}

\newcommand{\mnox}{MnO$_{\mbox{x}}$\xspace}
\newcommand{\bof}{$\beta_{14}$\xspace}
\newcommand{\costs}{$\cos\theta_\odot$\xspace}

\newcommand{\snoccfluxunc}{1.68^{+0.06}_{-0.06}\mbox{(stat)}^{+0.08}_{-0.09}\mbox{(syst)}} 
\newcommand{\snoesfluxunc}{2.35^{+0.22}_{-0.22}\mbox{(stat)}^{+0.15}_{-0.15}\mbox{(syst)}} 
\newcommand{\snoncfluxunc}{4.94^{+0.21}_{-0.21}\mbox{(stat)}^{+0.38}_{-0.34}\mbox{(syst)}} 

\newcommand{\snoccfluxconc}{1.72^{+0.05}_{-0.05}\mbox{(stat)}^{+0.11}_{-0.11}\mbox{(syst)}} 
\newcommand{\snoesfluxconc}{2.34^{+0.23}_{-0.23}\mbox{(stat)}^{+0.15}_{-0.14}\mbox{(syst)}} 
\newcommand{\snoncfluxconc}{4.81^{+0.19}_{-0.19}\mbox{(stat)}^{+0.28}_{-0.27}\mbox{(syst)}} 

\newcommand{\snoccncratiounc}{0.340\pm 0.023~\mbox{(stat)}~^{+0.029}_{-0.031}~\mbox{(syst)}}
\newcommand{\snoccncratiocon}{0.358 \pm 0.021~\mbox{(stat)~}^{+0.028}_{-0.029}~\mbox{(syst)}}

\newcommand{\snoccesratiounc}{0.712\pm 0.075~\mbox{(stat)}~^{+0.045}_{-0.044}~\mbox{(syst)}}
\newcommand{\snoccesratiocon}{0.736\pm 0.079~\mbox{(stat)}~^{+0.050}_{-0.049}~\mbox{(syst)}}

\newcommand{\snoncmcccon}{3.09\pm0.22~\mbox{(stat)}~^{+0.30}_{-0.27}~\mbox{(syst)}}
\newcommand{\snoncmccunc}{3.26\pm0.25~\mbox{(stat)}~^{+0.40}_{-0.35}~\mbox{(syst)}}

\newcommand{\snoesmcccon}{3.97\pm1.56~\mbox{(stat)}~^{+0.92}_{-0.89}~\mbox{(syst)}}
\newcommand{\snoesmccunc}{4.36\pm1.52~\mbox{(stat)}~^{+0.90}_{-0.87}~\mbox{(syst)}}

\newcommand{\nccfit}{2176\pm 78} 
\newcommand{\nesfit}{279\pm 26} 
\newcommand{\nncfit}{2010\pm 85}  
\newcommand{\nncbefit}{128\pm 42}

\def\nuc#1#2{\relax\ifmmode{}^{#1}{\protect\text{#2}}\else${}^{#1}$#2\fi}

\newcommand{\combinedpdth}{0.29 ^{+0.18}_{-0.18}} 
                                    
\newcommand{\exsitupdth}{0.42 ^{+0.23}_{-0.17}} 
				
\newcommand{\exsitupdu}{{<1}} 

\newcommand{\insitupdth}{0.22 ^{+0.16}_{-0.16}} 
				
\newcommand{\insitupdu}{0.28 ^{+0.04}_{-0.07}} 

\newcommand{\insitupdthd}{0.15 ^{+0.23}_{-0.18}} 
				
\newcommand{\insitupdud}{0.32 ^{+0.05}_{-0.05}}

\newcommand{\insitupdthn}{0.36 ^{+0.25}_{-0.25}} 
				
\newcommand{\insitupdun}{0.26 ^{+0.06}_{-0.06}}

\newcommand{\shwlabel}[1]{\label{#1}}

\renewcommand{\today}{\number\day\space\ifcase\month\or January\or 
 February\or March\or April\or May\or June\or July\or August\or 
 September\or October\or November\or December\fi\space\number\year}

\begin{document}

\preprint{XXX/XXX-XXX}

\title{Electron Energy Spectra, Fluxes, and Day-Night Asymmetries of \\ 
$^{\bm{8}}$B Solar Neutrinos from the 391-Day Salt Phase SNO Data Set}

\newcommand{\ubc}{Department of Physics and Astronomy, University of
British Columbia, Vancouver, BC V6T 1Z1, Canada}
\newcommand{\bnl}{Chemistry Department, Brookhaven National
Laboratory,  Upton, NY 11973-5000}
\newcommand{\carleton}{Ottawa-Carleton Institute for Physics, 
Department of Physics, Carleton University, Ottawa, Ontario K1S 5B6, 
Canada}
\newcommand{\uog}{Physics Department, University of Guelph,
Guelph, Ontario N1G 2W1, Canada}
\newcommand{\lu}{Department of Physics and Astronomy, Laurentian
University, Sudbury, Ontario P3E 2C6, Canada}
\newcommand{\lbnl}{Institute for Nuclear and Particle Astrophysics and
Nuclear Science Division, Lawrence Berkeley National Laboratory, 
Berkeley, CA 94720}
\newcommand{\lbla}{ Lawrence Berkeley National Laboratory, Berkeley, CA}
\newcommand{\lanl}{Los Alamos National Laboratory, Los Alamos, NM 87545}
\newcommand{\llnl}{Lawrence Livermore National Laboratory, Livermore, CA}
\newcommand{\lanla}{Los Alamos National Laboratory, Los Alamos, NM}
\newcommand{\oxford}{Department of Physics, University of Oxford,
Denys Wilkinson Building, Keble Road, Oxford OX1 3RH, UK}
\newcommand{\penn}{Department of Physics and Astronomy, University of
Pennsylvania, Philadelphia, PA 19104-6396}
\newcommand{\queens}{Department of Physics, Queen's University,
Kingston, Ontario K7L 3N6, Canada}
\newcommand{\uw}{Center for Experimental Nuclear Physics and Astrophysics,
and Department of Physics, University of Washington, Seattle, WA 98195}
\newcommand{\uta}{Department of Physics, University of Texas at 
Austin, Austin, TX 78712-0264}
\newcommand{\triumf}{TRIUMF, 4004 Wesbrook Mall, Vancouver, BC V6T 2A3, Canada}
\newcommand{\ralimp}{Rutherford Appleton Laboratory, Chilton, Didcot 
OX11 0QX, UK}
\newcommand{\iusb}{Department of Physics and Astronomy, Indiana 
University, South Bend, IN}
\newcommand{\fnal}{Fermilab, Batavia, IL}
\newcommand{\uo}{Department of Physics and Astronomy, University of 
Oregon, Eugene, OR}
\newcommand{\rcnp}{Department of Physics, Osaka University, Osaka, Japan}
\newcommand{\slac}{Stanford Linear Accelerator Center, Menlo Park, CA}
\newcommand{\mac}{Department of Physics, McMaster University, Hamilton, ON}
\newcommand{\doe}{US Department of Energy, Germantown, MD}
\newcommand{\lund}{Lund University, Sweden}
\newcommand{\mpi}{Max-Planck-Institut for Nuclear Physics, Heidelberg, Germany}
\newcommand{\uom}{Ren\'{e} J.A. L\'{e}vesque Laboratory, 
Universit\'{e} de Montr\'{e}al, Montreal, PQ}
\newcommand{\cwru}{Department of Physics, Case Western Reserve 
University, Cleveland, OH}
\newcommand{\pnnl}{Pacific Northwest National Laboratory, Richland, WA}
\newcommand{\uc}{Department of Physics, University of Chicago, Chicago, IL}
\newcommand{\mitt}{Department of Physics, Massachusetts Institute of 
Technology, Cambridge, MA }
\newcommand{\ucsd}{Department of Physics, University of California at 
San Diego, La Jolla, CA }
\newcommand{	\lsu	}{Department of Physics and Astronomy, 
Louisiana State University, Baton Rouge, LA 70803}
\newcommand{\imp}{Imperial College, London SW7 2AZ, UK}
\newcommand{\uci}{Department of Physics, University of California, 
Irvine, CA 92717}
\newcommand{\ucia}{Department of Physics, University of California, Irvine, CA}
\newcommand{\suss}{Department of Physics and Astronomy, University of 
Sussex, Brighton  BN1 9QH, UK}
\newcommand{\lifep}{Laborat\'{o}rio de Instrumenta\c{c}\~{a}o e 
F\'{i}sica Experimental de Part\'{i}culas, Lisboa, Portugal}
\newcommand{\tokyo}{Kamioka Observatory, Institute for Cosmic Ray
  Research, University of Tokyo, Kamioka, Gifu, 506-1205, Japan}

\affiliation{	\ubc	}
\affiliation{	\bnl	}
\affiliation{	\carleton	}
\affiliation{	\uog	}
\affiliation{	\lu	}
\affiliation{	\lbnl	}
\affiliation{	\lanl	}
\affiliation{	\lsu	}
\affiliation{	\oxford	}
\affiliation{	\penn	}
\affiliation{	\queens	}
\affiliation{	\ralimp	}
\affiliation{	\uta	}
\affiliation{	\triumf	}
\affiliation{	\uw	}

\author{	B.~Aharmim	}			\affiliation{ 
	\lu	}
\author{	S.N.~Ahmed	}			\affiliation{ 
	\queens	}
\author{	A.E.~Anthony	}			\affiliation{ 
	\uta	}
\author{	E.W.~Beier	}			\affiliation{ 
	\penn	}
\author{	A.~Bellerive	}			\affiliation{ 
	\carleton	}
\author{	M.~Bergevin	}			\affiliation{ 
	\uog	}
\author{	S.D.~Biller	}			\affiliation{ 
	\oxford	}
\author{	J.~Boger	} 
	\altaffiliation{Present Address: \doe}	\affiliation{	\bnl 
	}
\author{	M.G.~Boulay	}			\affiliation{ 
	\lanl	}
\author{	M.G.~Bowler	}			\affiliation{ 
	\oxford	}
% ...

\collaboration{SNO Collaboration}
\noaffiliation

\begin{abstract}
Results are reported from the complete salt phase of the Sudbury
Neutrino Observatory experiment in which NaCl was dissolved in the
\dto target.  The addition of salt enhanced the signal from neutron
capture, as compared to the pure \dto detector.  By making a
statistical separation of charged-current events from other types
based on event-isotropy criteria, the effective electron recoil energy
spectrum has been extracted.  In units of $ 10^6$ cm$^{-2}$ s$^{-1}$,
the total flux of active-flavor neutrinos from \be decay in the Sun is
found to be $\snoncfluxunc$ and the integral flux of electron
neutrinos for an undistorted \be spectrum is $\snoccfluxunc$;  the
signal from ($\nu_x$,e) elastic scattering is equivalent to an
electron-neutrino flux of $\snoesfluxunc$.  These results are
consistent with those expected for neutrino oscillations with the
so-called Large Mixing Angle parameters, and also with an undistorted
spectrum.  A search for matter-enhancement effects in the Earth
through a possible day-night asymmetry in the charged-current integral
rate is consistent with no asymmetry.  Including results from other
experiments, the best-fit values for two-neutrino mixing parameters
are $\Delta m^2 = (8.0^{+0.6}_{-0.4} ) \times 10^{-5}$ eV$^2$ and
$\theta = 33.9 ^{+2.4}_{-2.2}$ degrees.
\end{abstract}

\maketitle

\section{\label{sec:intro}Introduction}

Results from the completed second phase of the Sudbury Neutrino
Observatory (SNO)~\cite{bib:sno_nim} are presented in this paper.  The
second phase began in June of 2001 with the addition of $\sim$2000~kg
of \nacl~\cite{bib:mgcl_footnote} to SNO's $\sim$1000~tonnes of \dto,
and ended in October 2003 when the \nacl was removed.  The addition of
the salt enhanced SNO's ability to detect solar \be neutrinos in three
ways.  First, the neutron capture efficiency increased by nearly
three-fold, allowing a statistically precise measurement of the
neutral-current (NC) disintegration of deuterons by solar neutrinos.
Second, the total energy of the $\gamma$-rays from the neutron capture
on \cl is 2.32~MeV above the energy of the single $\gamma$ from the
capture on deuterons.  This higher $\gamma$-ray energy approximately
corresponds to a 1 MeV upward shift in the observed energy peak for
neutrons and allows a precise measurement that is well
above the low energy radioactive backgrounds to be made.  Lastly, and
perhaps most importantly, the isotropy of the photomultiplier tube
(PMT) hit distribution on the geodesic array from multiple
$\gamma$-rays emitted after neutron capture on \cl is significantly
different from that produced by \Ckv light emitted by a single
relativistic electron.  Hence, neutrons from the NC reaction and
electrons from the charged-current (CC) interaction can be separated
statistically without any assumptions about the underlying neutrino
energy spectrum.

Results from the first SNO phase~\cite{bib:snocc,bib:snonc,bib:snodn},
using pure \dto in the target volume, confirmed earlier measurements
\cite{bib:homestake,bib:sage,bib:gno,bib:superk} of an observed
deficit of solar electron-type neutrino flux compared to solar model
expectations~\cite{bib:bp2000,bib:bp2004} but additionally
demonstrated, through measurement of the total active \be solar flux,
that neutrinos undergo flavor transformation in transit to terrestrial
detectors.

Initial measurements of the total active \be solar flux, based on the
first 254.2 live days of the salt data set, have been
published~\cite{bib:saltprl} and confirm and improve on results from
the first \dto phase
measurements~\cite{bib:snocc,bib:snonc,bib:snodn}.  In particular, the
statistical discrimination of CC and NC events with salt enabled an
independent measure of the total active $^8$B flux.  The measured flux
is in very good agreement with solar model
calculations~\cite{bib:bp2000,bib:bp2004,bib:bs05,bib:tc04} .

The favored interpretation of these results is that neutrinos undergo
oscillations between flavor states given by linear combinations of the
non-degenerate mass eigenstates as described in the
Maki-Nakagawa-Sakata-Pontecorvo (MNSP) framework~\cite{bib:mnsp}.  For
the case of $^8$B solar neutrinos, the measurements support the
Mikheyev-Smirnov-Wolfenstein (MSW) \cite{bib:msw} hypothesis of matter
enhanced oscillation, where electron neutrinos experience an
additional interaction, compared with muon or tau neutrinos, in the
presence of matter that can enhance neutrino oscillations.  The SNO
data, when combined with other solar neutrino measurements and reactor
antineutrino results from the KamLAND experiment~\cite{bib:kamland},
show that neutrino oscillations are the dominant cause of flavor
transformation and significantly restrict the allowed range of the
relevant neutrino mixing parameters.

In terms of neutrino mass and mixing parameters $\Delta\mbox{m}^2$ and
$\tan^{2}{\theta}$, solar neutrino data favor the so-called Large
Mixing Angle (LMA) region.  Maximal mixing is ruled out with a
high-degree of confidence.  

The present paper extends the analysis to a total of 391 live days of
SNO data from the salt phase, provides new results for the integral
fluxes, and provides the CC energy spectrum and day-night spectral
asymmetries. The values for $\tan^{2}{\theta}$ and $\Delta\mbox{m}^2$
are updated using a two-neutrino oscillation analysis. The paper also
provides a more detailed description of the full data analysis process
for the SNO salt phase.

The layout of the paper is as follows: Sec.~\ref{sec:sno_det} of the
paper describes details of the detector hardware and software
simulation.  The data set, live time determination and event selection
are discussed in Sec.~\ref{sec:dataset}.

The following sections discuss detector response and backgrounds,
along with their associated systematic uncertainties on the neutrino
measurements.  Detailed systematic uncertainty evaluations are
presented in the discussions of optical and energy calibration in
Sec.~\ref{sec:sno_calib},~ event vertex and direction reconstruction
and isotropy in Sec.~\ref{sec:recon},~ and neutron response in
Sec.~\ref{sec:neutron}.  Complete analyses of the many potential
background sources are given in Sec.~\ref{sec:backgrounds} followed by
the evaluation of specific systematic uncertainties associated with
the day-night asymmetry measurement in Sec.~\ref{sec:dnspecific}.

The procedure for analyzing the solar neutrino signal is discussed in
Sec.~\ref{sec:sigex}.  Solar neutrino results are presented with
particular emphasis on the CC energy spectrum with the evaluation of
differential energy systematic uncertainties in
Sec.~\ref{sec:spectrum}, integral flux in Sec.~\ref{sec:intflux}, and
day-night asymmetry in Sec.~\ref{sec:dnresults}.  Interpretation of
the results in the context of the MSW framework for SNO data only and
for the combined solar and reactor analysis can be found in
Sec.~\ref{sec:physint}.  A summary provided in Sec.~\ref{sec:summary}
concludes the paper.

\section{\label{sec:sno_det}SNO Detector and Simulation}

\subsection{ Detector}

The Sudbury Neutrino Observatory is a real time heavy water \Ckv
detector located in the Inco Ltd., Creighton mine near Sudbury,
Ontario, Canada.  The center of the detector is at a depth of 2092~m,
or 6010~meters of water equivalent.  At this depth, approximately 65
muons enter the detector per day.  The neutrino target is 1000 tonnes
of 99.92\% isotopically pure \dto contained inside a 12-m diameter
acrylic vessel (AV).  An array of 9456 20-cm Hamamatsu R1408 PMTs,
which is mounted on an 18-m diameter stainless steel geodesic
structure, is used to detect \Ckv radiation in the target.  A
non-imaging light concentrator is mounted on each PMT to increase the
effective photocathode coverage by the complete array to approximately
54\% of 4$\pi$. 

To minimize the effects of radioactive backgrounds on the detection of
solar neutrinos, materials with low intrinsic radioactivity were
selected for the construction of the detector.  The acrylic vessel and
the geodesic sphere are immersed in ultra-pure \hto to provide
shielding against radioactive backgrounds from the geodesic structure
and the cavity rock.  An additional 91 PMTs are mounted looking
outwards on the geodesic sphere and 23 PMTs are suspended facing
inwards in the outer \hto volume to act as cosmic veto counters.  Four
PMTs that are installed in the neck region of the acrylic vessel
provide veto signals to reject certain classes of instrumental
background events (Sec.~\ref{sec:backgrounds}).  Further details of
the detector can be found in~\cite{bib:sno_nim}.
Note that for analysis purposes Cartesian coordinates are defined such
that the center of the vessel is at ($x$,$y$,$z$) = (0,0,0), and the neck
region is located symmetrically about the positive $z$ axis.

The SNO experiment detects solar neutrinos through the charged-current
(CC) and neutral-current (NC) interactions on the deuteron, and by
elastic scattering (ES) on electrons:
\begin{displaymath}
\begin{array}{ll}
  \mbox{CC:} & \nu_e + d \, \rightarrow\, p + p + e^-
  -1.442\,\mbox{MeV} \\ \mbox{NC:} & \nu_x + d \, \rightarrow\, p + n
  + \nu_x - 2.224\,\mbox{MeV} \\ \mbox{ES:} & \nu_x + e^-
  \,\rightarrow\, \nu_x + e^-
\end{array} 
\end{displaymath}
where $\nu_x$ refers to any active flavor of neutrinos.  The NC
channel has equal sensitivity to all active neutrinos, while the ES
channel is sensitive primarily to electron-neutrinos.  Hence, the NC
measurement can determine the total active solar neutrino flux even if
electron-neutrinos transform to another active flavor~\cite{bib:chen}.

In the first phase of the experiment with pure \dto, NC interactions were
observed by detecting the 6.25-MeV $\gamma$-ray following capture of
the neutron by the deuteron.  For the second phase of data taking,
(0.196$\pm$0.002)\% by weight of purified \nacl was added to the \dto
in May 2001 to increase the capture and the detection efficiencies of
the NC neutron.  The thermal neutron capture cross section of
$^{35}$Cl is 44~b, which is significantly higher than that of the
deuteron at 0.5~mb.  When a neutron captures on $^{35}$Cl, the total
energy released is 8.6~MeV.  The combination of the increased cross
section and the higher energy released results in a larger neutron
detection efficiency at the same analysis threshold.

Neutron capture on $^{35}$Cl typically produces multiple $\gamma$-rays
($\sim$2.5 per capture), while the CC and ES reactions produce single
electrons. Each $\gamma$-ray predominantly interacts through Compton
scattering, producing an energetic electron.  The \Ckv light from
neutron capture events, compared to that from CC and ES events, is
more isotropic as the light is typically from several
electrons rather than one.  This greater isotropy, together with the
strong directionality of ES events, allows good statistical separation
of the event types. 

A precise measurement of the total active solar neutrino flux can be
made through the NC channel without assumptions about the underlying
neutrino energy spectrum.  This is relevant because the neutrino energy
spectrum can be distorted from the generated $^8$B spectrum via
oscillation effects.

\subsection{SNO Monte Carlo Simulation}

The {\bf SNO} {\bf M}onte Carlo and {\bf an}alysis (SNOMAN) code is used
for off-line analysis of the SNO data and provides an accurate model of
the detector for simulating neutrino and background events.  The
Monte Carlo (MC) processor in SNOMAN provides processors for the
generation of different classes of events, propagation of the primary
particles and any secondary particles (such as Compton electrons) that
are created, detection of the signal by the PMTs and simulation of the
electronics response.  With the exception of a few physics simulations
(such as optical photon propagation), widely used packages such as
EGS4~\cite{bib:egs}, MCNP~\cite{bib:mcnp} and FLUKA~\cite{bib:fluka}
are used in SNOMAN to provide accurate propagation of electromagnetic
showers, neutrons, and hadrons.

Detailed models of all the detector components and calibration sources
are implemented in SNOMAN.  Generators for neutrino and calibration
source signals, radioactive backgrounds and cosmic rays are also
provided.  Input parameters such as optical attenuation coefficients
are determined from detector calibration.  Calibration and detector
parameters are input to SNOMAN and probability density functions
(PDFs), used in the neutrino analysis, are generated.  These features
allow a direct assessment of the systematic uncertainties in physics
measurements by comparing the detector responses for various
calibration sources with the predictions of SNOMAN.

For the analysis of SNO data, SNOMAN provides various processors to
unpack the data, to provide charge and time calibration of the PMT
hits for each event, to reconstruct event position and
direction, and to estimate the event energy.

\section{\label{sec:dataset}Data Set and Event Selection}

\subsection{\label{sec:data_set}Data Set and Live Time}

The measurements reported here are based on analysis of $391.432 \pm
0.082$ live days of data recorded between July 26, 2001 and August 28,
2003.  As described below, 176.511 days of the live time were recorded
during the day and 214.921 days during the night.

The selection of solar neutrino data runs for analysis is based upon
the evaluation of detector operation logs and outputs from an
automated off-line data evaluation processor in SNOMAN.  The automated
processor checks the validity of event times, the trigger thresholds,
the status of the electronics channels, and other detector parameters
to ensure the quality of the runs selected for analysis.  In addition,
neutrino runs that are known to have elevated levels of radioactive
background were removed from the data set.  For instance, some
neutrino runs have an elevated level of Rn ingress while others
contain residual $^{24}$Na radioactivity resulting from neutron
activation during detector calibration.

The bulk of the time that the detector was not live for neutrino data
acquisition was used for detector calibration and maintenance
activities.  Table~\ref{tab:det_time} provides a tabulation of the detector
operational status.
\begin{table}
\caption{\label{tab:det_time} Tabulation of detector operational status.}
\begin{center}
\begin{tabular}{ld} \hline
\hline 
Data & \multicolumn{1}{c}{Percentage of } \\  
& \multicolumn{1}{c}{ run period } \\ \hline 
Detector maintenance, \\
\hspace{0.1in} $\nu$ runs rejected by data quality checks, &  \\ 
\hspace{0.1in} or elevated Rn or $^{24}$Na levels & 22.6 \\ 
Calibration activities & 20.2 \\
Detector off & 6.2 \\
Selected $\nu$ runs for analysis & 51.0 \\
\hline\hline
\end{tabular}
\end{center}
\end{table} 

The raw live time of the data set is calculated using a GPS synched 10
MHz clock on a run-by-run basis from the difference in times between
the first and last triggered events in each run.  These calculated
time intervals are verified by comparing the results against those
measured independently with the 50 MHz detector system clock.
Separate day and night live times are determined by splitting each run
based on solar zenith angle $\theta_z$ where
$day\equiv(\cos{\theta_z}>0)$ and $night\equiv(\cos{\theta_z}<0)$.
The ratio of day to night live times is 0.82128 with an uncertainty of
less than $5\times 10^{-7}$.

Several data selection cuts remove small periods of time from the data
set.  These include time intervals following high-energy cosmic-ray
events and intervals containing time-correlated instrumental events To
calculate the final live time for the neutrino data set, the total
time removed by the full set of data selection cuts is subtracted.
Data selection cuts remove a combined total of 1.8\% of the raw live
time for the neutrino data set.  The dominant effect is caused by the
cosmic ray veto cut.
                                                                                
The live time calculation, including the corrections due to data
selection cuts, is checked with an analysis of data from the detector
diagnostic trigger.  This pulsed global trigger (PGT) is a
detector-wide trigger issued at a frequency of 5Hz based on timing
from the 50 MH clock.  Systematic uncertainties in live time are
evaluated by comparing the PGT measurement to the 10MHz clock
measurement, and by analyzing electronics and data acquisition effects
that could prevent the detector from being live to neutrino data for
short times.  The total live time uncertainty is calculated to be
$\pm0.021\%$.

\subsection{\label{sec:event_selection}Event Selection}

The first step in selecting the solar neutrino candidate events
involves the rejection of instrumental backgrounds and residual
backgrounds from cosmic rays.  A typical data rate was $\sim$20~Hz,
which was dominated by low-energy radioactive backgrounds in the
detector and the PGT.  The primary contributors to the instrumental
backgrounds include events generated by static discharge inside the
PMTs, known as flashers, events produced when light is emitted from
the neck region of the acrylic vessel, known as neck events, and
electronic pickup events. The typical combined rate for these events
is approximately one per minute compared to the $\sim$10 per day rate
of solar neutrino events.  These instrumental background events are
identified and removed based on analysis of the charge and timing
distributions of the triggered PMTs, event geometry in detector and in
electronics space, signals from the cosmic veto counter
outward-looking PMTs, PMTs installed to identify neck events, and time
between events.  Additional cuts using reconstruction information,
referred to here as ``high-level" cuts (Sec.~\ref{sec:backgrounds}),
are used to remove events that do not possess the timing and isotropy
characteristics of \Ckv light from either single or multiple $\beta$s.

After the instrumental background and high-level cuts, two cuts are
applied to remove cosmic-ray events.  The first cut removes events
that occur in a 20~s time interval after each event identified as a
muon. The principal feature of muon identification is the requirement
that there be at least 5 PMT hits in the outward-looking PMTs and 150
PMT hits (equivalent to $\sim$20~MeV) in the inner detector.  A
second, simpler cut removes any event following within 250~ms of any
event with at least 150 PMT hits. This cut removes neutrons from muons
missed by muon identification and neutrons induced by most
atmospheric neutrino interactions inside the detector.

The primary background sources are low in energy and external to the
\dto volume.  Hence, to minimize the systematic uncertainties
associated with backgrounds, all events selected for the results
described here are required to have a reconstructed vertex position
within 550 cm of the center of the detector.  In terms of the
volume-weighted variable $\rho\equiv (R/R_{\rm AV})^3$, where $R_{\rm
AV}$=600.5~cm is the radius of the acrylic vessel that contains the
\dto, the fiducial volume cut is $\rho<0.77$.  Candidate events are
required to have an effective electron kinetic energy $\teff$ (see
Sec.~\ref{sec:sno_calib}) greater than 5.5 MeV.  Details of all
background sources are provided in Sec.~\ref{sec:backgrounds}.

\begin{figure}
\begin{center}
\includegraphics[width=3.73in]{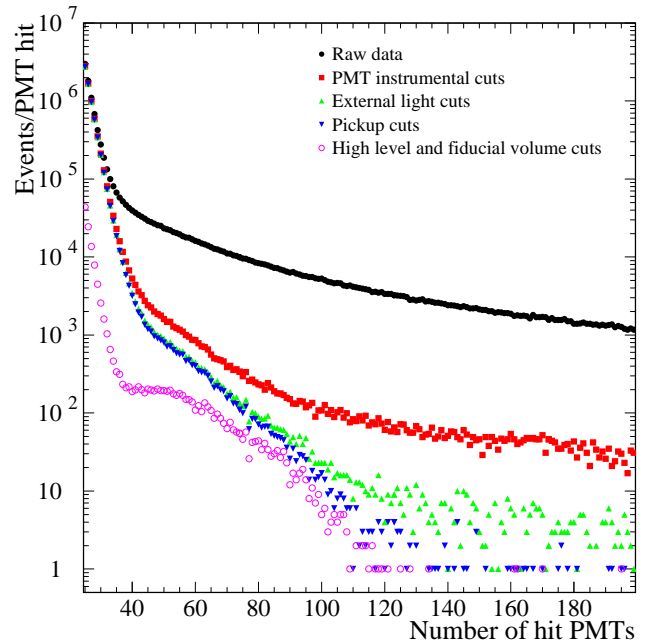}
\caption{Reduction of the data set as successive cuts are applied.
The PMT instrumental, external light and pickup cuts remove
instrumental backgrounds originating from the detector hardware.  The
high-level cuts (Sec.~\ref{sec:backgrounds}) further reduce the
instrumental backgrounds by rejecting events that do not possess the
characteristics of \Ckv light emission from  single or multiple
$\beta$s.  The fiducial volume cut, which selects events 
reconstructed with $\rho<0.77$, removes most of the radioactive
background events that originate outside the \dto target.  }
\label{fig:nhit_red}
\end{center}  
\end{figure}

The results of sequentially applying the cuts to the data set are
shown in Fig.~\ref{fig:nhit_red}. The instrumental cuts are applied in
steps to show the removal of various classes of backgrounds. The
application of the high-level cuts then reduces the data set further,
leaving the set of neutrino candidates.  Some of the cuts
remove individual events based upon their characteristics, while
others remove periods of time from the data set.

The efficiency of the event-based cuts is measured independently for
electron and neutron events.  For neutrons, a software tagging
approach (Sec.~\ref{sec:neutron}) is used to select neutron events
from a $^{252}$Cf fission source. These tagged neutron events are then
used to measure the efficiency of all cuts.  For electrons, the
efficiencies of the instrumental cuts and high-level cuts are measured
independently. The efficiency of the instrumental cuts is measured
using tagged $\beta$ events from a $^8$Li source~\cite{bib:li8_nim},
while 6.13-MeV $\gamma$-rays from a tagged $^{16}$N
source~\cite{bib:n16_nim} are used to measure systematic
uncertainties.  The efficiency of the high-level cuts is established
by MC simulations.  Tagged $^{16}$N events are again used to measure
systematic uncertainties and corrections.  The estimated signal loss
for each class of event, integrated over the expected distributions,
is shown in Table~\ref{tab:signal_loss}.
\begin{table}
\caption{\label{tab:signal_loss} Signal loss of the instrumental and
high-level cuts for each signal class.  These measurements are
averaged over the energy spectra above $\teff=5.5$ MeV.  For CC
and ES signals an undistorted $^8$B neutrino spectrum has been assumed.}
\begin{center}
\begin{tabular}{l@{\hspace{1in}}c} \hline\hline
Signal Class & Signal Loss, \% \\ \hline \\ [-2.5mm] Charged current &
$0.57^{+0.16}_{-0.11}$ \\[1mm] Neutrons & $0.68^{+0.16}_{-0.11}$
\\[1mm] Elastic scattering & $0.86^{+0.21}_{-0.17}$ \\[1mm]
\hline\hline
\end{tabular}
\end{center}
\end{table} 

\begin{figure}
\begin{center}
\includegraphics[width=3.73in]{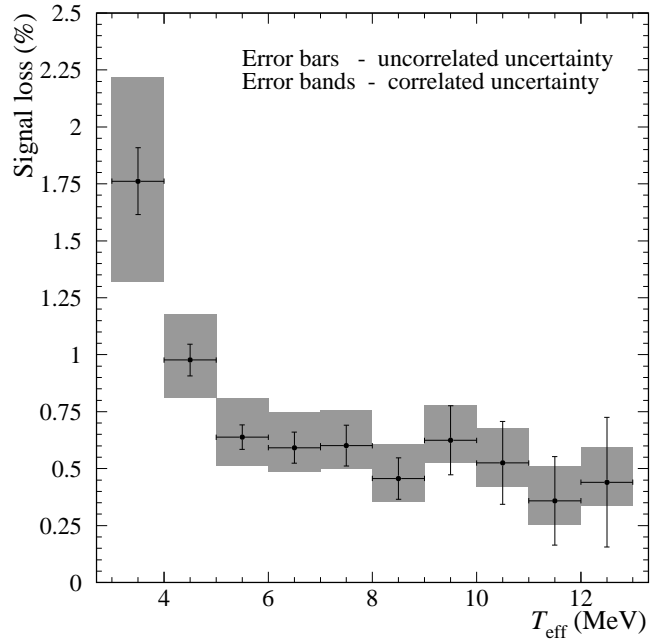}
\caption{Signal loss as a function of energy for CC electrons assuming
an undistorted $^8$B spectrum.  Signal
loss uncertainties are divided into two classes: uncorrelated (error
bars) and correlated (error bands).  Correlated uncertainties arise from
systematic uncertainties in the measurement of signal loss and
uncorrelated uncertainties arise from statistical uncertainties in the
calibration data used in this measurement.}
\label{fig:cc_cut_eff}
\end{center}  
\end{figure}

Signal loss as a function of energy for CC electrons is shown in
Fig.~\ref{fig:cc_cut_eff}. This figure shows separately the systematic
uncertainties that are correlated between energy bins and those that
are independent.

Neutrons produced by cosmic-ray muons and atmospheric neutrinos are
used to measure the day-night asymmetry in the cut efficiency.
Because muon-induced neutrons are generated throughout the live time
of the experiment, they correctly sample any variations in cut
efficiency as a function of time.  Candidate neutron events are
identified by selecting events with a reconstructed vertex radius less
than 550~cm, energy in the range from 6 to 10~MeV, and inside a time
window between 4~$\mu$s to 40~ms after any muon.  Note that the mean
neutron capture time is 5.3~ms at the center of the detector and
essentially all neutrons are captured by 40~ms.

The instrumental background cuts are applied to produce a clean
sample of muon-induced neutrons, and the efficiency of the high-level
cuts is estimated for the sample.  Alternatively, the high-level cuts
are applied and the cut efficiency of the instrumental background cuts
on the remaining events is estimated.  The loss of neutron events from
all cuts, as measured with muon-induced neutrons, is $0.94\pm0.17$\%
and is consistent with the results from neutron calibration sources.
The measured day-night asymmetry, defined as the difference between
the night and the day signal loss divided by their average, is $0.18
\pm 0.33$.

After all data selection cuts have been applied, including the energy
threshold and fiducial volume cuts, 4722 candidate neutrino events
remain.

\section{\label{sec:sno_calib}Detector Calibration}

Interpretation of SNO's signals requires measurement and
calibration of the detector components and response. Many of the
details of the calibration of the detector components are described in
\cite{bib:sno_nim} and will not be discussed here.

A variety of calibration sources are deployed in the heavy and light
water regions to characterize the detector response.  Source
deployment is achieved with a manipulator system~\cite{bib:sno_nim}
that is able to maneuver sources to various positions in two
orthogonal planes within the \dto and along 6 vertical lines in the
\hto. The positional accuracy of the manipulator system is $\sim$2~cm
along the central axis and $\sim$5~cm off-axis in the \dto. In the
\hto the accuracy is $\sim$2~cm.

Table~\ref{tab:cal_sources} lists the primary calibration sources
used.  These include pulsed nitrogen laser light for optical
calibration and PMT timing, \ns $\gamma$-rays to produce a reliable
energy calibration, $^{8}$Li for energy and reconstruction
calibration, a $^3$H($p,\gamma$)$^4$He source (``pT source") to test
linearity of the energy scale, $^{252}$Cf and Am-Be sources of
neutrons, and U, Th, Rn, neutron activated $^{24}$Na, and $^{88}$Y to
test detector response to backgrounds.  Note that the ``Rn spike'' was a
controlled release of a measured quantity of Rn gas into the \dto.

The following section describes the optical and energy calibrations
and our evaluation of the systematic uncertainties associated with
these assessments.  Later sections provide details on event
reconstruction, and determination of the detector response to neutrons
and backgrounds.
\begingroup
\squeezetable
\begin{table}[H]
\caption{\label{tab:cal_sources} Primary calibration sources.}
\begin{center}
 \begin{tabular}{lllc}
\hline
\hline
Calibration source & Details & Calibration & Ref.  \\
\hline 
Pulsed nitrogen laser & 337, 369, 385,       & Optical \& & \cite{bib:sno_nim}  \\
             & 420, 505, 619~nm       & \hspace{0.1in} timing calibration  &  \\
\ns          & 6.13-MeV $\gamma$-rays & Energy \& reconstruction & \cite{bib:n16_nim}  \\
\lie         & $\beta$ spectrum    & Energy \& reconstruction & \cite{bib:li8_nim} \\
\cf          & neutrons            & Neutron response & \cite{bib:sno_nim} \\
Am-Be & neutrons             & Neutron response & \\
$^3$H$(p,\gamma)^4$He  (``pT'') & 19.8-MeV $\gamma$-rays & Energy linearity & \cite{bib:pt_nim} \\
U, Th      &   $\beta-\gamma$ & Backgrounds & \cite{bib:sno_nim} \\
$^{88}$Y & $\beta-\gamma$ & Backgrounds & \\
Dissolved Rn spike &  $\beta-\gamma$ & Backgrounds & \\
\textit{In-situ} $^{24}$Na activation & $\beta-\gamma$ & Backgrounds &
\\
\hline
\hline
\end{tabular}
\end{center}
\end{table}
\endgroup

\subsection{Optical Calibration}

The detector optical response parameters are required for MC
simulations and for the energy reconstruction processor.  The optical
parameters are determined by analyzing the detector response to
photons generated with a pulsed nitrogen laser.  Laser light is
transmitted from the laser through optical fibers to a diffusing ball
that can be positioned at various locations in the detector.

A complete optical scan typically consists of measurements taken at
approximately $\sim40$ different positions and at six wavelengths
(337, 369, 385, 420, 505, and 619~nm).  This set of wavelengths spans
SNO's detectable \Ckv light spectrum.

The laser plus diffuser ball system produces short (0.8 ns) pulses of
light in the detector. For any given position of the diffuser ball,
the PMT array measures a time distribution of light.  The difference
between the measured time of each PMT hit and the expected
time-of-flight from the source position to PMT is called the
``time-residual."  The time-residual distribution exhibits a large
peak originating from ``prompt", or unscattered, light.  The full
width at half maximum of this peak, after all timing corrections have
been applied, is $\sim1.8$~ns.  Smaller peaks occur between 10 and
100~ns later. These ``late" peaks arise from light reflected from
various surfaces within the detector and from late pulsing of the
PMTs.  The optical model analysis is restricted to a $\pm 4$~ns window
centered on the prompt peak to reduce sensitivities to the complicated
structure of the late light distribution.  Additional analyses with a
$\pm 10$~ns prompt peak window are conducted to verify the results and
evaluate sensitivities to the timing window cut.

The optical model is used to predict the number of prompt PMT counts
in each laser calibration run based on operational parameters of the
laser source and the detector optical parameters.  The relevant
operational parameters of the laser source are its light intensity and
angular distribution.  The optical parameters of the detector are the
\dto attenuation length, the combined acrylic and \hto attenuation
length, and the relative efficiency of the PMT-reflector assembly as a
function of incidence angle.  These optical parameters are extracted
by fitting the data collected during a multi-position scan in the \dto
using a $\chi^2$ minimization method.  Note that the fits return the
inverse of the attenuation lengths or ``attenuation coefficients''.
This technique is not sensitive to the separate AV and \hto
attenuation coefficients, only the combined AV+\hto attenuation
coefficients.  The AV attenuation coefficients, given in Table
\ref{tab:atten}, were obtained from {\em ex-situ}
measurements~\cite{bib:atten}-\cite{bib:atten2} and the \hto
attenuation coefficients are determined from the difference between
the AV attenuation coefficients and the measured sum.
\begin{table}[H]
\caption{\label{tab:atten}Acrylic vessel attenuation coefficients.}
\begin{center}
 \begin{tabular}{cc}
\hline \hline Wavelength (nm) & Attenuation coefficient (10$^{-3}$
cm$^{-1}$)\\ \hline 337 & 56.4\\ 369 & 23.0\\ 385 & 12.2\\ 420 &
7.70\\ 505 & 7.09\\ 619 & 7.09\\ \hline \hline
\end{tabular}
\end{center}
\end{table}

The measured \dto and H$_2$O+acrylic attenuation coefficients include
the effect of Rayleigh scattering which removes a fraction of the
light from the prompt time window.  Because the MC simulation must
model both absorption and scattering, the scattering contribution is
subtracted from the measured coefficient and the resulting absorption
coefficient is used as an input to the MC. 
The scattering coefficient is determined using laserball data taken with a
collimating mask over the ball and measuring the hit probabilities for PMTs
outside of the angular acceptance of the collimated beam.

\begin{figure}
\begin{center}
\includegraphics[width=3.73in]{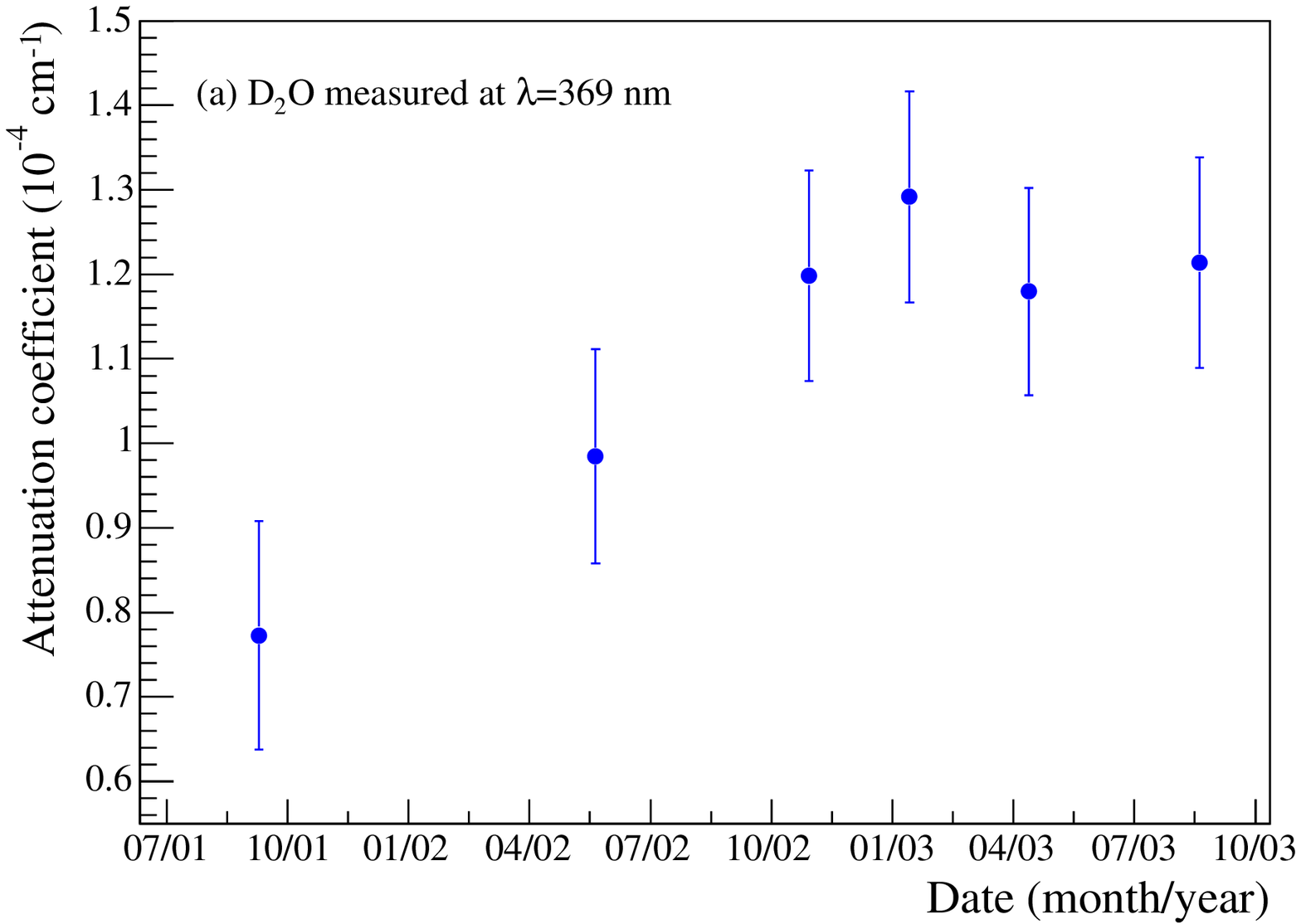}
\includegraphics[width=3.73in]{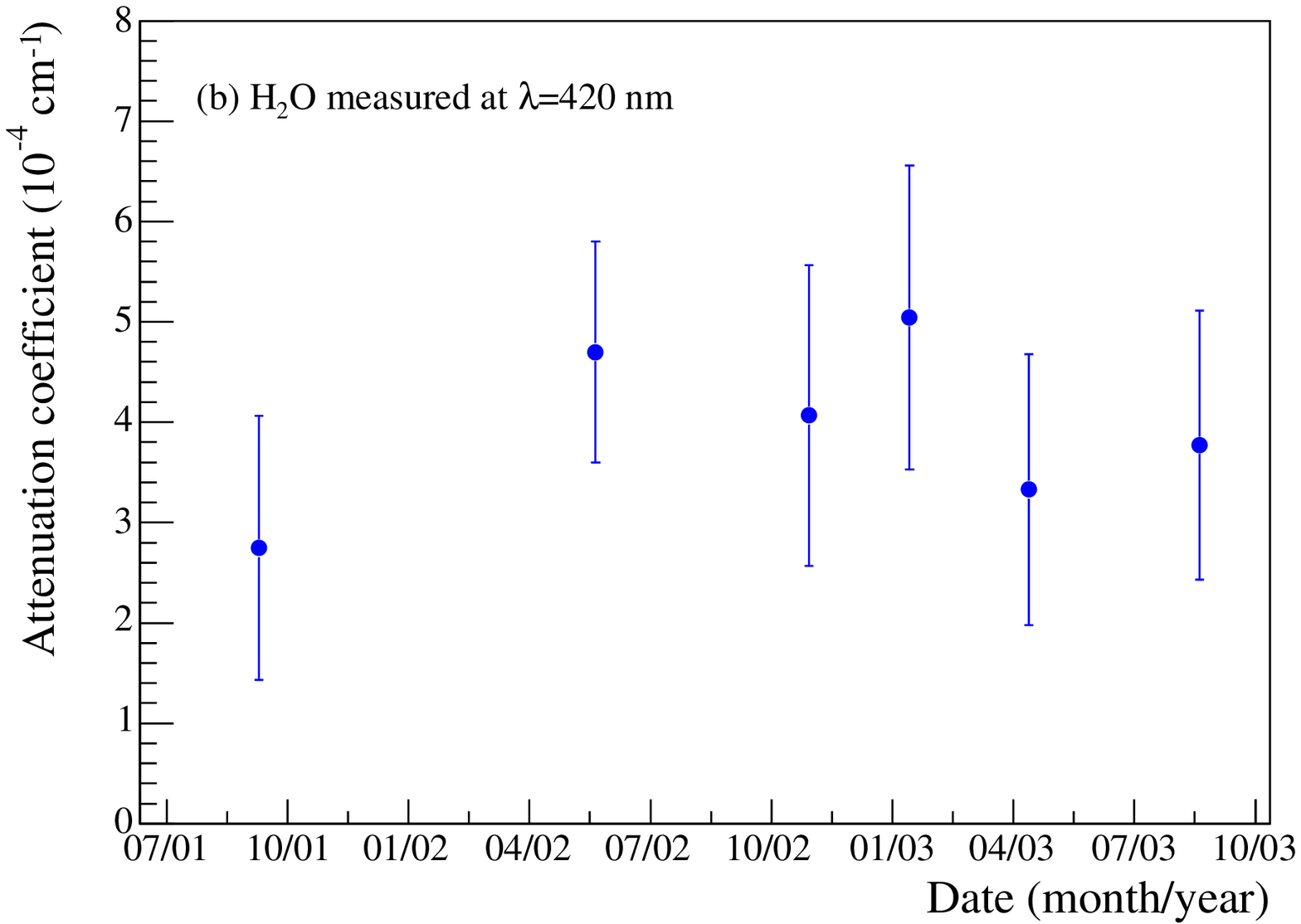}
\caption{\label{fig:d2o_420} Attenuation coefficients of (a) \dto and
(b) H$_2$O as a function of date for 369 and 420~nm pulsed laser
scans.  The data points are correlated by systematics that are common
to each of the measurements.  Note that the \hto values are determined
by subtracting the acrylic vessel {\em ex-situ} measured attenuation
coefficients from the measured H$_2$O+acrylic values.}
\end{center}
\vspace{-4ex}
\end{figure}

Figure~\ref{fig:d2o_420} shows sample \dto and H$_2$O attenuation
coefficients, measured at 369 nm and 420 nm respectively, for six
scans taken during the salt phase.  The H$_2$O coefficients are
constant within the accuracy of the measurements, while the \dto
values exhibit a steady increase until late in the salt phase.  For
both MC simulation and energy reconstruction, the \dto attenuation
coefficients are determined, based on the date at which the given run
was taken, from a linear fit to the measured attenuation coefficients
as a function of time.

Chemical assays of the \dto indicate that the change in attenuation
can be attributed to trace levels of contaminants.  In particular, the
possible presence of organic complexes in the salt phase, as well as
measured increases in Mn contamination, likely associated with the
\mnox assays discussed below, are correlated with the increase in the
attenuation coefficients.  As described below, the changing
response was independently measured by the laser and the \ns sources,
and was corrected for in the data processing.  Following
desalination by reverse osmosis and water purification after
completion of the salt phase, the attenuation levels returned to those
measured in June 2000 prior to salt deployment.

\begin{figure}
\begin{center}
\includegraphics[width=3.73in]{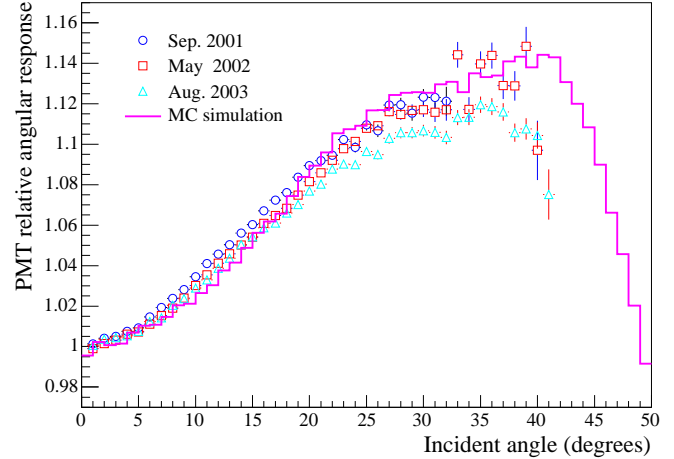}
\caption{\label{fig:angres386} Measured angular response curves for
PMT-reflector assemblies from three optical scans taken at 385 nm
during the salt phase of operation.  The incident angle for photons
originating from the fiducial volume ($\rho<0.77$) is confined to less
than 35 degrees.  The average response from all the laser
scans is used as the input to the MC simulation. Note that the y axis zero
is suppressed.}
\end{center}
\vspace{-4ex}
\end{figure}

In addition to the attenuation coefficients, the relative average
response of the PMT-reflector assemblies as a function of photon
incidence angle is measured.  A typical angular response distribution,
given in Fig.~\ref{fig:angres386}, presents angular response
measurements for 385-nm light from three laser scans taken during the
salt phase.  The response is normalized to unity at normal incidence
angle and, as can be seen, the response increases by $\sim12\%$ at
35$^\circ$.  The angular response curves are reasonably constant over
the salt running period and the average of the response functions from
all laser scans is used as input to the MC simulation and the energy
reconstruction.

\subsection{Energy Calibration}

Once the optical calibration constants are determined, the energy
calibration and systematic uncertainty evaluations are carried out.

The simplest energy estimator in SNO is the number of PMTs that trigger
in an event (\nhits).  However, for a given event energy, the corresponding mean
number of triggered PMTs varies with event position and direction
owing primarily to the effects of the D$_2$O, AV, and \hto attenuation
coefficients and the varying PMT angular response.

\begin{figure}
\begin{center}
\includegraphics[width=3.5in]{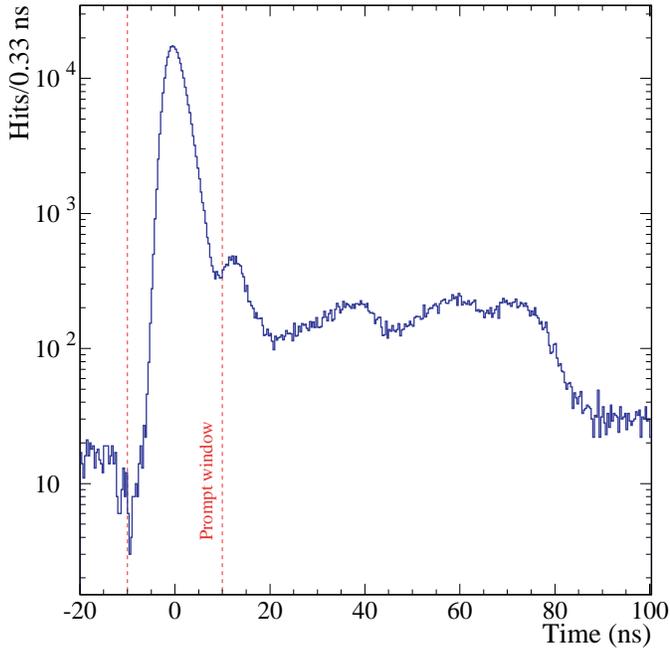}
\caption{\label{fig:timeres} PMT time-residual spectrum for a \ns run
taken with the source positioned at the center of the detector.  The
time residual is calculated without any walk correction to the PMT hit
time.  Only hit PMTs with time residuals within the 20-ns ``prompt
window" centered at the ``prompt peak" are used to estimate the energy
of an event.}
\end{center}
\vspace{-4ex}
\end{figure}

The late light, described above, is difficult to accurately model.
In order to minimize systematic uncertainties associated with the late
light, only the number of PMTs that  fire within the ``prompt"
range of $\pm10$ ns centered on the time-residual peak is used to
estimate the event energy.  Figure~\ref{fig:timeres}, produced from
\ns events generated at the center of the detector, illustrates the
shape of a typical time-residual distribution for \Ckv light.

The number of prompt PMT hits ($N_{\rm prompt}$) is corrected for
noise ($N_{\rm noise}$), optical response ($\epsilon_{\rm response}$)
relative to response at the detector center ($\epsilon_{\rm 0}$),
and for the fraction of working PMTs at the time of the event ($
PMT_{\rm working}$) compared to the total number of PMTs ($PMT_{\rm
total}=9456$) to produce the corrected variable $N_{\rm corrected}$:
\begin{equation}\label{eq:ncor}
N_{\rm corrected} = \frac{(N_{\rm prompt}-N_{\rm
noise})}{\epsilon_{\rm response}/\epsilon_{\rm 0}} \frac{PMT_{\rm total}} {PMT_{\rm
working}}.
\end{equation}
This is the effective number of prompt PMT hits that would have fired
in an ideal detector with the event vertex at the center of the \dto
volume.  During the salt phase of operation, the number of working
PMTs was generally between 8800 and 8600.  The correction for PMT dark
noise, measured typically to be $\sim$0.1~hits/event, is small when
compared to the average response for 5.0-MeV kinetic energy electrons
of $\sim$35 corrected hits/event.

The optical response function corrects for the relative effects of
path length through \dto, AV, and \hto and for incidence angle onto
the PMT-reflector assembly compared to an event at the center of the
detector.  The response function is given by
\begin{eqnarray}
\epsilon_{\rm response} & = &
\sum_{\theta^{\prime}}\sum_{\phi^{\prime}}\sum_{\lambda}
\frac{\epsilon_{PMT}(\lambda)}{\lambda^{2}}
P(r,\theta,\theta^{\prime},\phi^{\prime},\lambda)\\ && \times
M(r,\theta^{\prime},\phi^{\prime}) g(\theta^{\prime},\phi^{\prime})
e^{-\mu_{1}d_{1}} e^{-\mu_{2}d_{2}} e^{-\mu_{3}d_{3}} \nonumber
\end{eqnarray}
where the sums are over 10 polar ($\theta^{\prime}$) and 10 azimuthal
($\phi^{\prime}$) angle bins relative to the reconstructed event
vertex and direction ($\theta^{\prime}$=0), and wavelengths $\lambda$
in a range (220-710 nm) that encompasses the wavelengths to which the
detector is sensitive.  $\epsilon_{PMT}(\lambda)$ is the average
wavelength response of the PMT-reflector assembly,
$P(r,\theta,\theta^{\prime},\phi^{\prime},\lambda)$ represents the
angular response function, $g(\theta^{\prime},\phi^{\prime})$ the \Ckv
light weighting distribution, $M(r,\theta^{\prime},\phi^{\prime})$ a
correction for multiple photon hits in the PMTs, and
$e^{-\mu_{i}d_{i}}$ are $\lambda$ dependent attenuation factors for
the three media ($1\equiv$~\dto, $2\equiv\mbox{AV}$, $3\equiv$~\hto). 

The next step in the energy calibration is to translate $N_{\rm
corrected}$ into an energy.  A combination of data and MC simulated
high-rate ($\sim200$ Hz) 6.13-MeV $\gamma$-ray events from the \ns
source and MC simulated electrons are utilized for this purpose.

\begin{figure}
\begin{center}
\includegraphics[width=3.73in]{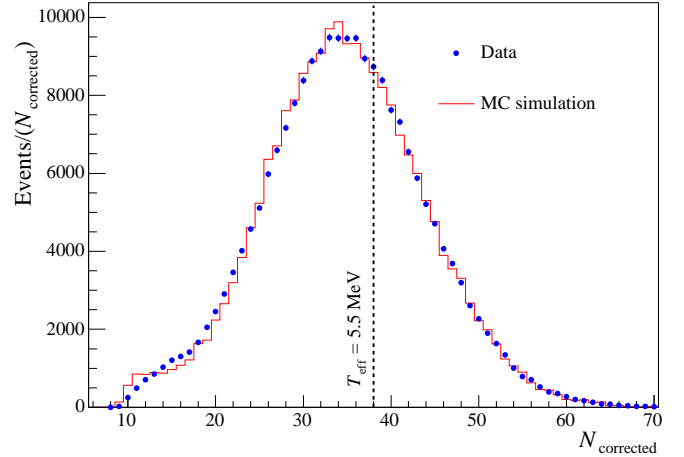}
\caption{\label{fig:n16ncor}  $N_{\rm corrected}$ distributions for data
(points) and MC (histogram) for a \ns run taken with the source
positioned at the center of the detector. The dashed line corresponds
to the energy threshold cut of $\teff=5.5$~MeV.}
\end{center}
\vspace{-4ex}
\end{figure}

First, MC simulated \ns events are generated with the source at the
center of the detector.  The optical constants described above are
used as inputs to the simulation, leaving an overall global PMT
efficiency factor free which is tuned by matching the mean of the MC
simulated $N_{\rm corrected}$ distribution to the corresponding
distribution for \ns data, as shown in Fig.~\ref{fig:n16ncor}.  This
sets the MC simulated $N_{\rm corrected}$ scale to that of the SNO
detector.

With the global efficiency set, electron MC events are then generated
at a series of fixed energies to produce a ``look-up" table that
translates $N_{\rm corrected}$ into equivalent electron kinetic energy $\teff$.

\subsection{Systematic Uncertainties}

Among the most important systematic uncertainties on the measured
fluxes and energy spectra are the detector energy scale and energy
resolution uncertainties.  Near the analysis energy threshold of
$\teff=5.5$~MeV, these uncertainties are extracted from comparisons
between \ns data and MC simulations. \ns data are recorded at regular
intervals throughout the salt phase.  This consists of approximately
monthly deployments at the center of the vessel and periodic scans
throughout the $x$-$z$ and the $y$-$z$ planes in the \dto volume.
Determinations of energy scale and energy resolution systematics are
tested and extended with analyses of $^8$Li source data taken during
the salt phase and pT source measurements taken during the \dto phase.

The factor $PMT_{\rm total}/PMT_{\rm working}$ in equation
\ref{eq:ncor} directly affects the energy scale and requires accurate
identification of PMTs exhibiting normal working behavior.  Normal PMT
operation is assessed from analysis of PMT charge and timing
distributions from high-statistics laser calibration runs.  Analysis
of PMT occupancy distributions from \ns data is used evaluate the
average fraction of improperly functioning PMTs missed in analysis of
the laser data.  The estimated energy scale uncertainty associated
with misidentification of improperly working PMTs is conservatively
estimated to be 0.20\%

The energy scale response of the detector is sensitive to the
electronic threshold and PMT gain.  Cross-talk measurements and the
shape of PMT charge distributions are both sensitive to such changes
and are evaluated with \ns calibration data.  Special \ns runs were
taken during the salt phase in which the PMT high voltages were
varied, and separately with the data acquisition threshold settings
varied, to simulate the effects of gain and threshold changes on the
charge and cross-talk distributions.  Comparison to the standard set
of \ns data places limits on possible threshold and gain effects on
the energy scale of 0.20\% and 0.40\% respectively.

Differences in the detector response as a function of event rate can
directly affect comparisons between \ns source runs (high-rate) and
neutrino data (low-rate).  \ns runs, with the source event rate tuned
to simulate the neutrino mode rate, are recorded at approximately
monthly intervals.  Rate effects are evaluated by comparing low-rate
and high-rate \ns data and are estimated to be less than 0.1\%.
Additional energy scale uncertainties associated with timing
resolution are evaluated to be less than 0.1\%.

The data span approximately two years of detector operation. To
accurately extract the integral flux and day-night energy spectra, it
is critical to evaluate and model the time-dependence of the detector
state.  As indicated above, the \dto attenuation coefficients
increased slowly throughout most of the salt phase.

\begin{figure}
\begin{center}
\includegraphics[width=3.73in]{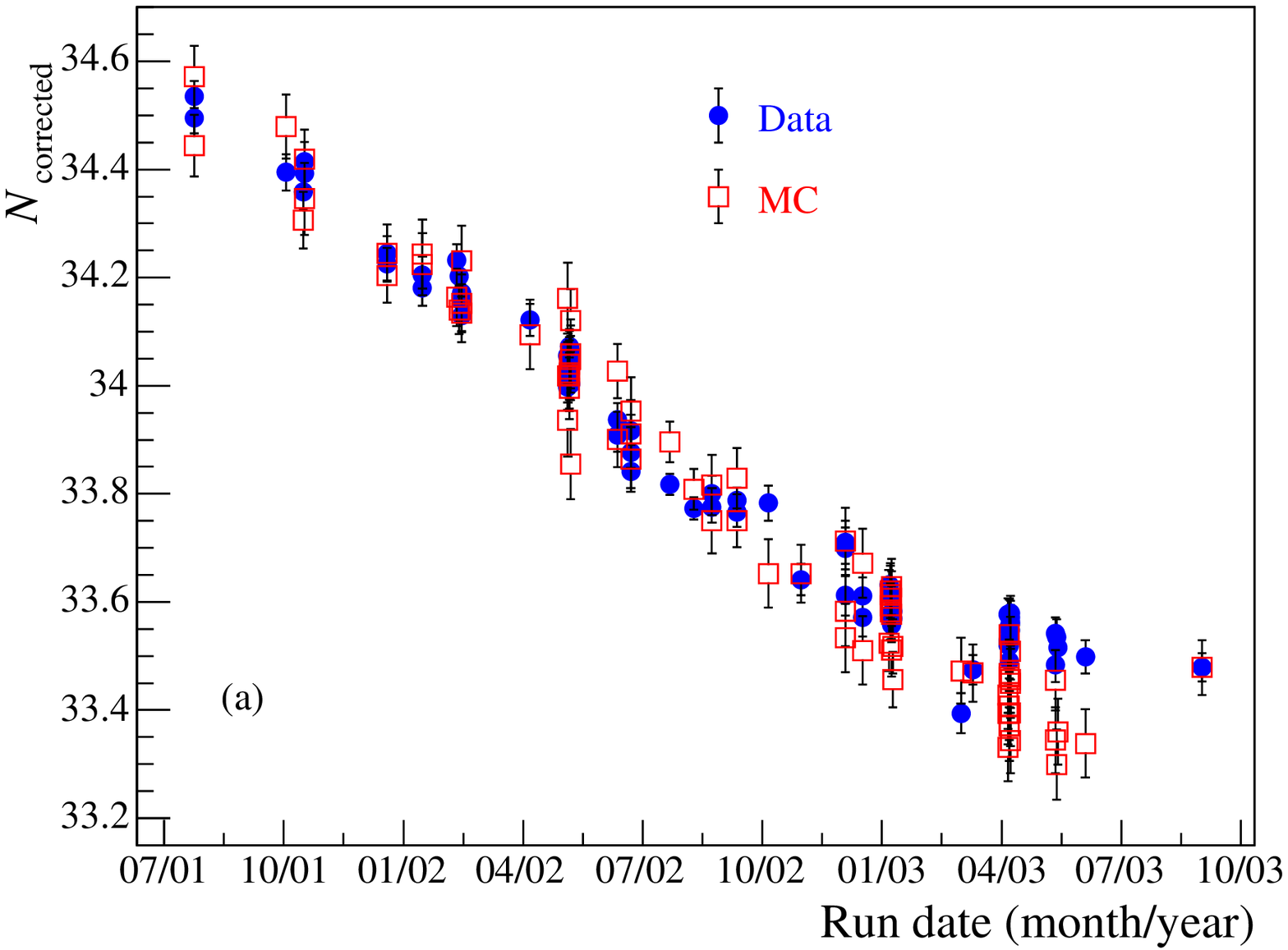}
\includegraphics[width=3.73in]{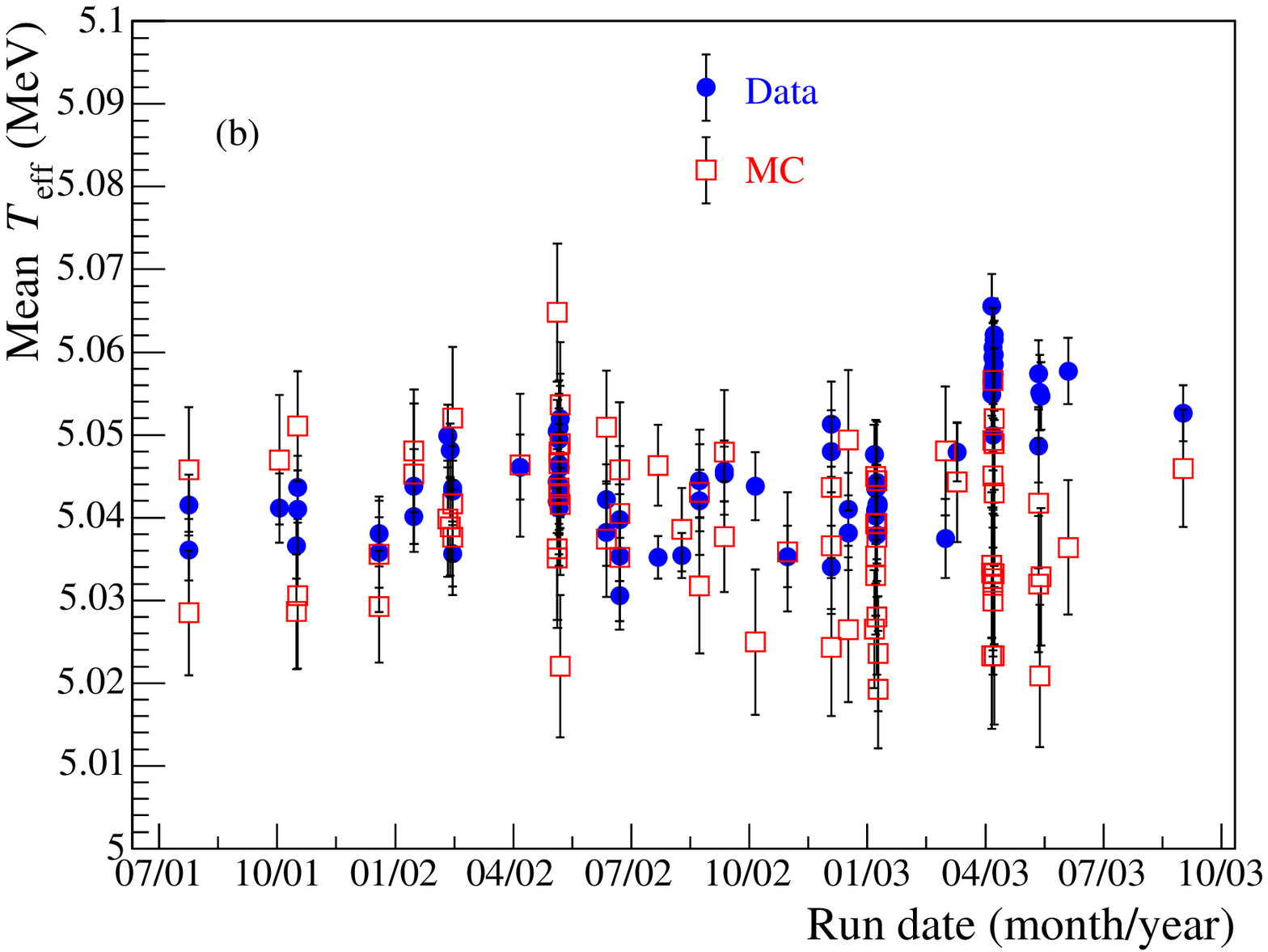}
\caption{\label{fig:ncor} (a) Mean $N_{\rm corrected}$ and (b) mean
$\teff$ versus date for data and MC high-rate \ns calibrations
runs with the source at the center of the detector.  Note that the
full $y$-axis ranges are $\sim4\%$ and $\sim2\%$ of the average
$N_{\rm corrected}$ and $\teff$ values respectively.  Error bars are
statistical only, and the spread of the variation between data and
Monte Carlo provides the measure of the energy scale uncertainty
arising from temporal variations in detector response.  }
\end{center}
\vspace{-4ex}
\end{figure}

For each \ns data and MC simulated run, estimates of the mean and
width of the $N_{\rm corrected}$ and energy distributions are
generated by fitting a Gaussian function to the central portion of the
spectra.  Figure~\ref{fig:ncor} shows the distributions of mean
$N_{\rm corrected}$ and reconstructed energy as functions of date for
\ns runs with the source positioned at the center of the detector.
The MC simulated runs have been generated with \dto attenuation
coefficients increasing as shown in Fig.  \ref{fig:d2o_420}.  As is
seen, the decreasing MC simulated response in $N_{\rm corrected}$
matches the slope observed in the \ns data.

During the final period of the salt phase, no \mnox assay
(Sec.~\ref{sec:backgrounds}) was taken and measurements indicate that
the detector energy response also ceased to change.  The mean energy
distributions shown in Fig.~\ref{fig:ncor}(b) have been corrected for
the changing attenuation. The run-by-run differences between data and
MC simulation mean and width estimates are taken as the temporal
stability systematic uncertainties on energy scale and resolution
respectively.

The energy scale uncertainty arising from the temporal stability
evaluation is 0.15\%.  The energy resolution stability uncertainty is
determined to be 1.8\% and is dominated by an average offset between
the data and the MC simulation.  This offset is attributed to a
combination of effects from electronic cross-talk and from
tube-to-tube variation in PMT efficiencies not modeled in the MC
simulation.

Modeling the detector and the calibration sources involves a
variety of simplifications and uncertainties.  For the \ns source,
uncertainties in the \ns decay branching ratios, minor differences
between pure \dto and the salt brine in the production and propagation
of \Ckv light, exact details in describing the source geometries, the
finite step size in the EGS4 simulation, neglecting the minor velocity
dependencies in the wavelength spectra of \Ckv light, and the wavelength
dependence of the index of refraction are estimated, in total, to contribute a
0.65\% systematic uncertainty to the energy scale.

\begin{figure}
\begin{center}
\includegraphics[width=3.73in]{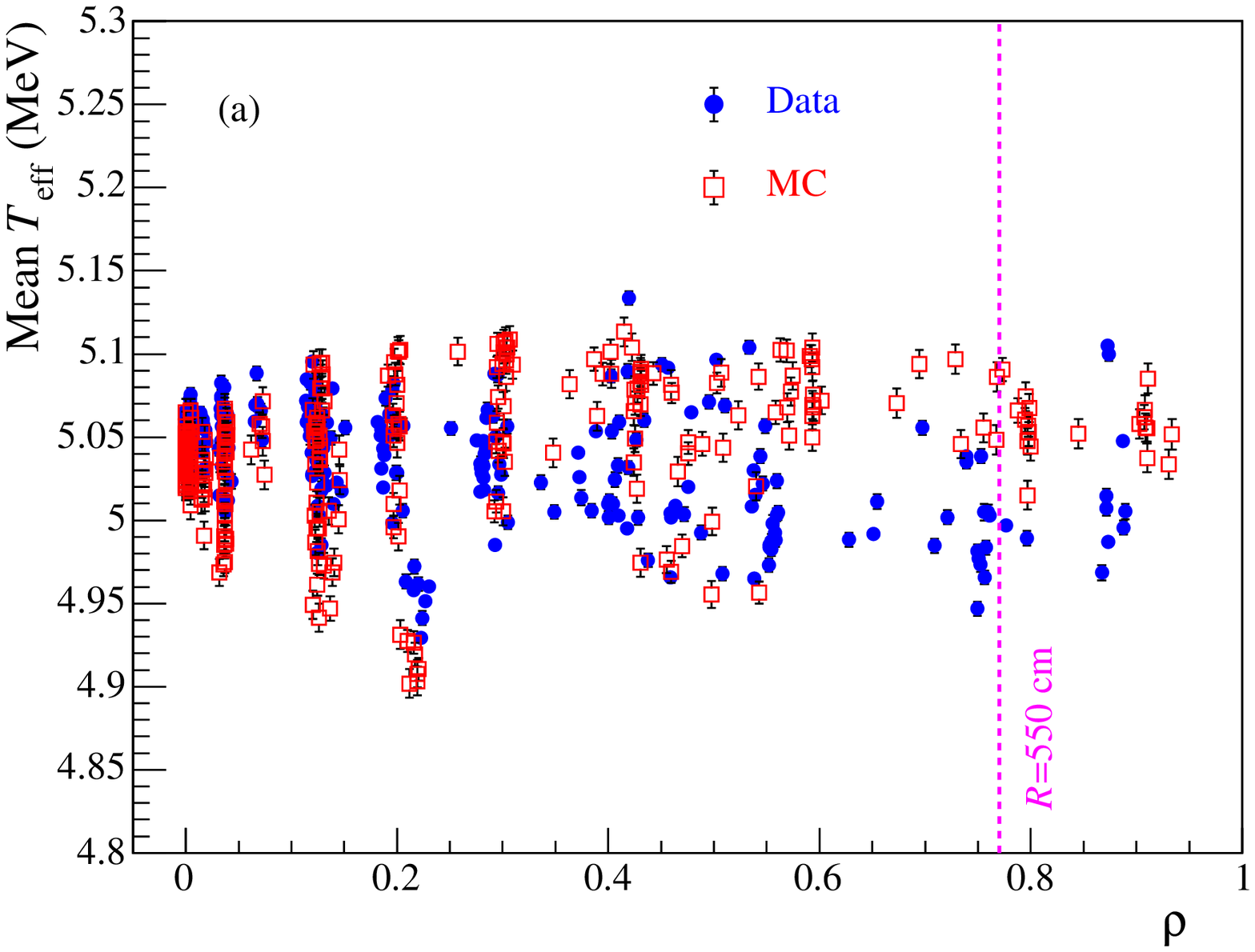}
\includegraphics[width=3.73in]{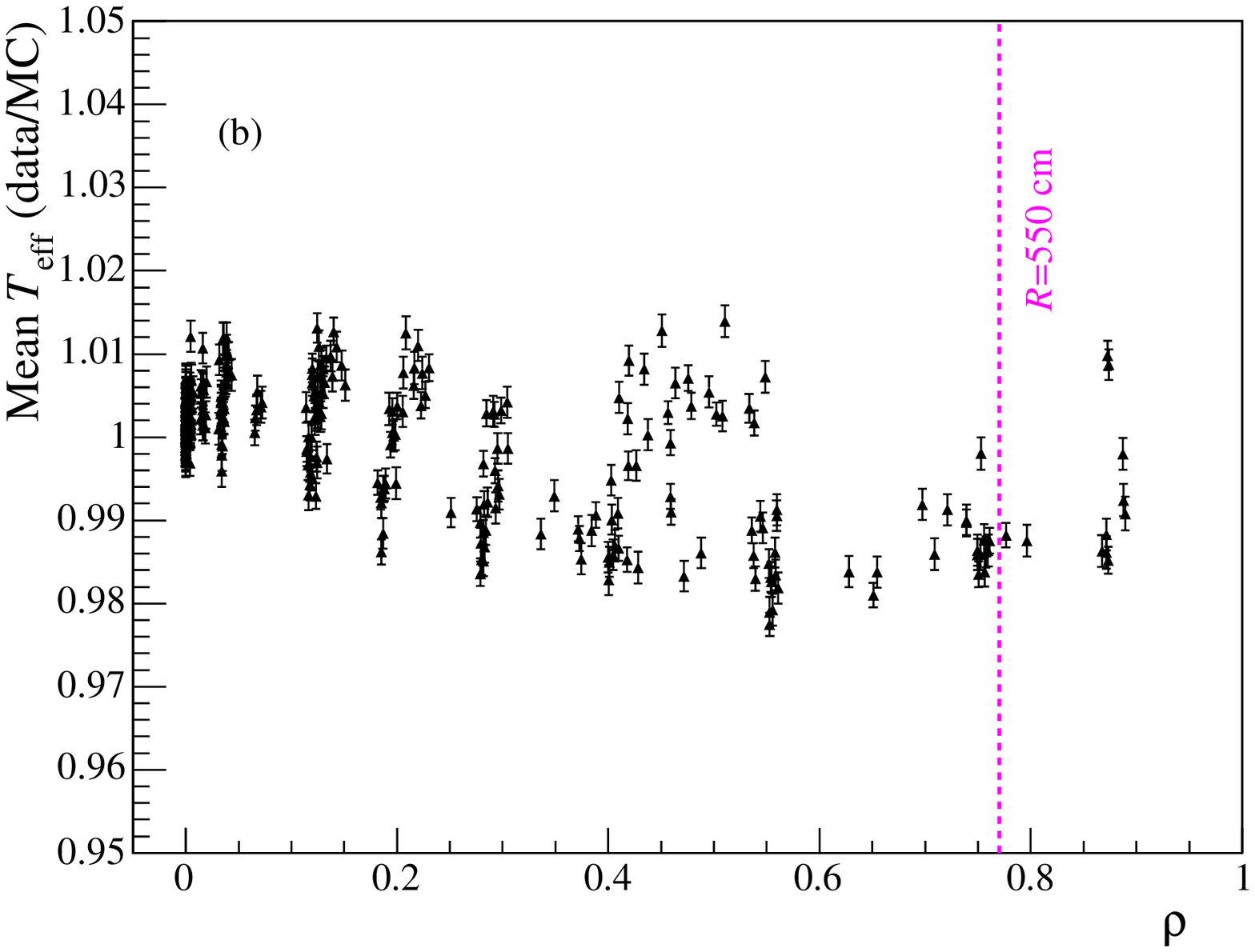}
\caption{\label{fig:n16rad} (a) Data and MC mean energy versus $\rho$
distributions and (b) the run-by-run ratio of data to MC mean energy
versus $\rho$ are shown for \ns calibration runs.  The dashed
vertical line corresponds to the fiducial volume cut at $R=550$ cm.
Points at the same value of $\rho$ can have differing energy response
in data or Monte Carlo due to local point-to-point non-uniformities in
detector response.}
\end{center}
\vspace{-4ex}
\end{figure}

Significant contributions to the energy scale uncertainty arise from
evaluations of the radial response and detector asymmetries.
Figure~\ref{fig:n16rad} shows data and MC simulation mean energy
values and their ratio versus source radial position $\rho$.  The
volume-weighted mean difference between data and MC simulation of
0.45\% is taken as the radial response energy scale uncertainty contribution.

The detector asymmetry component means point-to-point non-uniformities
in detector response due to asymmetric detector features such as
support ropes and the neck of the AV.  It is evaluated by determining
the volume-weighted average standard deviation in each of seven radial
bins of the data/MC simulation $\teff$ ratio distribution and is
calculated to be 0.59\%.  The corresponding energy resolution radial
and detector asymmetry systematic uncertainties are evaluated in an
analogous fashion and found to be 1.4\% and 0.78\% respectively.

These estimates of energy scale and resolution uncertainties are
obtained near the detector threshold.  The cross-talk and multiphoton
effects of higher energy events are probed with high-intensity pulsed
laser data.  In addition, the pT source, which generates 19.8 MeV
$\gamma$-rays, enables a direct test of the higher energy scale
systematic uncertainties.  However, high-rate of neutrons emanating
from the pT source precluded its deployment during the salt phase of
the experiment.  Comparison of the pure \dto and salt laser runs
indicate no additional unmodeled effects and supports the application
of the pT data to the salt data set.  It is found that the energy
scale uncertainty, evaluated from pT data, is not greater than that
evaluated with the \ns source.  The effects of cross-talk and noise on
the energy scale are estimated to be $<0.25$\%.  The energy resolution
uncertainty was determined from pT data to be 10\% at 19.8 MeV.  Hence
the energy resolution uncertainty is applied as the function
\begin{eqnarray}
\teff<\mbox{5 MeV}&:&\Delta\sigma_{T} = 3.4\%\\ \teff >\mbox{5 MeV}&:&
\Delta\sigma_{T} = [3.4 +0.478(\teff-5)]\%.
\end{eqnarray}
The energy response for electrons was characterized as a Gaussian
function with resolution $\sigma_T = -0.131+0.383\sqrt{T_e}+0.03731
T_e$, where $T_e$ is the true electron kinetic energy in MeV.  Table
\ref{tab:esys_summary} summarizes all contributions to energy scale
and resolution systematic uncertainties.  Energy scale contributions
are added in quadrature giving a total uncertainty of 1.15\%.  The
radial and detector-asymmetry energy resolution components are added
together in quadrature and then added linearly to the data-MC offset
to produce a 3.4\% total uncertainty.

\begin{table}[H]
\caption{\label{tab:esys_summary}Summary of energy scale and resolution systematic
uncertainties.}
\begin{center}
 \begin{tabular}{ld}
\hline
\hline
\multicolumn{2}{c}{Scale uncertainty}\\
\hline
Source & \multicolumn{1}{c}{Uncertainty} \\
\hline
Detector PMT status               & 0.20\%  \\
Electronics threshold                    & 0.20\%  \\
Electronics gain                         & 0.40\%  \\
Electronics rate effects              & 0.10\%   \\
Time calibration             & 0.10\%   \\
 Time drift/stability: data-MC    & 0.15\%   \\
Radial distribution: data-MC& 0.45\%  \\
Detector asymmetry           & 0.59\% \\
\ns source modeling                  & 0.65\%  \\
Cross-talk/pickup non-linearity & 0.25\% \\
\hline
Total                        & 1.15\%   \\
\hline
\multicolumn{2}{c}{Resolution uncertainty}\\
\hline
Source &  \multicolumn{1}{c}{Uncertainty} \\
\hline
Central \ns runs:  data-MC & 1.8\% \\
Detector asymmetry              & 1.4\% \\
Radial dependence                 & 0.8\%  \\
\hline
Total                  & 3.4\% \\
\hline
\hline
\end{tabular}
\end{center}
\end{table}

\section{\label{sec:recon}Event Reconstruction}

For each event, PMT trigger times and positions are used to reconstruct
the event vertex, direction, and isotropy. The following sections outline the
algorithms and the determination of systematic uncertainties
associated with these reconstructed variables.

\subsection{Event Vertex and Direction}

Event positions and directions are reconstructed by analysis of
the times and positions of triggered PMTs in each
event.  The time-of-flight corrected PMT trigger
time is used to define the PMT time-residual 
\begin{equation}
T^{\rm{res}}_{i} = t_{i}-t_{\rm fit} - \frac{|\vec{r}_{\rm
fit}-\vec{r}_{i}|}{u_{\rm{eff}}}
\end{equation}
where $t_{i}$ and $\vec{r}_{i}$ are the trigger time and the position
of the $i^{th}$ PMT in the event, and $t_{\rm fit}$ and $\vec{r}_{\rm
fit}$ are the fit time and the reconstructed position of the
event vertex.  The effective photon velocity, $u_{\rm eff}= 21.87$
cm/ns, is the group velocity of the mean detected photon wavelength at
380 nm in D$_2$O.

For each event a likelihood is constructed from the PMT time-residual
probabilities where MC simulated events are used to derive the
reference PDF.  The PDF is approximated as a constant for
time-residual greater than 15~ns because the time peaks associated
with reflected photons strongly depend on the event location.  A time
cut of $\pm$50~ns relative to the median PMT hit time is imposed to
reduce the effects of reflected photons and PMT dark noise.  Seed
vertices are chosen randomly from the detector volume and the negative
log likelihood function is minimized with respect to $t_{\rm fit}$ and
$\vec{r}_{\rm fit}$ until a global minimum is found.

Event direction is reconstructed independently after the best-fit
vertex has been found.  It is estimated based on the assumption that
the events produce \Ckv light emitted in the characteristic cone
shaped pattern.  A likelihood is constructed based on the distribution
of directions from the reconstructed event position to the triggered
PMTs relative to the corresponding MC calculated distribution for \Ckv
light events.  A ``prompt" time cut ($\pm$10~ns) on the time-residual
distribution excludes most Rayleigh scattered and reflected photons.
The negative log likelihood is minimized to determine the best fit
direction.  For multiple electron events the reconstructed direction
tends to be weighted towards the direction of the most energetic
electron(s) in the event.

\begin{figure}
\begin{center}
\includegraphics[width=3.73in]{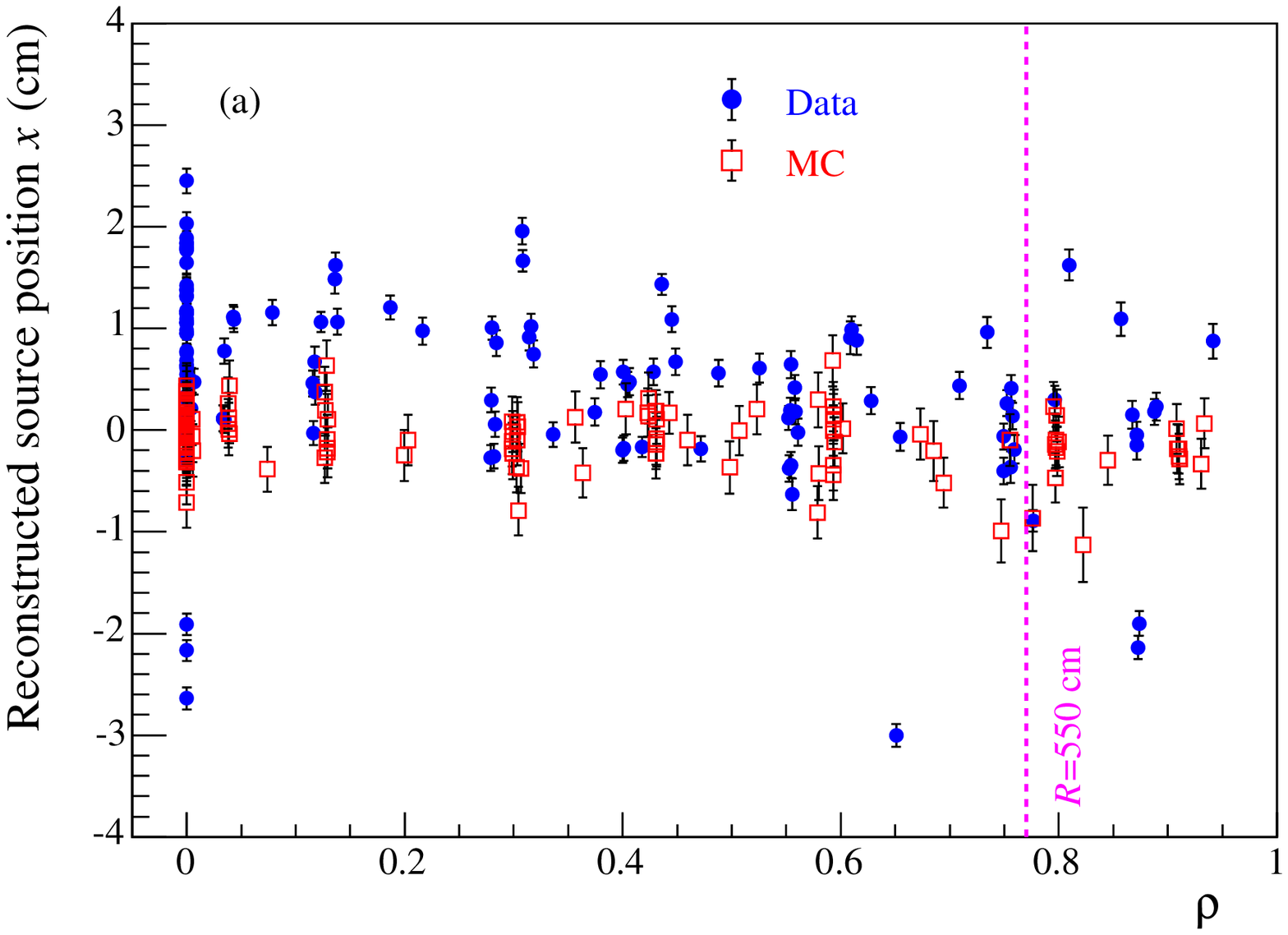}
\includegraphics[width=3.73in]{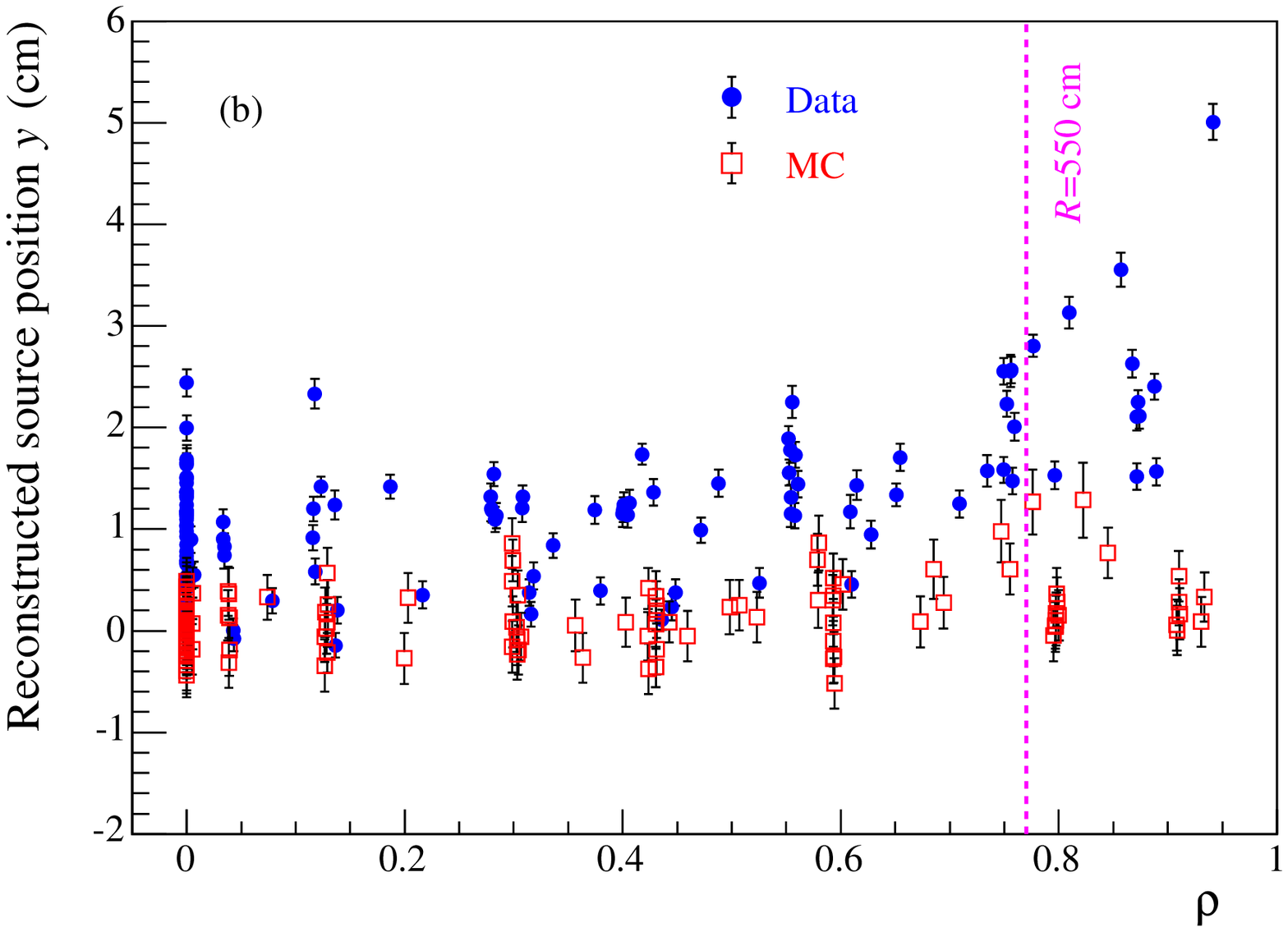}
\includegraphics[width=3.73in]{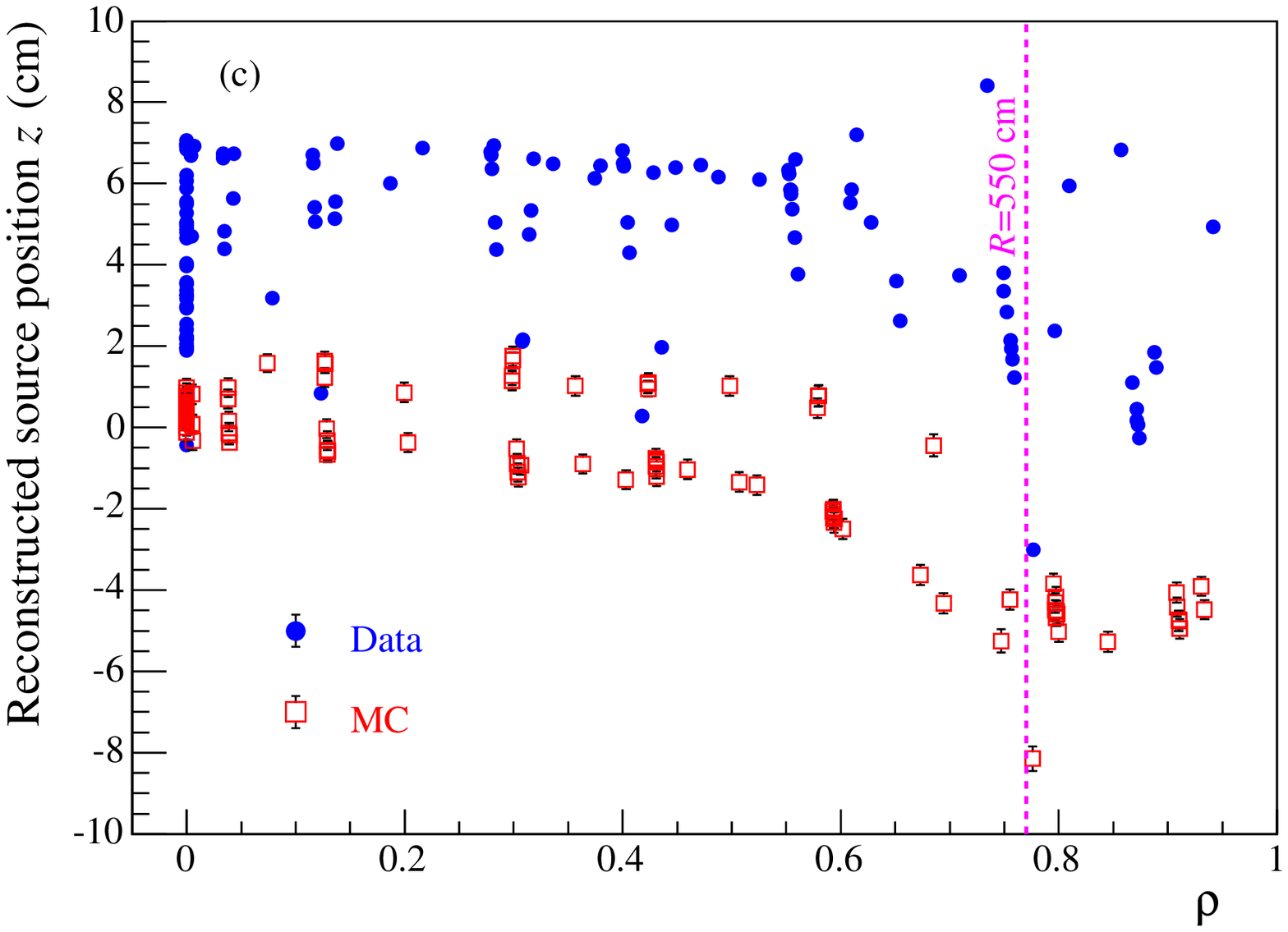}
\caption{\label{fig:vertex} Difference between mean reconstructed (a)
$x$, (b) $y$, and (c) $z$ coordinates and the coordinates of the
deployed \ns source versus $\rho$.  The \ns source was deployed along
the central ($z$) axis for the data shown in this figure.  From these
data, systematic uncertainties on $x$ and $y$ coordinate vertex
reconstruction are evaluated to be 2~cm and the uncertainty on $z$
coordinate vertex reconstruction is evaluated to be 6~cm.}
\end{center}
\vspace{-4ex}
\end{figure}

Vertex reconstruction uncertainties are evaluated by comparing average
reconstructed event positions of ${}^{16}$N calibration data with
${}^{16}$N MC simulations and by comparing the average reconstructed
source position to manipulator estimated position for \ns calibration
data.  The manipulator source position measurement is most accurate
when operating in single axis mode along the $z$-axis of the detector.
Figure~\ref{fig:vertex} shows the difference between mean
reconstructed $x$, $y$, and $z$ positions and manipulator estimated
source position as functions of $\rho$.  These figures indicate that
the $x$ and $y$ vertex reconstruction uncertainties are not more than
2 cm.  A larger difference in $z$ is apparent and the $z$ vertex
reconstruction uncertainty is taken to be 6 cm.

In addition to coordinate shifts, fiducial volume uncertainty is also
evaluated.  Radial scaling bias could occur through reconstruction
biases or timing calibration uncertainty.  Reconstructed radial
distributions of events from calibration and neutrino signal data near
the AV are compared to MC simulations to evaluate this uncertainty.
The radial uncertainty is estimated to be 1.0\% of the radius
(i.e., 5.5 cm at 550 cm).

\begin{figure}
\begin{center}
\includegraphics[width=3.73in]{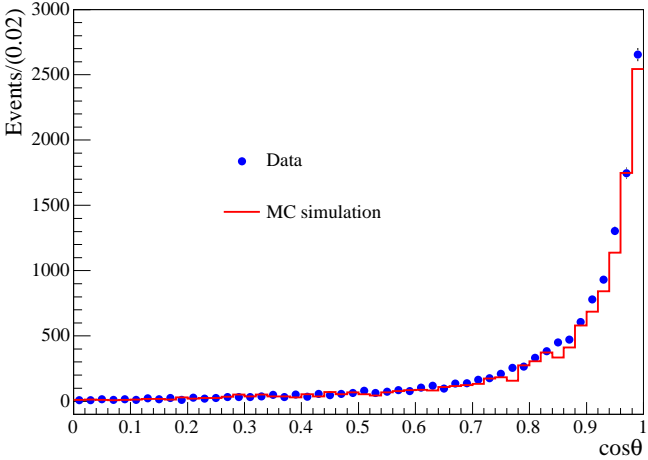}
\caption{\label{fig:angres} Data and MC distributions of the cosine of
the angle between generated event direction and reconstructed
direction for a \ns source located at the center of the detector are
shown.  The events are selected such that the reconstructed vertex is
at least 120 cm away from the source position.}
\end{center}
\vspace{-4ex}
\end{figure}

Angular response uncertainty is determined from analysis of \ns data
and \ns MC simulated events.  The $\gamma$-rays produced by the \ns
source travel an average of 30 cm in D$_2$O before a Compton scatter
occurs.  The vector from the ${}^{16}$N $\gamma$-ray source to the
reconstructed event vertex provides a good estimate of the generated
electron direction, since for events above $\teff=5.0$~MeV the \Ckv
light is dominated by that from Compton electrons that are
forward-peaked. Gammas that travel at least 120 cm from the source
before scattering are employed to ensure the estimate of the event
direction is reliable.  For each data and MC simulated \ns run, the
distribution of the cosine of the angle $\theta$ between the electron
direction and the reconstructed direction is fit with the function
\begin{equation}
R = N [ \mbox{e}^{{\beta_{S}(\cos{\theta}-1)}} + \alpha_{M}
\mbox{e}^{\beta_{M}(\cos{\theta}-1)}]
\label{eq:one}
\end{equation}
where N is the overall normalization, $\beta_{S}$ parametrizes the
distribution for electrons scattered only a small amount, $\beta_{M}$
accounts for those scattered through large angles, and the ratio of
these components is $\alpha_{M}$. Shown in Fig.~\ref{fig:angres} are
sample data and MC simulated distributions of $\cos{\theta}$ for an \ns run at
the center of the detector.
The angular resolution systematic uncertainty, determined by comparing
the average difference between data and MC simulation fit parameters
for all \ns runs taken during the salt phase, is 16\% of $\beta_{S}$.

\subsection{Event Isotropy}

In the context of SNO event analysis and signal extraction, isotropy
refers to the uniformity of the distribution of triggered PMTs on the
PMT array.

For data collected during the SNO pure \dto phase, the CC and ES
signal events produced a single primary electron while the NC events
produced a single 6.25-MeV capture $\gamma$-ray.  For a NC event above
$\teff = 5.0$~ MeV the Cherenkov light is dominated by that from a
single forward-scattered Compton electron, so all signal event types
in the pure \dto phase had similar isotropy distributions.  With salt
added to the SNO detector the characteristic response to neutrons is
multiphoton.  Event isotropy is measured from the spatial distribution
of triggered PMTs and is an effective signal separation tool in this
circumstance.

Several variables constructed to measure isotropy are found to have
comparable separation power between the single electron (CC and ES)
and the neutron (NC) signals.  The variable used, which could be
simply parameterized and facilitate systematic uncertainty
evaluations, is $\beta_{14}\equiv\beta_1+4\beta_4$ where
\begin{equation}
\beta_l = \frac{2}{N(N-1)}\sum_{i=1}^{N-1} \sum_{j=i+1}^N
P_l(\cos\theta_{ij}).
\end{equation}
In this expression $P_l$ is the Legendre polynomial of order $l$,
$\theta_{ij}$ is the angle between triggered PMTs $i$ and $j$ relative
to the reconstructed event vertex, and $N$ is the total number of
triggered PMTs in the event.

Initial comparisons of the isotropy distributions for \ns data and MC
simulated \ns showed that the mean value of \bof\ for \ns data was
$\sim$2.5\% larger than for the \ns MC simulation. This was caused
primarily by approximations used in the description of electron
scattering in the MC simulation.  Electron transport is handled within
the MC simulation by EGS4~\cite{bib:egs} in which elastic collisions
with atomic nuclei are modeled using Moli\`ere's theory of
multiple-scattering as formulated by Bethe~\cite{bib:bethe}.  This
neglects the 
effect of the spin of the electrons which slightly reduces the amount
of large-angle scattering.  The MC simulation therefore overestimates
the amount of \Ckv\ light emitted in the backward hemisphere,
producing a light distribution which is slightly too isotropic, i.e.,
a \bof slightly too small.  There are also other approximations in the
treatment of multiple-scattering in EGS4, that affect the
distribution, but the effect of spin is the most significant.

Including the effect of spin multiplies the Rutherford cross section
by a factor $M(\theta,\beta,Z )$, first calculated by
Mott~\cite{bib:mott}, given by~\cite{bib:mckinley}:
\begin{equation}
M(\theta,\beta,Z) = 1 - \beta^{2}\sin^{2}\frac{\theta}{2} +
\pi\beta\frac{Ze^{2}}{\hbar c}\sin\frac{\theta}{2}(1
-\sin\frac{\theta}{2}),
\end{equation}
where $\theta$ is the single scattering angle, $\beta$ is the electron
speed divided by $c$, and $Z$ is the charge of the scattering
nucleus. In EGS4, an electron is propagated a step length $x$ at which
point its direction is changed by an angle $\varphi$.  This angle is
drawn from a probability density distribution for electrons
multiple-scattered by a screened Coulomb potential without the Mott
terms.  A simple MC program was written to evaluate the probability
density distributions with and without the Mott terms, and correction
constants were generated giving the amount $\Delta\varphi$ to be
subtracted from the EGS4 angle $\varphi$ for a number of different
step lengths and electron kinetic energies. For large angles the
correction is approximately step length independent and close to that
obtained when assuming single scattering dominates, while for small
angles and longer step lengths it is smaller.

This correction to the EGS4 multiple-scattering angles was
parameterized as a function of energy using the average EGS4 step
length for a given energy. The fraction of electrons that scatter into
the forward hemisphere after passing through a layer of water was
compared to that obtained with the updated version of EGS4,
EGSnrc~\cite{bib:egsnrc} which includes the Mott terms as well as
other improvements.  For electrons with a kinetic energy of 5.0~MeV
passing through 1~cm of water, the percentage increase in the fraction
that scatters into the forward hemisphere over that obtained with EGS4
is ~0.9\% when applying the average step length correction, and 1.2\%
using EGSnrc. The correction to EGS4 was tuned to give the same
percentage as EGSnrc.

\begin{figure}
\begin{center}
\includegraphics[width=3.4in]{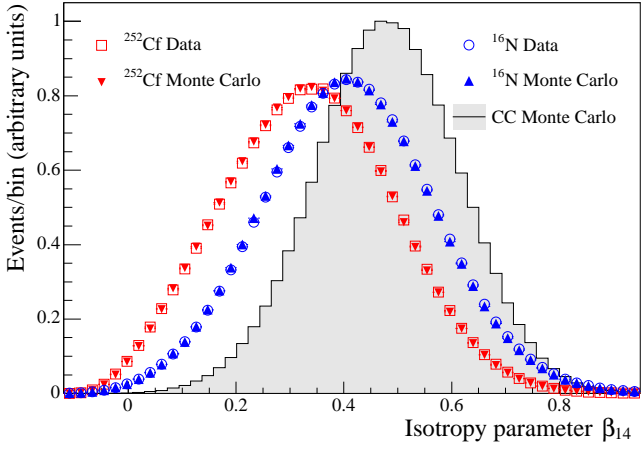}
\caption{\label{fig:isotropy1}
$\beta_{14}$ isotropy distributions for \cf data and MC, \ns data and
  MC, and simulated CC events. }
\end{center}
\vspace{-4ex}
\end{figure}

Figure~\ref{fig:isotropy1} shows distributions of $\beta_{14}$ using
data from \cf and \ns sources and from corresponding MC simulations.
Also shown is a MC simulated distribution for CC events. The \cf and
CC events have an imposed kinetic energy threshold of 5.5~MeV, while
the \ns events have kinetic energies between $\sim4$~MeV and
$\sim6$~MeV. The \ns source emits 6.13-MeV $\gamma$-rays, which
undergo Compton scattering and produce more than one lower energy
electron.  The CC electrons undergo relatively less
multiple-scattering per unit path length than the Compton scattered
electrons from the $^{16}$N source, as the CC electrons have higher
energy.  The average $\beta_{14}$ for \ns events is therefore smaller
(more isotropic) than for CC events.

The \bof parameter is correlated with event energy and
to a lesser extent with radius. Multi-dimensional PDFs are therefore
used in the analysis, as described in the signal extraction section.

\begin{figure}
\begin{center}
\includegraphics[width=3.73in]{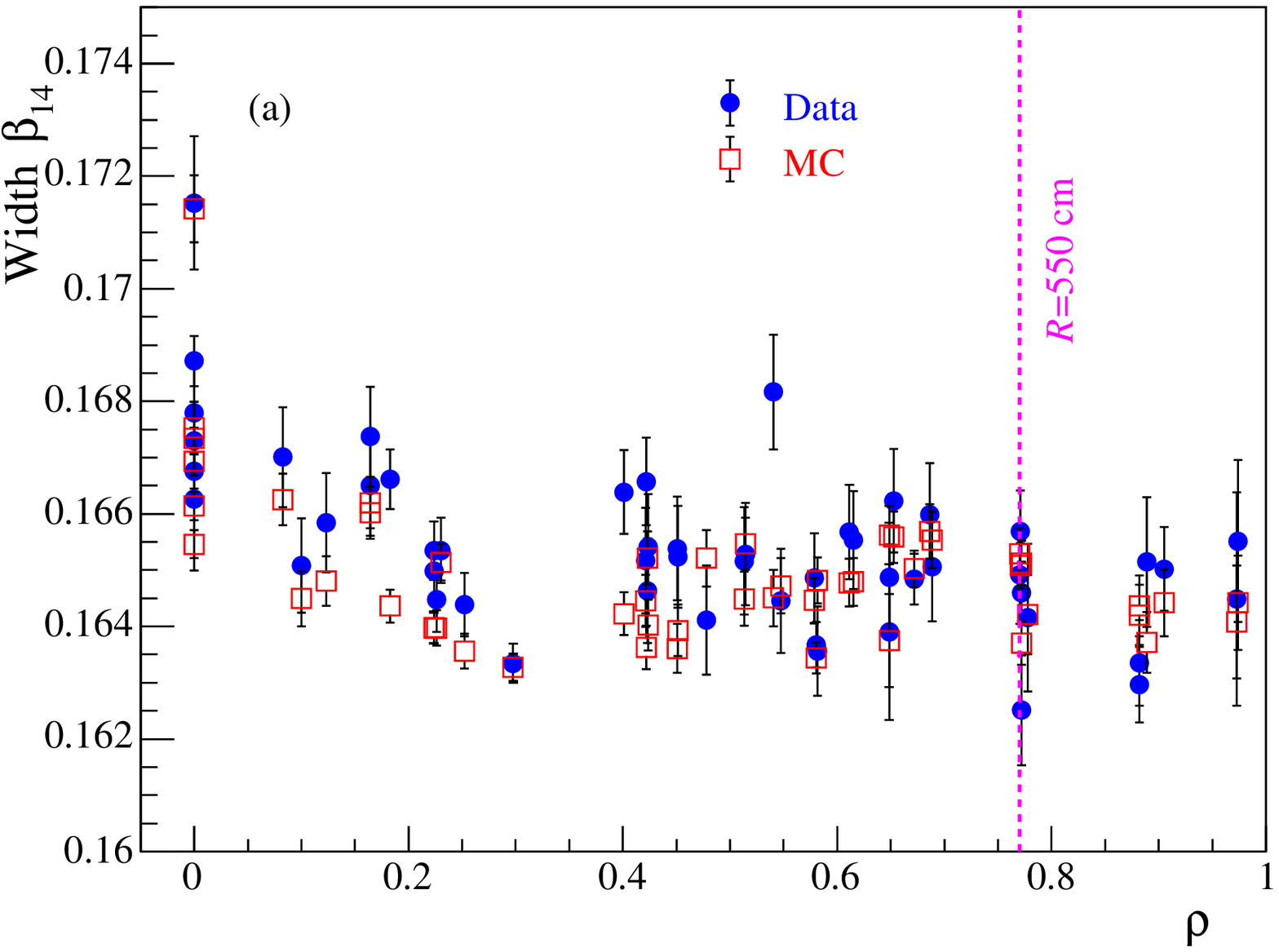}
\includegraphics[width=3.73in]{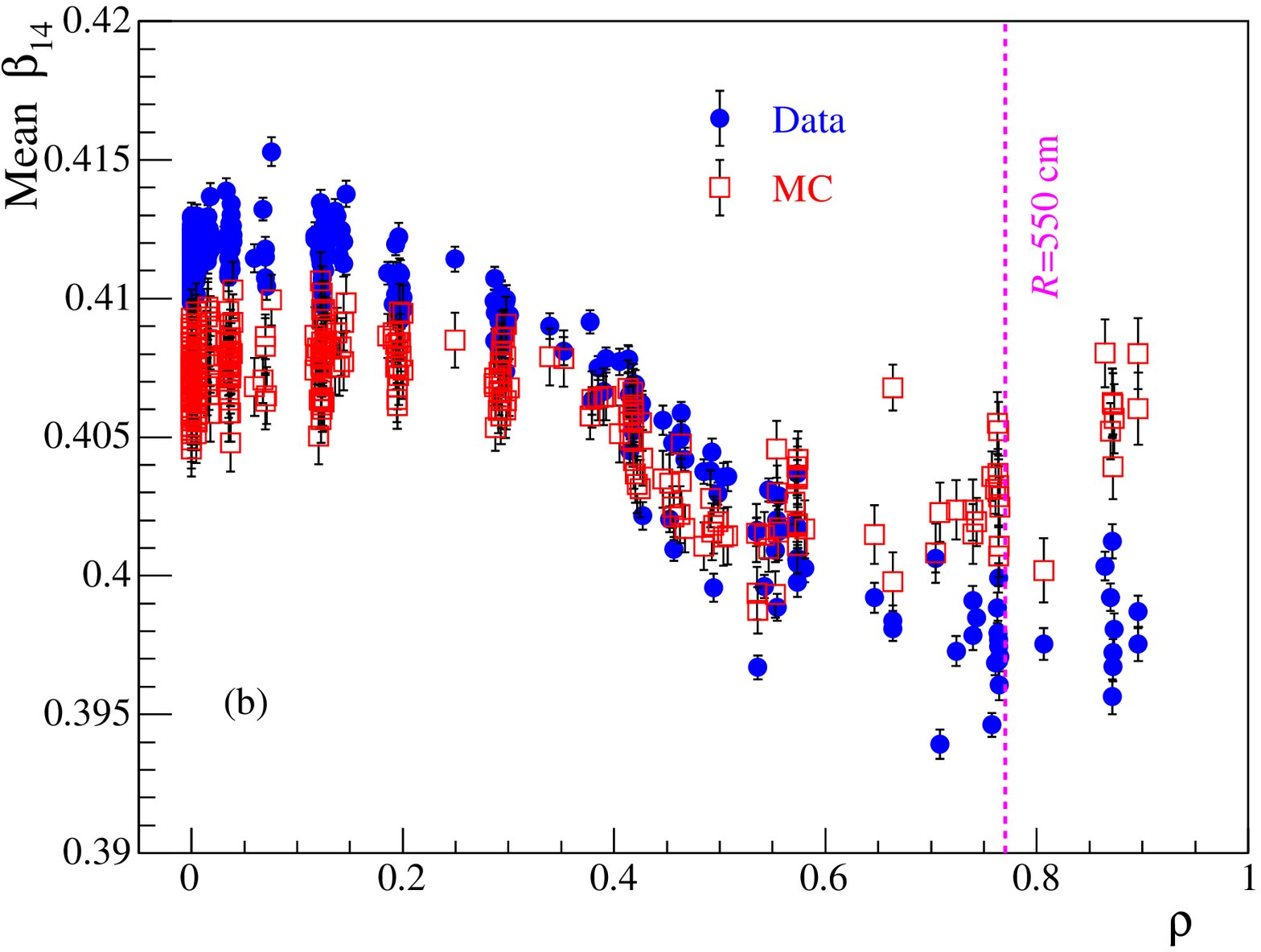}
\caption{\label{fig:isotropy2} $\beta_{14}$ isotropy distributions of
  (a) \cf data and MC width versus $\rho$ and (b) \ns data and MC mean
  versus $\rho$.  Systematic effects observable in the distributions
  are attributed to temporal and spatial non-uniformities in detector
  response and give the estimates on the \bof systematic
  uncertainties. }
\end{center}
\vspace{-4ex}
\end{figure}

Systematic uncertainties for $\beta_{14}$ are evaluated through
data-MC comparisons of \cf and \ns calibration runs.  For
mono-energetic data, $\beta_{14}$ can be well approximated by a
Gaussian function.  Calibration data for \cf and \ns and the
corresponding MC simulated distributions are fit run-by-run.  The
difference in the extracted means and widths of the fits is used to
characterize the uncertainties.  Temporal stability is measured with
runs taken with the sources at the center of the detector.  Radial
uncertainty is determined from multi-axis scans.
Figure~\ref{fig:isotropy2} shows the distributions of $\beta_{14}$
width for \cf data and MC as a function of $\rho$, and $\beta_{14}$
mean for \ns data and MC versus $\rho$.
The average temporal and volume-weighted radial data-MC differences
are evaluated separately and then added in quadrature.
Table~\ref{tab:isotropy} gives the estimated systematic mean and width
$\beta_{14}$ uncertainties from both \cf and \ns sources.  These
uncertainties are propagated through the signal extraction by shifting
the means and smearing the $\beta_{14}$ PDFs.
\begin{table}[H]
\caption{\label{tab:isotropy} Summary of $\beta_{14}$ scale and
resolution systematic uncertainties.}
\begin{center}
 \begin{tabular}{l@{\hspace{0.3in}}d@{\hspace{0.1in}}d}
\hline \hline & \multicolumn{2}{c}{Uncertainty}\\ \hline Source &
\multicolumn{1}{c}{Scale} & \multicolumn{1}{c}{Resolution} \\ \hline
\cf & 0.48\% & 0.67\%\\ \ns & 0.85\% & 0.94\% \\ \hline \hline
\end{tabular}
\end{center}
\end{table}

The energy dependence of the systematic uncertainty on \bof for CC
events was evaluated using $^8$Li calibration data in addition to \ns.

\section{\label{sec:neutron}Neutron Response}

With salt added to the \dto volume, neutron capture is dominated by
capture on \cl and losses due to capture on $^1$H and $^{17}$O are
significantly reduced. For a $^{252}$Cf source at the center of the
detector, the probabilities of neutron capture are 90\%(\cl), 4\%
($^2$H), 2.5\%($^1$H ), with the remaining 3.5\% absorbed by oxygen,
sodium or other isotopes (only 0.3\% capture on $^{37}$Cl).  In \dto
with no salt additive, the capture probabilities are 49\% ($^2$H) and
30\%($^1$H), with 14.5\% absorbed by oxygen isotopes in the heavy
water and the remaining 6.5\% escaping the \dto volume.  For neutrons
generated uniformly in the heavy water, the probability of capture on
\cl in the salt phase is about three times larger than that of capture
on deuterium in the absence of salt.

\begin{figure}
\begin{center}
\includegraphics[width=3.73in]{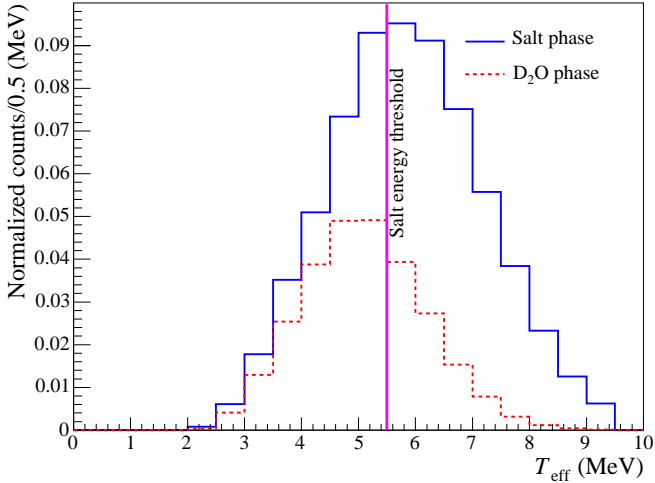}
\caption{\label{fig:cf_energy}
Neutron energy response during the \dto (dashed) and salt (solid)
running periods.  The vertical line represents the analysis energy
threshold of $\teff=5.5$~MeV in the salt period.  For the \dto period
the analysis energy threshold was $\teff=5.0$~MeV.  The distributions
shown here are normalized to the neutron detection efficiency in the
two phases for R$<550$ cm.}
\end{center}
\vspace{-4ex}
\end{figure}

Not only is the capture efficiency increased, but the energy deposited
in the detector is also increased.  As shown in Fig.~\ref{fig:cf_energy},
the peak of the energy distribution moves to higher energy so that,
for the same energy cut, the salt phase has improved neutron detection
efficiency compared to the \dto phase.  This allowed a higher energy
threshold for the salt phase and hence less low-energy background
contamination.
\begin{figure}
\begin{center}
\includegraphics[width=3.73in]{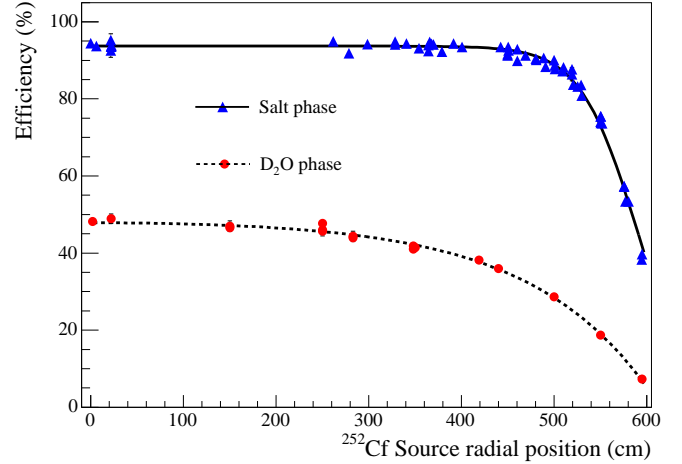}
\caption{\label{fig:cf_eff}
Neutron capture efficiency versus radial position of the \cf source
for the pure \dto and salt phase.  The solid line is a fit
of the salt phase data to Eq.~(\ref{eq:n_er}), and the dotted line is
a fit of the \dto phase data to a neutron diffusion model.
}
\end{center}
\vspace{-4ex}
\end{figure}

Neutron response is calibrated primarily with neutrons produced by a
\cf source with secondary checks made by analysis of neutrons
generated by an Am-Be source and by MC simulations.  To determine the
neutron detection efficiency using the \cf fission source, the
absolute neutron production rate (source strength) has to be
determined.  Shown in Table \ref{tab:cfsource} are the results of four
different techniques used to evaluate the source strength and
presented as inferred strength of the source on June 12, 2001.  The
\cf activity decays with a half-life of 2.645 years, and this
decay, together with that of a small $^{250}$Cf contamination, is
taken into account in evaluating the source strength at the time of a
given calibration run.

The Frisch Grid and triggered Si(Li) methods both use an array of
calibrated $^{3}$He detectors to detect neutrons with the Si(Li)
method being triggered on the fission daughter products of \cf.  These
methods provide independent measures of the source strength prior
to deployment in the SNO detector. 

\textit{In-situ} measurements of the source strength and efficiency
have also been made in the \dto and dissolved-salt phases.  The \dto
multiplicity method is an {\textit{in-situ}} method used in the pure
\dto phase of SNO operation to determine the detection efficiency and
fission rate.  In this method the distribution of the number of
neutrons detected in 2-second time windows is plotted and then fit to
the multiplicity function.  For a model in which the neutron capture
time is negligible compared to this time window, the probability of
detecting $d$ neutrons in a given time window for $l$ neutrons
generated is given by
\begin{eqnarray}
P(d) = e^{-\lambda T}\delta_{d,0} &&+ \sum_{l=d}^{\infty}\frac{l!}{d!
         (l-d)!}\epsilon^{d}(1-\epsilon)^{l-d}\nonumber \\
         &&\times\sum_{N=1}^{\infty}
         \frac{e^{\frac{-(l-N\mu)^2}{2N\sigma^{2}}}}{(2\pi
         N\sigma^{2})^{1/2}}e^{-\lambda T}\frac{(\lambda T)^{N}}{N!}.
\label{eqn:multi}
\end{eqnarray}
The neutron detection efficiency $\epsilon$ and the fission rate
$\lambda$ are the free parameters in the fit.  The factor
$\delta_{d,0}$ is 1 for $d=0$ and is 0 otherwise.  The time window is
$T$. The multiplicity of the \cf source, $\mu$, is taken to be
3.7676$\pm$0.0047 neutrons per fission, and the width of the
multiplicity distribution, $\sigma$, is 1.57~\cite{bib:cfmult}.  The
corrections due to the finite neutron capture lifetime in the \dto
volume were estimated by MC simulations.

A time-series method was used in the salt phase to extract the neutron
detection efficiency and the fission rate by using the time separation
between fission daughter $\gamma$-rays and neutrons in a 5-parameter
fit to an analytical distribution that generalizes the model on which
Eq.~(\ref{eqn:multi}) is based to explicitly include the neutron
capture lifetime and the effect of detecting a small fraction of the
fission $\gamma$-rays that accompany neutron production.  As can be
seen in Table \ref{tab:cfsource}, the various techniques are in
agreement and we calculate a weighted mean as our best estimate of the
neutron source strength.
\begin{table}[htp]
\caption{\label{tab:cfsource}Results of the various methods for
determining the \cf source strength. The source strength is determined
for June 12, 2001.  The $\chi^2$ for the tabulated source strengths 
is 5.6.}
\begin{center}
\begin{tabular}{lc}
\hline \hline Method & Source strength (neutrons per second)\\ \hline
LANL Frisch Grid & 16.75 $\pm$ 0.14\\ 
LANL triggered Si(Li) & 17.08 $\pm$ 0.43\\ 
\dto multiplicity & 16.33 $\pm$ 0.18\\ 
Salt time-series & 16.46 $\pm$ 0.12 \\ \hline
Weighted mean & 16.55 $\pm$ 0.08 \\
\hline \hline
  \end{tabular}
\end{center}
\end{table}

The neutron efficiency is determined by comparing the number of
neutrons detected to the number produced, as a function of the
position of the \cf source and the energy threshold.  In this
comparison it is important to take into account events from fission
$\gamma$-rays that are emitted in the spontaneous decay of \cf.  First
the small fraction due to $\gamma$-rays above $\teff= 6.5$~MeV is
determined; then the neutron energy spectrum in the salt phase
(Fig.~\ref{fig:cf_energy}) is used to extrapolate down to zero
threshold to give the capture efficiency (Fig.~\ref{fig:cf_eff}), and
to $\teff= 5.5$~MeV (Fig.~\ref{fig:cf_energy}) to give the detection
efficiency.

To extract the shape of the \cf $\gamma$ spectrum above
$\teff=6.5$~MeV, a \cf source run was recorded, during the short pure
\dto phase following the removal of salt.  In \dto, neutrons can be
separated easily from the source $\gamma$-rays by requiring that
neutron candidate events reconstruct more than 150 cm from the source.
The fission $\gamma$-ray energy distribution is then obtained by
subtracting the energy distribution of the selected neutron events
from the energy distribution of the events that reconstruct within 150
cm of the source.  The $\gamma$-ray energy distribution obtained from
this measurement and the neutron energy distribution obtained from the
salt phase (see below) are then fit to the raw \cf energy distribution
in salt with the scaling on the neutron and $\gamma$ distributions
left as free parameters to obtain the $\gamma$ background correction
above $\teff=6.5$~MeV.  The time series analysis also gives an
independent estimate for the gamma fraction. The combined result from
these two methods yields ($1.34^{+1.05}_{-0.56}$)\%.

To determine the neutron energy spectrum in the salt phase, a clean
neutron sample is required.  As the capture distance for neutrons in
the salt phase is similar to the attenuation length of fission
$\gamma$-rays, a radial cut cannot be used to select neutron events
from the calibration data.  Instead a ``burst cut" was developed to
select neutron events from calibration data using the coincidence
between fission $\gamma$-rays and neutrons.  In salt, the mean capture
time for neutrons from \cf data at the detector center is measured to
be $5.29\pm0.05$~ms and after approximately 40~ms almost all neutrons
have been captured.  The mean time between fission bursts for the \cf
source used in SNO is about 250~ms.  In the burst cut a fission
$\gamma$-ray candidate event is selected by choosing events with no
preceding events within a 50-ms time interval, and events in a time
interval of 40~ms after the selected first event are tagged as
neutrons.  The burst cut has a selection efficiency for neutrons of
40\%, but less than 0.1\% of the selected neutron candidate events are estimated
to be fission $\gamma$-rays above the threshold ($\teff=5.5$~MeV) used
in this analysis.

The capture efficiency is obtained for each \cf run with source radial
position $r$ (start radius of the neutrons).  These point source
efficiency results are fit to the empirical model
\begin{equation}
\epsilon(r) = A \left \{ \tanh \left [ B \left (r-C \right ) \right ]
- 1 \right \},
\label{eq:n_er}
\end{equation}
where $\epsilon(r)$ is the neutron capture efficiency at source
position $r$ and requiring neutrons to be captured (reconstructed
vertex) inside R$<550$ cm.  $A$, $B$, and $C$ are the fit parameters
of the model.  The volume-weighted capture efficiency
$\overline{\epsilon}$ is then obtained from the ratio of integrals
\begin{equation}
  \overline{\epsilon} = \frac{\int_{0}^{R_{\rm AV}}r^{2}\epsilon(r) dr
  }{\int_{0}^{R_{\rm AV}}r^{2}dr},
\end{equation}
where $\epsilon(r)$ is plotted in Fig.~\ref{fig:cf_eff}.

\begin{figure}[htp]
  \includegraphics[width=3.73in]{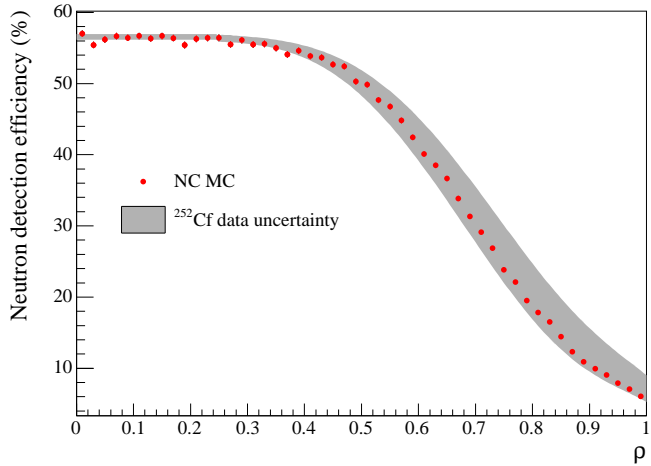}
\caption{Comparison of neutron detection efficiency ($\teff>5.5$~MeV)
for MC simulated NC events (data points) and that derived from \cf
calibration data (shaded band) as a function of volume-weighted radius
$\rho$.  The band represents the statistical and systematic
uncertainties summarized in Table~\ref{tab:cf_sys}.  An additional
1.0\% radial reconstruction uncertainty that is assigned to the solar
neutrino flux is also included in the band.  The volume-weighted NC MC
efficiency is within the systematic uncertainty assigned to the \cf
measurement.  }
\label{fig:cf_mccomp}
\end{figure}

Figure~\ref{fig:cf_mccomp} shows the comparison between detection
efficiency distributions obtained from calibration data and from the
NC MC simulation as functions of $\rho$ after the $\teff>5.5$~MeV
selection criterion has been applied.  The neutron detection
efficiency along with its uncertainty derived from $^{252}$Cf
calibration data are shown as the shaded band in this figure.  The
ratio of the neutral current MC simulation efficiency to the
efficiency obtained from calibration data is within 2.2\% of one;
which is within the estimated systematic uncertainty.

A list of corrections applied to the \cf efficiency measurement is
summarized in Table \ref{tab:cf_cor}.  Apart from the gamma fraction
correction discussed previously, these corrections are calculated
through MC simulation studies The source sampling correction is needed
due to detector asymmetry.  The \cf calibration data were
predominantly collected with the source positioned in the bottom half
of the detector, and therefore do not sample the whole detector.  This
correction is calculated by doing MC simulation studies at the same
\cf source positions as the data, and comparing the efficiency from
the point source MC simulation studies with the efficiency derived
from a MC simulation study of \cf neutrons uniformly distributed in
the detector.  The uncertainty on this correction is included in
Table~\ref{tab:cf_sys}.  A correction is applied to the calibration
efficiency measurement to account for the $^2$H$(n,2n)^1$H reaction
with fission energy neutrons.  This correction is determined from a MC
simulation study.  The $\gamma$ fraction correction is the previously
discussed correction to account for \cf fission $\gamma$-rays
contaminating the \cf neutron data.  A correction is also applied to
account for the $^{16}$O(n,$\alpha$)$^{13}$C reaction with fission
energy neutrons.  Finally, a source geometry correction is applied to
account for neutron captures on the steel and acrylic of the source
holder.

The systematic uncertainties on the \cf neutron detection efficiency
measurement are listed in Table~\ref{tab:cf_sys}.  Some of the
uncertainties arise from the corrections listed in
Table~\ref{tab:cf_cor}.  The source strength uncertainty is derived
from the results summarized in Table \ref{tab:cfsource}.  Source
position uncertainty is obtained by shifting the estimated source positions by
$\pm$2~cm or $\pm$10~cm in radius, depending on the detector region,
and then re-calculating the volume-weighted efficiency. The AV
position uncertainty is taken from a $\pm$6~cm $z$ shift in the
acrylic vessel position.  An estimate of the uncertainty in the
interpolation shown in Fig.~\ref{fig:cf_eff} was taken as the
difference between a high-order polynomial fit and the empirical fit
to the data.  The volume-weighted neutron detection efficiency for the
analysis threshold of $\teff=5.5$~MeV and a fiducial volume of 550~cm
after applying the corrections listed above is (40.7 $\pm$ 0.5
$^{+0.9}_{-0.8}$)\% where the first uncertainty is statistical and the
second the combined systematic uncertainty.  This is the efficiency
for detecting neutrons uniformly generated within the whole AV and
with reconstructed vertex less than $\rho=0.77$.
\begin{table}[htp]
\caption{\label{tab:cf_cor} Corrections applied to the neutron 
efficiency measurement from the calibration data.}     
\begin{center} 
\begin{tabular}{l@{\hspace{0.15in}}l}
\hline 
\hline 
 Source & Correction, \%  \\\hline 
Source sampling & $-$(2.4 $\pm$ 1.0) \\ 
(n,2n) &  $-$(0.58 $\pm$ 0.10)\\ 
(n,$\alpha$) &  $+$(0.66 $\pm$ 0.13)\\ 
Gamma fraction &  $-$(1.34$^{+0.56}_{-1.05}$)\\
Source geometry & $+$(2.03$\pm$0.53)\\ \hline
Total  & $-$1.73$^{+1.3}_{-1.6}$ \\
\hline
\hline
\end{tabular}
\end{center}
\end{table} 

\begin{table}[htp]
\caption{\label{tab:cf_sys} Systematic uncertainties on the neutron 
efficiency  measurement from the calibration data.}   
\begin{center}   
\begin{tabular}{lc}
\hline 
\hline 
 Source & Uncertainty, \% \\\hline 
Source strength & $\pm$0.5 \\ 
Source position &  ${+1.7},\ {-1.0}$ \\ 
Gamma fraction & ${+0.56}, \ {-1.05}$\\ 
AV position &  $\pm$0.3\\ 
(n,2n) &  $\pm$0.10\\ 
(n,$\alpha$) &  $\pm$0.13\\ 
Empirical fit - polynomial fit &  $+$0.4\\  
Source sampling & $\pm$1.0 \\ 
Source geometry & $\pm$0.53 \\ \hline 
Total uncertainty &  ${+2.3},\ {-2.0}$\\ 
\hline
\hline
\end{tabular}
\end{center}
\end{table}

\section{\label{sec:backgrounds}Backgrounds}

Several sources of backgrounds are present in the data.  These include
instrumental backgrounds, backgrounds from the natural \tttt and
\utte radioactivity chains, and backgrounds associated with
products of cosmic muon spallation and atmospheric neutrino
interactions in the detector.  While some of these background types can be
eliminated by analysis cuts, most cannot be distinguished from the
solar neutrino signals.  Table~\ref{tab:letable} provides a summary of the
estimated contributions from these backgrounds.  In the following subsections, the
identification and the determination of the contributions of these
backgrounds is discussed.
\begin{table}
\caption{\label{tab:letable}Summary of backgrounds.  The second column
gives the event rate in the SNO detector D$_2$O volume (r=600.5 cm)
when an average production rate was used.  The last column gives the
estimated number of events remaining in the 391-day salt phase data
set after cuts. The internal neutron and $\gamma$-ray backgrounds are
determined from independent information and constrained in the
analysis. The external-source neutrons are extracted along with the
signal estimated from the energy-unconstrained signal extraction fit.  For backgrounds for
which only an upper limit can be determined, the 68\% CL upper limit
is used as a 1-standard-deviation uncertainty in the error
propagation.}
\begin{center}
\begin{tabular}{lclr}
\hline\hline Source & \multicolumn{2}{l}{Average rate}   &  Counts in \\
		  &            &         &  data set \\ \hline
\multicolumn{4}{l}{ Neutrons generated inside D$_{\bm 2}$O:} \\
\multicolumn{3}{l}{\hspace{1em}$^2$H photodisintegration [U, Th]}  & 
$91.3^{+30.4}_{-31.5}$ \\  
\multicolumn{3}{l}{\hspace{1em}$^2$H photodisintegration [${}^{24}$Na]}  & 
$10.2 \pm 2.5$ \\  
\hspace{1em}n from fission [U] & 0.43 &  n $\mu$g$^{-1}$U y$^{-1}$ &  $0 \pm 0$ \\
\hspace{1em} $^2$H($\alpha$,$\alpha$n)$^1$H [Th] & 1.9 & n $\mu$g$^{-1}$ Th
y$^{-1}$ &   $0.93 \pm 0.50$
\\
\hspace{1em}$^2$H($\alpha$,$\alpha$n)$^1$H
[$^{222}$Rn] & 0.80 & n $\mu$g$^{-1}$ U y$^{-1}$  & $2.89 \pm 0.47$
\\
\hspace{1em}$^{17,18}$O($\alpha$,n)$^{20,21}$Ne [Th] & 0.09 & n $\mu$g$^{-1}$
Th y$^{-1}$  & $0.03 \pm 0.02$
\\
\hspace{1em}$^{17,18}$O($\alpha$,n)$^{20,21}$Ne
[$^{222}$Rn] & 0.20 & n $\mu$g$^{-1}$ U y$^{-1}$  & $0.72 \pm 0.12$ 
\\
\hspace{1em}n from atmospheric $\nu$    & &  &  $15.8 ^{+21.3}_{-4.6}$ \\
\hspace{1em}$^{24}$Na from muons & 0.33 & n y$^{-1}$  & $0.14 \pm 0.14$
\\
\hspace{1em} muons in SNO & 11240 & n y$^{-1}$ &  $\leq 1$
\\
\hspace{1em} muons in rock & 0.14 &n y$^{-1}$ &  $0.08 \pm 0.01$
\\
\hspace{1em}$\overline{\nu_e}$ ``ccp'' & 0.03 & n y$^{-1}$ &  $0.01 \pm 0.01$
\\
\hspace{1em}$\overline{\nu_e}$ ``ccd'' & 1.43 &n y$^{-1}$ &  $0.6 \pm 0.1$
\\
\hspace{1em}$\overline{\nu_e}$ ``ncd''-reactor & 3.24 & n y$^{-1}$  & $1.4 \pm 0.3$
\\
\hspace{1em}$\overline{\nu_e}$ ``ncd''-terrestrial & 1.2 & n y$^{-1}$ &  $0.5 \pm 0.1$
\\
\hspace{1em}CNO $\nu$ & 1.0 & n y$^{-1}$ & $0.4 \pm 0.4$ \\
\hline
\multicolumn{3}{l}{\em Total internal-source neutrons} & 
$125.1^{+37.3}_{-32.0}$  \\
\hline\hline
\multicolumn{4}{l}{ ${\gamma}$-rays generated uniformly inside D$_{\bm 2}$O:}\\
\hspace{1em}$\gamma$ from fission [U]  & 0.04 &  $\gamma$ $\mu$g$^{-1}$U y$^{-1}$ 
& $0 \pm 0$ \\
\hspace{1em}$\gamma$ from atmospheric $\nu$    & &  &  $3.2 ^{+4.6}_{-4.4}$ \\
\hline
\multicolumn{3}{l}{\em Total internal-source $\gamma$-rays} & $3.2 ^{+4.6}_{-4.4}$   \\
\hline\hline
\multicolumn{4}{l}{ Decays of spallation products throughout D$_{\bm 2}$O:}\\
\hspace{1em}$^{16}$N following muons & & $^{16}$N y$^{-1}$  & $< 1.3$ 
\\
\hspace{1em}Other spallation & 1.2 & $^{A}$Z y$^{-1}$  & $\leq 0.8$
\\
\hline\hline
\multicolumn{4}{l}{ \Ckv events from radioactivity inside D$_{\bm 2}$O:}\\
\multicolumn{3}{l}{\hspace{1em} $\beta\gamma$ decays (U,Th,$^{24}$Na)}& 3.6$^{+1.0}_{-0.9}$  \\
\hline\hline
\multicolumn{4}{l}{ Backgrounds produced outside D$_{\bm 2}$O:}\\
\multicolumn{3}{l}{\hspace{1em}Externally generated neutrons (from fit)} & 128.5$\pm$42.4 \\
\multicolumn{3}{l}{\hspace{1em}$\beta\gamma$ decays (U, Th) in AV, \hto, PMTs} &  $<$ 18.5    \\
\multicolumn{3}{l}{\hspace{1em}Instrumental contamination} & $<$3  \\
\multicolumn{3}{l}{\hspace{1em}Isotropic acrylic vessel events}  & $<6.55$  \\
\hline\hline
\end{tabular}
\end{center}
\end{table}

\subsection{Instrumental Backgrounds}
A significant portion of the events comprising the raw data are
the instrumental backgrounds discussed in
Sec.~\ref{sec:event_selection}.  The instrumental background cuts and
the high-level cuts are very efficient at removing these events.  The
residual contamination of instrumental backgrounds in the data set is
measured by using a bifurcated analysis~\cite{bib:bergbusch}. In this
analysis, each set of cuts is used to calibrate the acceptance for
background of the other set, allowing the leakage through the
combination of both sets to be calculated.  The two sets of cuts must
be orthogonal (uncorrelated) for the bifurcated analysis to work and
are chosen appropriately.  Orthogonality is demonstrated using a
technique known as ``relaxing the box" in which the bifurcated
analysis correctly estimated the increase in residual background as
cuts are relaxed when applied to the data set.

\begin{figure}
\begin{center}
\includegraphics[width=3.5in]{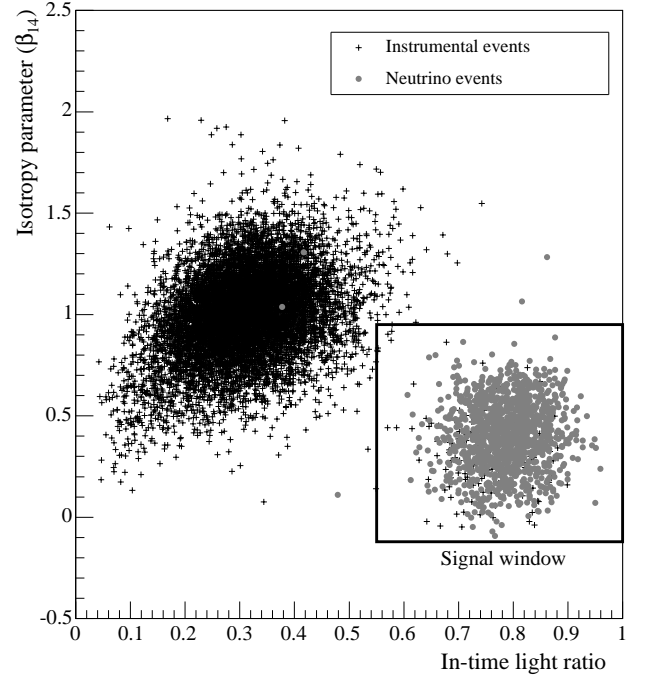}
\caption{The distribution of high-level cut parameters for
instrumental backgrounds and neutrino candidates.  The two cut
parameters are the isotropy \bof and the fraction of PMT hits within
the prompt time window.  The \ns calibration source was used to
generate the sample "neutrino candidate" events 
and thereby establish the signal window for \Ckv events and to
calibrate the cut efficiency.}
\label{fig:contam}
\end{center}  
\end{figure}

The bifurcated analysis provides an upper limit on the residual
instrumental contamination of 3.0 events in the neutrino data which is
treated as a 68\% CL limit in subsequent analyses.  This analysis is
represented in Fig.~\ref{fig:contam} where the distribution of the
high-level cut parameters is presented for the neutrino data set
before application of these cuts, and for instrumental backgrounds
rejected by the instrumental cuts.  In this figure, the two parameters
are the isotropy of the light distribution \bof and the fraction of
PMT hits within the prompt light time window.

One key point regarding the bifurcated analysis is that both
sets of cuts must be sensitive to each background class in order for that
class to be included in the background estimate. This is true of all
known instrumental backgrounds, except for the isotropic events from
the acrylic vessel. For this reason a separate estimate of this
background is required.

\subsection{Isotropic Acrylic Vessel Background}

A class of background events of uncertain origin was identified in the
\dto phase. These events reconstruct near the acrylic vessel and are
characterized by a nearly isotropic light distribution.  They may
result from triboluminescence as stresses are relieved in acrylic. In
the analysis of the data from the \dto phase, the events were removed
by the isotropy and fiducial volume cuts.  However, the isotropy cut
has been relaxed for the salt phase because it would eliminate too
many neutrons, whose \Ckv light distribution is more isotropic in the
presence of salt.

Two independent analyses were performed to estimate the background
remaining after the fiducial volume selection cut of $R<$550~cm was
applied.  In the first analysis the isotropy distribution of the salt
data was fit to the expected shapes for neutrons and electrons.  The
difference between this fit result and the actual signal in the
high-isotropy region of the spectrum was attributed to isotropic
background events in the data set.

In the second method, a cut was applied on the isotropy parameter to
the pure \dto data set, where the neutrino signals are clearly
defined.  The difference between the actual number of events removed
by the cut and the predicted loss of neutrino events yields an
estimate of the background in the \dto data set. This estimate was
then scaled to obtain a limit for the salt data.

The combination of the analyses gives a 68\% upper limit of 6.55
events inside the fiducial volume of $R\leq$550~cm and with a kinetic
energy above 5.5~MeV in the salt data set.  The same analyses were
repeated separately on the day and night data sets, yielding a
day-night asymmetry, defined as the difference between the night and
the day signal rates normalized by their average, of 0.68$\pm$0.31.

\subsection{Photodisintegration Backgrounds from Internal Radioactivity}

The deuteron can be dissociated by a $\gamma$-ray above the binding
energy of 2.22~MeV.  The $\beta-\gamma$ decays of \nuc{208}{Tl} and
\nuc{214}{Bi} from the \nuc{232}{Th} and \nuc{238}{U} chains emit
$\gamma$-rays that are above this binding energy.  Neutrons
produced through photodisintegration are indistinguishable from those
produced by the NC reaction, therefore measurement of the
levels of backgrounds inside the detector is crucial for an accurate
measurement of the total \nuc{8}{B} flux.  A concentration of $3.8
\times 10^{-15}$~g~Th/gD$_2$O or $30 \times 10^{-15}$~gU/gD$_2$O in
the heavy water would each contribute one photodisintegration neutron
per day.

Two independent approaches were developed to measure these
backgrounds, which can be classified as
{\textit{ex-situ}} and {\textit{in-situ}} techniques.

Three {\textit{ex-situ}} techniques have been developed to assay
precursor radioisotopes of \nuc{208}{Tl} and \nuc{214}{Bi}
in the \dto and the H$_2$O.  The decays of the parent Rn and Ra
isotopes are counted in a system external to the SNO detector.  Two of
these techniques extract \nuc{224}{Ra} and \nuc{226}{Ra} using beads
coated with manganese oxide (MnO$_{\mbox{x}}$)~\cite{bib:mnox} or
membranes loaded with hydrous titanium oxide (HTiO)~\cite{bib:htio}.
Radioassays typically involve the circulation of 500~tonnes of \dto of
which approximately 400~tonnes is flowed through these media.  In the
\mnox technique, Rn daughters from the Ra decays are identified by
$\alpha$ spectroscopy.  In the HTiO technique, the extracted Ra atoms
are concentrated and identified by $\beta-\alpha$ coincidences of
their decay products.  Because the ingress of long-lived
(T$_{1/2}$=~3.8~d) \nuc{222}{Rn} (e.g., by emanation from materials or
ingress from laboratory air) breaks the equilibrium with
\nuc{226}{Ra}, this background in the water is obtained by degassing
and cryogenically concentrating the dissolved gas from assays of
approximately 5~tonnes of water~\cite{bib:rn}.  The \nuc{222}{Rn}
decays are subsequently counted in a ZnS(Ag) scintillation cell.

Over the entire salt phase, 16~\mnox and 6~HTiO assays were conducted
at regular intervals.  The addition of salt to the heavy water
affected the radium assay techniques in two ways: 1) a somewhat
reduced radium extraction efficiency was observed on \mnox (81\%
compared to 95\% in pure D$_2$O) and 2) dissolved manganese steadily
increased from a constant 2~ppb in pure \dto to 12~ppb at the end of the
salt phase, causing interference with the measurement of the
concentration of radium in the HTiO assays.  During the salt phase,
the manganese could not be removed by the reverse osmosis method, as
that would have also removed the salt.  Since the \mnox technique did
not suffer from interference with dissolved manganese in the D$_2$O,
its results were used as the \textit{ex-situ} measurement of the
\nuc{224}{Ra} concentration.

A small amount of activity was
observed in all the elements outside the acrylic vessel and
corresponded to a total production rate of
$816\pm165$~\nuc{224}{Ra}~atoms per day, with approximately a quarter
of these produced before the columns. This activity caused a
correction of $0.65\times10^{-15}$~gTh/gD$_2$O to the \textit{ex-situ}
measurements. An assay showed no significant change in activity when
the salt was added to the \dto and after the salt was removed most of
the activity in the elements outside the acrylic vessel went away. The
increase in the salt phase (from an upper limit of 16
\nuc{224}{Ra}~atoms per day in the \dto phase) is interpreted as
primarily due to a displacement of the equilibrium between radium in
solution and radium bound to its Th sources in the water system,
arising from the presence of Na$^{+}$ ions in the \dto.  The time
constant associated with the removal of this activity showed that it
originated from flowing as well as stagnant segments of the water
circulation and assay systems. Components of the systems can be
isolated and the $^{224}$Ra contribution from each component was measured by
circulating the enclosed water over a \mnox
column. Table~\ref{tab:exsitu_source} provides a summary of the
measured production rate of $^{224}$Ra in various components in the assay
and water circulation systems. The amount in the piping outside the AV
is the difference between the total and the sum of the other individual
elements. The piping within the AV is substantially shorter and
thinner than that outside and any contribution from this section was
assumed negligible.
\begin{table}
\caption{\label{tab:exsitu_source}
  \nuc{224}{Ra} assay results for elements of the \dto systems, compared
  to the total activity measured in the complete system.
  These measurements do not include the small piping section in the detector.
}
 \begin{center}
 \begin{tabular}{lc} \hline\hline
Element of heavy water system & Th bkgd. (\nuc{224}{Ra}/day)  \\
\hline \\ [-2.5mm]
Ultrafiltration unit (assay system) &  $28^{+27}_{-23}$  \\[1mm]
Heat exchanger                      &  $233^{+50}_{-48}$ \\[1mm]
Ultrafiltration unit (main stream)  &  $78^{+37}_{-31}$  \\[1mm]
Main recirculation pump             &  $72^{+33}_{-30}$  \\[1mm]
Process degasser                    &  $75^{+31}_{-29}$  \\[1mm]
Filtration unit (main stream)       &  $<36$  \\[1mm]
Piping			   	    &  $330 \pm 184$  \\[1mm]
\hline\\ [-2.5mm]
Complete system                   &  $816\pm165$  \\[1mm]
\hline\hline
\end{tabular}
\end{center}
\end{table}

During a normal assay, the activity measured on the \mnox assay
columns is the sum of contributions from the \dto and from the piping
leading to the columns.  The effect of this source of Th on the levels
in the heavy water target over time was modeled, given the source
distribution in the water piping and the water circulation paths and
times.  The model divided the piping external to the acrylic vessel
into sections, and traced the amount of \nuc{224}{Ra} activity added
during the assays.

The mean level of Th concentration in the \dto target was
$(1.76^{+0.41}_{-0.68})\times 10^{-15}$~gTh/gD$_2$O, where the quoted
uncertainty reflects the possible distribution of this background
activity.  Combining this in quadrature with the systematic
uncertainties associated with the \mnox assay technique of
$^{+32}_{-37}$\% gives $(1.76\pm0.44(\textrm{stat})^{+0.70}_{-0.94}
(\textrm{syst})) \times 10^{-15}$~gTh/gD$_2$O for the \textit{ex-situ}
analysis.

The {\textit{in-situ}} technique uses pattern recognition on the \Ckv\
light distribution to determine the equivalent concentration of
$^{232}$Th and $^{238}$U in the water.
 The decays of \tltze produce a more isotropic light distribution than
\bitof decays because of a more complex decay scheme.  In the energy
window $4<\teff<4.5$~MeV, decays of \tltze and \bitof
in the Th and the U chains are the dominant components of the \Ckv
signal.  By studying events that reconstruct with $R<$450~cm,
\tltze and \bitof are separated statistically by their differences in
the light isotropy \bof.  In this energy interval, the \Ckv light from \bitof
decays is primarily from the direct ground state $\beta$ decay with an
endpoint of 3.27~MeV, while almost every \tltze decay emits
a 2.614-MeV $\gamma$, accompanied by one or more low-energy $\gamma$-rays and
a $\beta$ with an endpoint energy up to 1.8~MeV.  Therefore \tltze
decays produce a more isotropic light distribution than \bitof decays.
The {\textit{in-situ}} technique also has the advantage of providing direct
determination of the background levels during data taking without
any assumptions regarding temporal variation between assays.

\begin{figure}
\begin{center}
\includegraphics[width=3.73in]{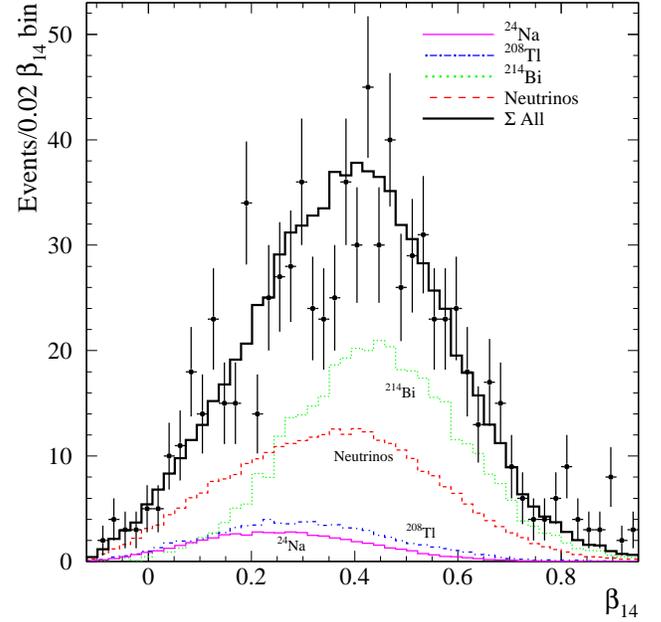}
\caption{\label{fig:infinal} \textit{In-situ} determination of the
  low-energy background in the \dto.  The data points represent
  low-energy events selected by the criteria described in the text.
  The data isotropy (\bof) distribution is fit to a combination of
  \tltze and \bitof distributions.  Also shown are the $^{24}$Na background and
  solar neutrino contributions that were constrained in the fit.  The
  fit result is shown as the sum histogram. }
\end{center}
\end{figure}

The statistical separation of the \tltze and \bitof decay signals
using \bof is shown in Fig.~\ref{fig:infinal}, where the probability
density functions (PDFs) used in the maximum likelihood analysis were
determined from MC simulations.  Solar neutrino and $^{24}$Na signals,
which are backgrounds to the {\textit{in-situ}} analysis, were
constrained in this analysis (an undistorted $^8$B neutrino spectrum was
assumed).  Some of the background $^{24}$Na was produced by neutron
activation of the \dto target during the deployment of calibration
sources.  Data runs that were known to have significant levels of
$^{24}$Na and Rn ingress were removed in the run selection
process described in Sec.~\ref{sec:data_set}.  Additional $^{24}$Na
was produced in the \dto in the chimney region of the acrylic vessel,
in a buffer tank used in the circulation of \dto, and in the water
circulation pipes.  These regions are not well shielded from fast
neutrons and $\gamma$-rays emitted from the rock in the underground
laboratory.  The residual contributions from these sources of
$^{24}$Na were tallied, and the photodisintegration neutron production
rate was found to be 0.064$\pm$0.016~d$^{-1}$.  The $^{24}$Na
background contribution introduced by water circulation was
calculated using a water flow model and constrained in the maximum likelihood fit.  The
\bof PDFs were calibrated by $^{222}$Rn and $^{24}$Na calibration
spikes in the \dto.  The latter has a \bof distribution similar to
\tltze decays.

The amplitudes of the \tltze and \bitof signals determined from the
{\textit{in-situ}} analysis were converted to the equivalent
concentration of \tttt and \utte under secular equilibrium.  The
equivalent concentrations integrated over the solar neutrino data set
are found to be
0.85$^{+0.44}_{-0.42}$(stat) $^{+0.42}_{-0.44}$(syst) $\times$ 10$^{-15}$~gTh/g\dto
and
8.28$^{+0.83}_{-0.81}$(stat) $^{+1.10}_{-1.94}$(syst )$\times$ 10$^{-15}$~gU/g\dto
respectively.

\begin{figure}
\begin{center}
\includegraphics[width=3.73in]{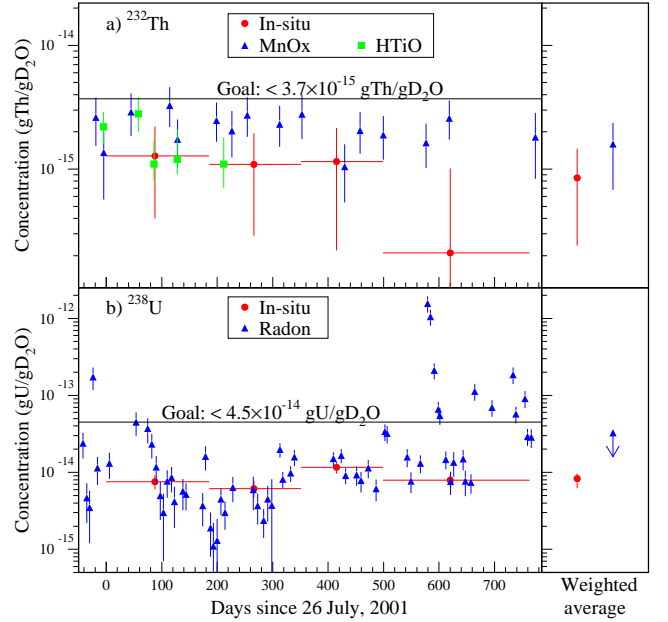}
\end{center}
\caption{\label{fig:inexfig} (a) Thorium and (b) uranium  backgrounds
  (equivalent equilibrium concentrations) in the D$_2$O deduced by
  {\textit{in-situ}} and {\textit{ex-situ}} techniques.  The
  MnO$_{\rm{x}}$ and HTiO radiochemical assay results, the Rn assay
  results, and the {\textit{in-situ}} \Ckv signal determination of the
  backgrounds are presented for the period of this analysis on the
  left-hand side of frames (a) and (b).  The right-hand side shows
  time-integrated averages including an additional sampling systematic
  uncertainty for the {\textit{ex-situ}} measurement.  The large
  $^{222}$Rn excess near day 580 is the decay of  Rn that was
  added for calibration purposes, and was excluded in calculating the
  mean \textit{ex-situ} results.  The weighted average of the
  \textit{ex-situ} $^{222}$Rn measurements appear as an upper limit
  only.  This is due to intermittent background appearing in the radon
  extraction process.}
\end{figure}

Results from the {\textit{ex-situ}} and the {\textit{in-situ}}
analyses of the \dto are shown in Fig.~\ref{fig:inexfig}.  They are
consistent with each other.  The {\textit{ex-situ}} and the
{\textit{in-situ}} techniques are independent and their
systematic uncertainties were independently assessed.  To obtain
the best measurement of the equivalent concentration of \tttt in the
\dto target during the 391-day live time period, the weighted mean of
the two techniques is used as the background input to the integral
neutrino flux measurement described in Sec.~\ref{sec:intflux}.  The \utte chain
activity is dominated by Rn ingress which is highly time dependent.
In addition, only an upper limit could be obtained for the weighted
average of the {\textit{ex-situ}} measurements because of intermittent backgrounds
in the radon extraction process.  Therefore the {\textit{in-situ}}
determination was used for the estimate of this activity as it provides the
appropriate weighting by neutrino live time.  The photodisintegration
neutron production rates from the natural chains were calculated by MC
simulations, which show that the equivalent of 3.79~$\mu$g~\tttt or
29.85~$\mu$g~\utte produces one photodisintegration neutron per day in
the \dto target.

The photodisintegration neutron production rates from decays in the
\tttt\ and \utte chains in the current data set are summarized in
Table~\ref{tab:pdtable}.  To determine the temporal variation of the
internal backgrounds, the data set was divided into four time bins.
An {\textit{in-situ}} determination was performed in each of the time
periods, and the equivalent concentrations of \tttt and \utte were
found to be relatively constant.  The {\textit{in-situ}} analysis was
repeated for the day and the night data set, and the results are also
summarized in Table~\ref{tab:pdtable}.  In the day-night asymmetry
measurement of the neutrino flux, the internal background asymmetry
was determined from the {\textit{in-situ}} analysis of the day and the
night data sets, as this gives the proper temporal variation.

Both the {\textit{ex-situ}} and {\textit{in-situ}} techniques were
also applied to the determination of radioactive backgrounds in
the \hto.  Throughout the salt phase, 86 radon, 30
\mnox and 13 HTiO radioassays of the \hto were performed. Results from
the radon assays performed on the same day were averaged.  The \mnox
and HTiO results were consistent with one another and the weighted
average, taking into account the neutrino live time, was used to
determine the mean concentration of radioisotopes.  The activities
were found to be 5.2$^{+1.6}_{-1.6}\times 10^{-14}$~gTh/gH$_2$O and
20.6$^{+5.0}_{-5.0}\times 10^{-14}$~gU/gH$_2$O by the \textit{ex-situ}
techniques.

In the {\textit{in-situ}} analysis of the \hto background, a
monitoring window for events with $4<\teff<4.5$~MeV in the \hto region
($650<R<680$~cm) was utilized, and the equivalent \tttt and \utte
concentrations were determined by fitting the isotropy distribution.
The radioactive backgrounds in the \hto deduced from the
\textit{in-situ} technique were 6.1$^{+4.1}_{-1.6}\times
10^{-14}$~gTh/gH$_2$O and 19.1$^{+11.1}_{-4.5}\times
10^{-14}$~gU/gH$_2$O, which are consistent with the results from the
\textit{ex-situ} technique.  As is discussed below, the neutron
background due to radioactivity external to the \dto target is
determined in the fit of the solar neutrino fluxes
(Sec.~\ref{sec:sigex}).  The photodisintegration background arising
from activity in the \hto is part of this external neutron background.

\begin{table}
\caption{\label{tab:pdtable}Photodisintegration neutron production rate
  from decays of daughters in the \tttt and \utte chains in the \dto.  The rates
  from the {\textit{ex-situ}} and the {\textit{in-situ}} techniques are
  consistent with each other.  The total rate from the last row is
  used for all the solar neutrino analyses that do not depend on the
  solar zenith angle.  The day (D) and night (N) {\textit{in-situ}} results are
  used  as inputs to the day-night solar neutrino flux asymmetry analysis.}
\begin{center}
\begin{tabular}{l@{\hspace{0.3in}}c@{\hspace{0.1in}}c} \hline\hline
   & \tttt & \utte \\
   & (n d$^{-1}$) & (n d$^{-1}$)  \\ \hline \\ [-2mm]
{\textit{in-situ}} (D) & $\insitupdthd$  & $\insitupdud$ \\ [1mm]
{\textit{in-situ}} (N) & $\insitupdthn$  & $\insitupdun$ \\ [1mm]
{\textit{in-situ}} (D+N) & $\insitupdth$ & $\insitupdu$ \\ [1.5mm] \hline 
{\textit{ex-situ}} (D+N) & $\exsitupdth$ & $\exsitupdu$ \\ [1.5mm] \hline 
Total (D+N) & $\combinedpdth$ & $\insitupdu$ \\ [1mm] \hline\hline
\end{tabular}
\end{center}
\end{table}

\subsection{\Ckv Backgrounds}

The broad energy resolution of the detector allows a small fraction of
the $\beta-\gamma$ decays in the natural radioactive chains to appear
in the neutrino data sample, even though their $Q$ values are lower
than the $\teff= 5.5$-MeV neutrino analysis threshold.  The number of
these {\it internal \Ckv} events originating within the D$_2$O target
is kept small primarily by ensuring low radioactivity levels.  Outside
the heavy water volume, however, the acrylic vessel, the light water,
and in particular the PMT array and support structure have higher
levels of radioactivity.  Most of these {\it external \Ckv} events are
eliminated by imposing a 550-cm fiducial volume cut.  These events can
``leak'' into the fiducial volume in two ways: $\gamma$-rays can travel
unscattered from their external origin inward and events whose origin
is outside the volume can have a misreconstructed vertex located inside.

\begin{figure}
  \begin{center}
    \includegraphics[width=3.5in]{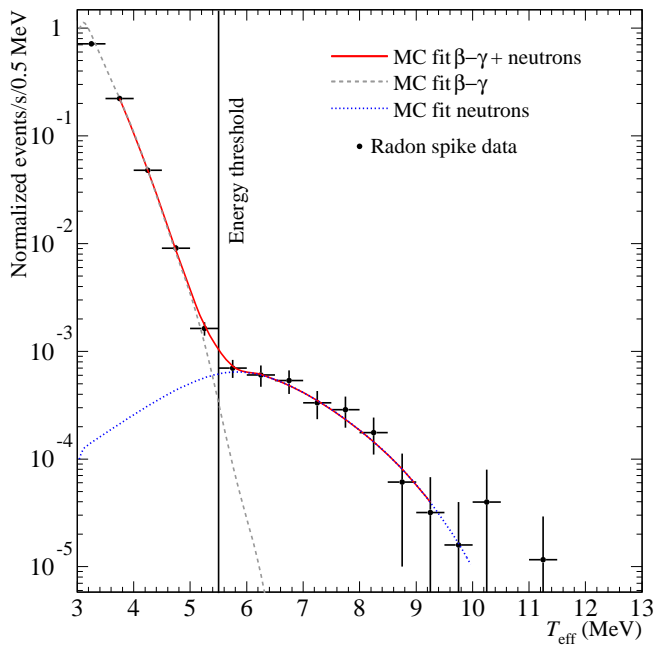}
  \caption{Fitted MC prediction to the radon spike data, with the
     steeply falling background \Ckv spectrum and neutron peaks shown
     separately.\label{fig:spikecomp}}
    \end{center}
\end{figure}

The internal \Ckv background was determined from MC simulations and
calibration with a controlled injection of 81$\pm$4~Bq of $^{222}$Rn
(a `radon spike') into the \dto target.  In the analyses of the internal \Ckv
background, the ratio between the number of internal \Ckv events and
the number of detected photodisintegration neutrons above the neutrino
analysis threshold was determined for backgrounds from the \tttt and
the \utte chains.  This ratio was then normalized by the measured
number of photodisintegration neutrons produced and the neutron
detection efficiency in the fiducial volume (described above).  For
\tltze decays in the Th chain and $^{24}$Na, MC simulations of their
\Ckv signals were used.  The systematic uncertainties were determined
by performing 10~000 simulated experiments, with the scale and
resolution of the energy response, vertex reconstruction, and \bof
drawn from distributions estimated by the respective analyses.  All
analysis cuts applied to the neutrino data sample were also applied
here, and their individual uncertainties included in the measurement.
For the \bitof decays in the U chain, the energy-differential
uncertainties of the Rn spike energy spectrum were first determined by
1000 simulated experiments. Each of the simulated experiments assumed
a different spatial distribution of radioactivity within the detector,
constrained by the reconstructed position of low-energy events in the
neutrino data set.  Uncertainties associated with the time variation
of the detector response over the course of the neutrino data set were
taken from the calibration analyses.  The \Ckv-to-photodisintegration
neutron ratio was determined by fitting the energy distribution of the
Rn spike (with its uncertainties determined from the simulated
experiments) to the simulated \Ckv background and neutron spectra.
The parameters which were allowed to float in the fit included the
energy scale, the energy resolution, and the amplitudes of both the
\Ckv events and the associated photodisintegration neutrons.
Figure~\ref{fig:spikecomp} shows the resultant fit.  The energy scale
and resolution uncertainties were both consistent with the analyses of
$^{16}$N and $^{252}$Cf data. The systematic uncertainties on the
ratio were determined by varying all the parameters over their allowed
uncertainties, including all covariances.

At $\teff>5.5$~MeV, the ratio of the number of \Ckv events to detected
photodisintegration neutrons was found to be 0.011$^{+0.005}_{-0.002}$
for \tltze, 0.090$^{+0.024}_{-0.018}$ for $^{24}$Na and
0.053$^{+0.011}_{-0.013}$ for \bitof, which correspond to a total of 3.6$^{+1.0}_{-0.9}$ 
observed events in the 391-day data set.  Since
the multiplicative factors used to scale the photodisintegration
neutron amplitude to the number of internal \Ckv events are
independent of the solar zenith angle, the internal \Ckv backgrounds in
the day and the night data set are obtained by scaling the
{\textit{in-situ}} results in Table~\ref{tab:pdtable}.

\begin{figure}
\begin{center}
\includegraphics[width=3.5in]{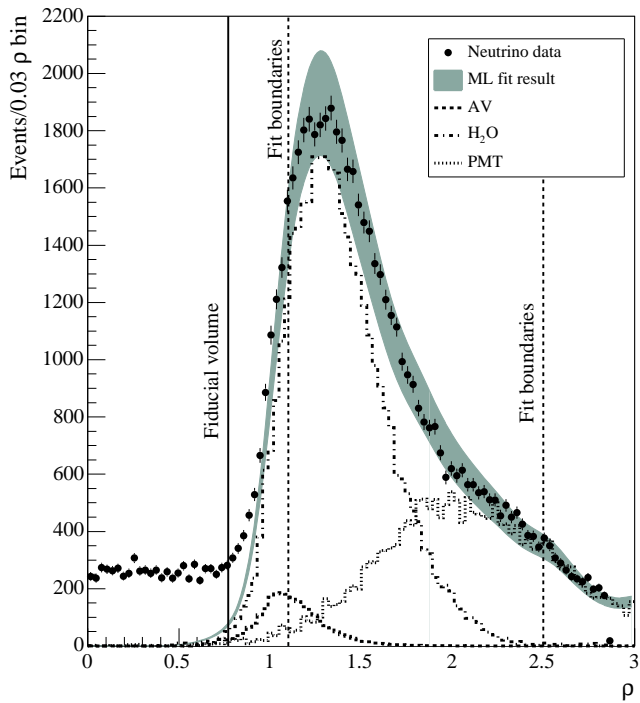}
\caption{\label{fig:exttail} Maximum likelihood fit of the data $\rho$
   distribution (data points) to the PDFs constructed from Th or U
   calibration sources.  Because the external Cherenkov backgrounds
   are small for the solar neutrino analysis threshold
   ($\teff>5.5$~MeV) and fiducial volume ($\rho<0.77$), this fit is
   performed for $\teff>4.5$~MeV in the region $1.1<\rho<2.5$ in order
   to enhance the statistics.  Contributions of these backgrounds are
   then obtained from extrapolating the fit results to the solar
   neutrino signal window.  The band represents the systematic
   uncertainties in this analysis.}
\end{center}
\end{figure}

The analysis methods for the internal \Ckv backgrounds are not
directly applicable to the external \Ckv backgrounds. Outside the
fiducial volume, the detector is not as well calibrated because of
greater optical complexity and limited accessibility for the primary
optical and energy calibration sources. To model the radioactivity in
this region, Th and U sources were deployed at various locations
external to the fiducial volume and used to generate
volume-weighted radial distributions for low-energy backgrounds
originating from the acrylic vessel, the \hto, and the PMT support
structure. The radial distributions were utilized since they
are relatively insensitive to Th and U differences. For the
acrylic vessel and the PMT support structure, the $\rho$ distributions were
based on a Th source encapsulated in acrylic. For the \hto region, a
calibrated $^{222}$Rn spike was used. Events with $\teff>4.5$~MeV and
$1.1< \rho<2.5$ in the neutrino data set were fit by a maximum
likelihood technique with radial PDFs generated with the same energy threshold
as the source data.  Figure~\ref{fig:exttail} shows the results
of this fit, where the width of the band in the figure represents the
systematic uncertainties.  These uncertainties include spatial and
temporal dependence of the reconstruction, subtraction of
contributions from photodisintegration neutrons near the acrylic
vessel, and the difference between the $\rho$ distributions of the decays of
U and Th daughters.  Due to limited statistics in the source PDFs, the
fit to the $\rho$ distribution for the neutrino data set could not be
done at the solar neutrino analysis energy threshold of
$\teff>5.5$~MeV. Therefore, the $\rho$ distributions for the source data
with $\teff>5.5$~MeV were normalized by the amplitudes obtained in the
fit to the data with $\teff>4.5$~MeV and extrapolated inside the
fiducial volume to estimate the number of background events in the
signal window. This analysis was repeated for the day and night
data sets. At the neutrino analysis energy threshold, the extrapolated
contributions of the backgrounds from different detector regions are
consistent with zero.  For the 391-day salt data set, the 68\% upper
limit for the acrylic vessel, \hto, and the PMT background
contributions are 7, 3, and 11 events respectively with a combined
upper limit of the total external \Ckv background of 18.5 events.  The
day-night asymmetry of this external background is -0.10$\pm$0.16.

\subsection{Other Backgrounds}

In addition to the main contributions stemming from deuteron
photodisintegration and low-energy \Ckv events from $\beta-\gamma$
decays, there are other sub-dominant backgrounds that must be
assessed.  These backgrounds include contributions from $\alpha$
reactions on elements in the water target and the construction
materials, $^{238}$U fission, cosmic ray spallation, and reactor and
atmospheric neutrinos.

A small source of neutrons can come from $\alpha$ reactions on
$^{2}$H, $^{17}$O, and $^{18}$O.  The $\alpha$s produced in the
uranium and thorium decay chains.  Neutrons can also be produced from
the spontaneous fission of $^{238}$U, which has a half-life of $(8.2
\pm 0.1) \times 10^{15}$ years.  The concentration of $^{238}$U has
been measured using \textit{ex-situ} HTiO radioassays.  Results from the assays
indicate a negligible contribution of neutrons and $\gamma$-rays from
the spontaneous fission of $^{238}$U in the D$_2$O.

Backgrounds from atmospheric neutrino interactions were estimated with
the aid of the NUANCE~\cite{bib:nuance} neutrino Monte Carlo
simulation package.  Atmospheric neutrino interactions can contribute
to the production of neutrons without other energy deposits to tag the
event, or via the production of untagged photons from the
de-excitation of $^{16}$O from neutral-current neutrino-nucleon
scattering.  The NUANCE simulation provides a comprehensive estimate
of various neutrino interactions and includes final state intranuclear
interactions.  The estimated contribution to this data set from such
events passing all selection cuts is $15.8^{+21.3}_{-4.6}$ neutrons
and $3.2 ^{+4.6}_{-4.4}$ $\gamma$-rays.

Events that produce two or more neutrons within short time intervals
can serve as a test of the background contributions from atmospheric
neutrinos, spontaneous fission, and $^2$H$(n,2n)^1$H.  Analysis of
burst data taken from the full data set compared to Monte Carlo
predictions show an excess of high-multiplicity bursts in the neutrino
data.  A burst is defined as two or more events, passing all neutrino
selection criteria, that occur within a 50 ms time interval.  A
likelihood calculation indicates that, if the MC accurately describes
the data, the probability of obtaining a worse likelihood is $1.6\%$.
Uncertainties in hadron transport and intranuclear reactions for
atmospheric neutrino interactions are considered the likely cause of
the deficit of high-multiplicity events in the MC prediction.  In
estimating the uncertainty associated with the single neutron events
from atmospheric neutrino interactions, the upper uncertainty has been
conservatively taken to encompass the difference between the data and
the Monte Carlo prediction.

Neutrons and $\gamma$-rays produced at the acrylic vessel and in the
light water can propagate into the fiducial volume.  During
construction of the acrylic vessel, Rn progeny accumulated on its
surfaces.  These daughters can initiate ($\alpha$,$n$) reactions on
$^{13}$C, $^{17}$O, and $^{18}$O.  External $\gamma$-rays originating
from ($\alpha,n\gamma$) and ($\alpha,p\gamma$) processes and
radioactivity in the construction material of the detector and the
rock cavity can enter the \dto target and photodisintegrate deuterons.
The radial distribution of these neutron sources differs from those of
the NC signal and of photodisintegration neutrons produced from
radioactivity in the \dto target.  The enhanced neutron detection
efficiency of the salt phase makes it possible to extract the
external-source neutron contribution in the neutrino signal window by
including an additional radial distribution function in the
statistical analysis of the solar neutrino flux.  Details of the
extraction of this background can be found in
Sec.~\ref{sec:sigex}.  Additional tests, including direct counting of
the $\alpha$ activity on the surface of the acrylic and the search for
coincident events generated by specific nuclear reactions associated
with the ($\alpha$,$n$) reaction, were performed.  The main source of
coincidence events is the $e^+$$e^-$ pairs from the excited state at 6.05-MeV
in $^{16}$O, in coincidence with a fraction of the neutrons produced by
$^{13}$C($\alpha$,$n$)$^{16}$O.  A weaker source is the two neutrons
from $^2$H$(n,2n)^1$H induced by fast neutrons from ($\alpha$,$n$)
reactions.

The results of these measurements are limited by statistics.  The sum of
this $\alpha$-induced neutron background and the photodisintegration
neutron background produced by radioactivity in the \hto and the AV is
consistent with the results from the radial fit technique.

At the depth of SNO, only neutrinos and muons from cosmic rays
survive.  Neutrons and other cosmogenic activity are produced from muon
capture, muon electrodisintegration, and from muon nuclear
spallation. The 20-second veto following a tagged muon event removes most
subsequent activity.  However, longer-lived spallation products, such
as $^{16}$N, can survive this cut.  A limit of less than $1.3$ \ns events
(68\% CL) was determined by analysis of long-lived activity present within 50
seconds after a muon event.  Activity from other spallation products
is estimated to contribute less than one event in the data set.

A small number of neutrons can also be created by $\bar{\nu}_e$
reactions from nuclear reactors.  The estimate of this background is
based on the average power output of all commercial reactors within
500~km of SNO and an average reactor $\bar{\nu}_e$ spectrum. Electron
antineutrinos from terrestrial radioisotopes do not contribute
significantly to the background because their energies are below
threshold for the CC reaction on $^2$H (``ccd'') and can only initiate
CC reactions on $^1$H (``ccp'') and NC reactions on $^2$H (``ncd'').
The total estimated background contributions from reactor and
terrestrial $\bar{\nu}_e$ interactions are 2.0$\pm$0.3 and 0.5$\pm$0.1
detected neutrons, respectively.

SNO is slightly sensitive to solar CNO neutrinos generated by the electron
capture decay of $^{15}$O and $^{17}$F, and this contributes $0.4\pm
0.4$ neutrons \cite{bib:stonehill}.

None of the backgrounds discussed in this section depend on the solar
zenith angle.  Thus, the contribution of these backgrounds to the day
and the night data sets can be determined by normalizing to the
respective live times.

\section{\label{sec:dnspecific}Systematics Associated with the 
Day-Night Asymmetry Measurement}

Differences in day and night neutrino fluxes are a prediction of
matter-enhanced neutrino oscillations.  Day-night results are reported
as asymmetry ratios in the measured neutrino fluxes.  The asymmetry
ratio for a flux is defined as $A = 2(\phi_N-\phi_D)/(\phi_N+\phi_D)$.
An advantage of the asymmetry ratio is that most systematics cancel and
only systematic effects that scale day and night fluxes by
different factors need to be considered.

Day-night systematics can be divided into four general classes.  {\em
Diurnal systematics} are variations in detector response over a 24-hour
timescale, such as might be caused by diurnal changes in the laboratory
environment.  Because the SNO detector is located far underground, it
is isolated from many diurnal effects.  Day-night differences in
detector response are therefore not expected, but limits
must be placed on their size.  {\em Directional systematics} arise
because the SNO detector is not completely spherically symmetric, and
because the directions of electrons from CC or ES neutrino
interactions are correlated with the time of day.  ES events in
particular are highly directional, and so ES events at night will
preferentially illuminate the upper half of the detector, while during
the day they illuminate the bottom half.  If there are differences in
the up-down response of the detector, these directional differences
can create effective differences in the day and night rates.
Directional systematics are expected to be important for ES events,
but are greatly suppressed for CC events, which have only a weak
directional correlation with the direction from the Sun.  Directional
systematics do not produce day-night systematic effects for directionally
isotropic events such as neutrons or backgrounds.  {\em Miscellaneous
systematics} include possible day-night differences in cut acceptance,
uncertainties in the live time calculation, and long-term variations
in detector response.  These can produce differences in the
time-averaged day and night detector responses if the long-term
variations are correlated with the seasonal variations in the
day-night live time exposure.  {\em Background systematics} reflect
uncertainties in the magnitude and day-night asymmetries of the
various background sources.  The following sections describe how
each class of systematics is measured.

\subsection{Diurnal Systematics}
\label{sec:diurnal}

Diurnal systematics are variations in detector response with a 24-hour
periodicity.  A strictly hypothetical example would be diurnal
variations in the laboratory's temperature, which in principle could
change the response of the SNO electronics if there were uncompensated
temperature dependencies.  Because the timescale for diurnal
variations is much shorter than the average interval between
calibrations, diurnal stability must be evaluated using classes of
events that are continually present in the detector.  These primarily
consist of secondary neutrons produced in the D$_2$O by through-going
muons and \Ckv events from low-energy $\beta$-$\gamma$ decays.

Muons traversing the acrylic vessel can produce secondary neutrons by
a variety of processes, including photodisintegration of deuterons and
nuclear spallation.  These neutrons are produced uniformly throughout
the D$_2$O and at an essentially constant rate on a diurnal timescale,
and so mimic the solar NC signal in spatial and temporal
distributions.  Small diurnal and seasonal variations of a few percent
in the predicted muon rate, expected as a result of variations in the
scale height of the atmosphere, are neglected.  Secondary neutrons
from muons are selected by identifying bursts of events inside the
D$_2$O that occur within a time window of $50~\mu$s - 20~ms following
a tagged muon event.  Spallation events can occasionally produce very
large bursts of neutrons.  To prevent such bursts from biasing the
spatial or temporal distributions of the neutrons, a multiplicity
selection requiring less than $15$ events in the bursts is applied to
the neutron selection.
\begin{table}[tb]
\caption{\label{tab:muonfollowers}Day-night differences for selected
muon-induced neutron distributions. }
\begin{center}
\begin{tabular}{lccc}
\hline \hline & & & Asymmetry \\ Quantity & Night value & Day value &
ratio (\%) \\ \hline Event rate (day$^{-1}$) & $3.50 \pm 0.09$ & $3.56
\pm 0.10$ & $-1.83 \pm 3.81$ \\ Mean energy (MeV) & $5.78 \pm 0.06$ &
$5.67 \pm 0.06$ & $1.96 \pm 1.49$ \\ Energy width (MeV) & $1.47 \pm
0.06$ & $1.47 \pm 0.06$ & $\phantom{-}0.00 \pm 4.69$ \\ Mean isotropy
($\beta_{14}$) & $0.311 \pm 0.004$ & $0.312 \pm 0.005$ & $-0.24 \pm
2.03$ \\ Capture time (ms) & $4.9 \pm 0.3$ & $4.6 \pm 0.3$ &
$\phantom{-}7.27 \pm 8.36$ \\ \hline \hline
\end{tabular}
\end{center}
\end{table}

\begin{figure}[tb]
\begin{center}
\includegraphics[width=3.5in]{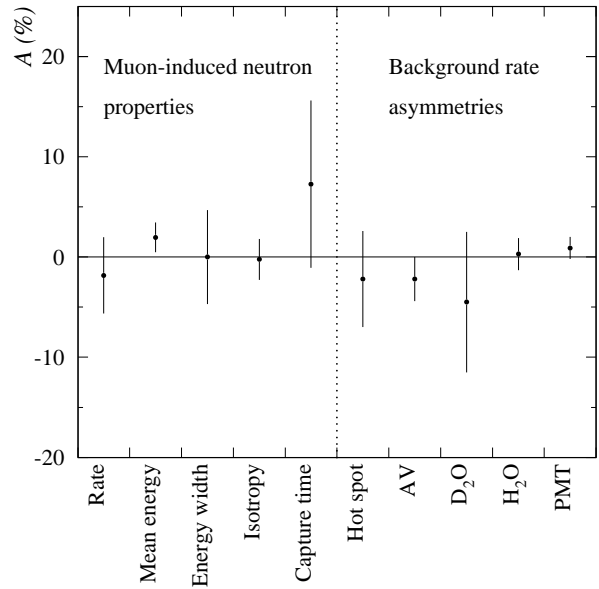}
\caption{\label{fig:dn_asymmetries}Day-night asymmetries of selected
 muon-induced neutron properties and background rates.  The day-night
 asymmetries for the muon-induced neutron properties are calculated
 for events in the full day and night data sets.  The day-night
 asymmetries of the background rates are determined by combining the
 calculated asymmetry of each individual data run.  Each data run
 lasted for less than 24~hours.}
\end{center}
\end{figure}

Table~\ref{tab:muonfollowers} shows the mean day and night values for
various distributions of muon-induced secondary neutrons.  The event
rate, mean event energy, width of the neutrons' energy distribution,
mean value of the $\beta_{14}$ isotropy parameter, and neutron capture
time are all consistent between day and night.  The left side of
Fig.~\ref{fig:dn_asymmetries} shows the day-night asymmetry on each
measured quantity.

While muon-induced secondary neutrons demonstrate the diurnal
stability of the detector, they are statistically limited by the low
muon rate ($\sim 3/$hour).  Better limits on most parameters may be
obtained from studies of intrinsic detector radioactivity.  These
include \Ckv events from $\beta$-$\gamma$ decays of low-energy
radioactivity in the D$_2$O, acrylic, H$_2$O, or PMTs.  

A localized region of higher-than-average background radioactivity was
discovered on the upper surface of the acrylic vessel.  The origin of
this ``AV hot spot'' is uncertain, but it was presumably introduced by
radioactive contamination during construction.  If composed of Th, the
total amount is approximately 10 $\mu$g, and is not a significant
neutron source given the location.  However, this hot spot provides an
excellent check of position reconstruction for a point source of
events.  Comparisons of the reconstructed day and night positions of
the AV hot spot show that its position is stable to within $\pm 0.3\%$
in radius between night and day, and indicate that the vertex
resolution for a point source differs by no more than $\pm 1.26$~cm
between night and day.

Because low-energy background events have steeply falling energy
spectra near and below the analysis threshold, small variations in
energy scale or energy resolution produce large changes in the
observed rate inside a low-energy window.  Measurements of diurnal
rate stability for low-energy radioactivity can thus be used to limit
diurnal variations in energy scale and energy resolution.

In order for this procedure to work, the actual level of radioactivity
must be diurnally constant.  Radioactivity inside the PMTs or acrylic
is immobile, and presumed to be constant except for possible slow
decay.  Radioactivity in the D$_2$O and H$_2$O can fluctuate over time
due to radon ingress, as is seen in the \textit{in-situ} radioactivity
measurements, but the timescale for these changes is generally long
compared to the 12-hour differences being sought in this analysis.
These longer-term variations in radioactivity can be mitigated by
calculating a day-night asymmetry for each individual run that
includes both day and night live time (implicitly assuming that the
radioactivity is constant over the several hour duration of a typical
run).  Assuming the real radioactivity level does not vary
significantly over this short time period, any observed rate variation
would indicate a diurnal difference in detector response. The
run-by-run asymmetries can be combined in a weighted average using a
maximum likelihood technique to determine an overall limit on the rate
asymmetry of each source.

\begin{table}[tb]
\caption{Combined day-night asymmetries for various background
regions.  The asymmetry ratio is formed by combining the measured
day-night asymmetries for each run with both day and night data.
\label{tab:bkgd_asymmetries}
}
\begin{center}
\begin{tabular}{l@{\hspace{0.1in}}c} \hline\hline
Background region & Combined asymmetry ratio \\
\hline
D$_2$O & $-0.045 \pm 0.070$\\
AV hot spot & $-0.022 \pm 0.048$\\
Shell around AV & $-0.023 \pm 0.022$\\
H$_2$O & $\phantom{-}0.003 \pm 0.016$\\
PMTs & $\phantom{-}0.009 \pm 0.011$\\ \hline\hline
\end{tabular}
\end{center}
\end{table}

The right side of Fig.~\ref{fig:dn_asymmetries} shows the combined
run-by-run asymmetries for radioactivity in five different regions of
the detector: in the AV hot spot, in a spherical shell around the AV
(excluding the hot spot), in the light and heavy water, and near the PMTs.
Events used to calculate these asymmetries were selected from an
energy region corresponding to $N_{\rm prompt}$ between 25 and 40
(approximately $4.1 < \teff < 6.2$~MeV).

Table~\ref{tab:bkgd_asymmetries} shows the asymmetries for each of
these background regions.  All are consistent with no asymmetry.  The
analysis was repeated, this time calculating day-night asymmetry ratios
for sets of 24-hour periods instead of for each run, and similar
results were obtained.  The insensitivity to the length of the
normalization period of the asymmetry, and the fact that asymmetries
in all regions of the detector are consistent with zero, make it very
unlikely there are diurnal variations in the detector response.

While the rate asymmetries on radioactivity in the H$_2$O and near the
PMTs have the smallest uncertainties, these regions are far from the
fiducial volume of the neutrino analysis.  A more representative
approach is to base the energy scale and resolution variation estimates on
the rate of radioactivity in the spherical shell around the AV.  This
region has adequate statistics, is at a similar radius compared to
neutrino events, and includes a significant fraction of acrylic
radioactivity which should be diurnally constant.  (This region also
contains events from thin shells of D$_2$O and H$_2$O adjoining the
AV.)  Allowing the rate asymmetry for this region to vary within its
measured limits, and assuming that the entire variation is due to
changes in energy scale, the diurnal energy scale variation is limited
to $\pm 0.4\%$.  Assuming instead that the entire variation is due
to changes in energy resolution, the diurnal energy resolution
variation can be limited to $\pm 0.5\%$.  These results are consistent
with limits derived from the other regions of radioactivity and with
the more direct measurements from muon-induced neutrons given in
Table~\ref{tab:muonfollowers}.

\textit{In-situ} radioactive background measurements can also place
limits on diurnal variations in event isotropy.  Comparison of the
mean day and night $\beta_{14}$ values for events in the spherical
shell around the AV results in the limit $|\beta_{14,\rm{night}} -
\beta_{14,\rm{day}}| < 0.006$.  This limit is comparable to the
independent limit of $0.0064$ derived from muon-induced neutrons
(Table~\ref{tab:muonfollowers}), and the two limits may be combined to
limit the diurnal shift in the mean $\beta_{14}$ value to $< 0.0043$.

\subsection{Directional Systematics}

As described near the start of this section, variations in detector
response with the direction of the event can produce a day-night
systematic for neutrino signals that have different directional
distributions for night and day.  The PDFs for the direction of
electrons produced by CC and ES interactions with respect to the Sun's
direction are shown in Fig.~\ref{fig:salt_pdfsb}(b).  Because the
$\gamma$-rays emitted by neutron captures have random directions,
detector asymmetries produce no day-night variations for NC events.

Variations in detector response with direction are measured with
6.13-MeV $\gamma$-rays from the $^{16}$N source.  Events are grouped
by bins in $\cos \theta$ and $\phi$ in detector coordinates.  For
each bin, the means of each of the eight event variables listed in
Table \ref{tab:dirsys} are calculated.  $^{16}$N calibration runs at
different positions in the heavy water are combined in a
volume-weighted average.  The result is a map of the detector response
versus directional bin for each systematic.

Monte Carlo calculations predict how many CC and ES events fall in
each directional bin for SNO's live time exposure for both night and
day.  By convolving the MC prediction for the directional distribution
in both time bins with the measured detector asymmetry, the mean value
of a detector response function is calculated for the night and day
data sets.  The volume-weighted difference between the night and day
values gives a measure of the directional systematic for each neutrino
signal.

\begin{table}[t]
\caption{Directional day-night systematics for CC and ES events.  The
day-night effect of directional systematics for NC and background
events is zero.  
\label{tab:dirsys}
}
\begin{center}
\begin{tabular}{lcc} \hline\hline
Systematic & Limit for CC events & Limit for ES events \\
\hline
Energy scale & $\pm 0.09\%$ & $\pm 0.79\%$ \\
Energy resolution & $\pm 0.13\%$ & $\pm 1.3\%$ \\
Radial shift & $\pm 0.02\%$ & $\pm 0.15\%$ \\
Vertex resolution & $\pm 0.13\%$ & $\pm 1.4\%$ \\
Angular resolution, $\alpha_M$ & $\pm 1.4\%$ & $\pm 11.6\%$ \\
Angular resolution, $\beta_M$ & $\pm 0.7\%$ & $\pm 6.1\%$ \\
Angular resolution, $\beta_S$ & $\pm 0.6\%$ & $\pm 5.2\%$ \\
Isotropy & $\pm 0.09\%$ & $\pm 0.82\%$ \\ \hline\hline
\end{tabular}
\end{center}
\end{table}
Table~\ref{tab:dirsys} gives the limits on the effective day-night
difference in detector response for CC and ES events resulting from
directional asymmetries.

\subsection{Miscellaneous Systematics}

The uncertainties on the day and night live times are calculated to be
$\pm0.021\%$.  These result in a $0.03\%$ uncertainty on the day-night
asymmetry.

The cuts used to reject instrumental backgrounds are described in
Sec.~\ref{sec:event_selection}.  The time variability in the fraction
of good events removed can be measured by applying the cuts to
muon-induced secondary neutrons.  The day-night asymmetry of the
fraction of good events removed by the cuts is measured to be
$A = 0.18 \pm 0.33$.  Note that this is an
asymmetry on a very small loss fraction.

In addition to diurnal variations in detector response (see
Sec.~\ref{sec:diurnal}), variations on longer timescales could exist.
Such variations can indirectly introduce differences in the day-night
detector response if they correlate with seasonal variations in the
day-night live time exposure.  For example, if the energy scale were
slightly mis-calibrated during the winter months, this miscalibration
would affect the night data set more than the day data, since during
the winter more night data is collected than day data due to the
seasonal variation in the lengths of night and day.  The effects of
these variations can be constrained by constructing worst-case models
that systematically over-estimate the measured response in summer and
under-estimate it during the winter, or vice versa.  These worst-case
models were constructed for energy scale and isotropy variations using
the regular calibration points taken with the $^{16}$N source and
their effect is shown in Tables~\ref{tab:dn_syserr_unconstrained} and
~\ref{tab:dn_syserr_unconstrained_anc0} as long term variations in
these parameters.

\subsection{Day-Night Background Systematics}

Uncertainties in backgrounds to the neutrino signals can produce
day-night systematics in two ways.  First, there may be
uncertainty in the day-night variation of the background (i.e., an
uncertainty on the rate asymmetry $A$ of the background).
This uncertainty will differentially affect the amounts of background
subtracted from the night and day data sets.  Second, the uncertainty
on the total amount of background (night+day) results in an
uncertainty on the average neutrino flux, which enters into the
denominator of the day-night ratio.  Generally, the uncertainty on the
asymmetry is a larger effect than the uncertainty on the amount of 
background, but both contribute.

Details of the day-night calculations of the background totals have
been given previously.  The reader is referred to the discussion of
the individual backgrounds in Sec.~\ref{sec:backgrounds}.

\section{\label{sec:sigex} Neutrino Signal Decomposition}

\begin{figure}
\begin{center}
\includegraphics[width=3.6in]{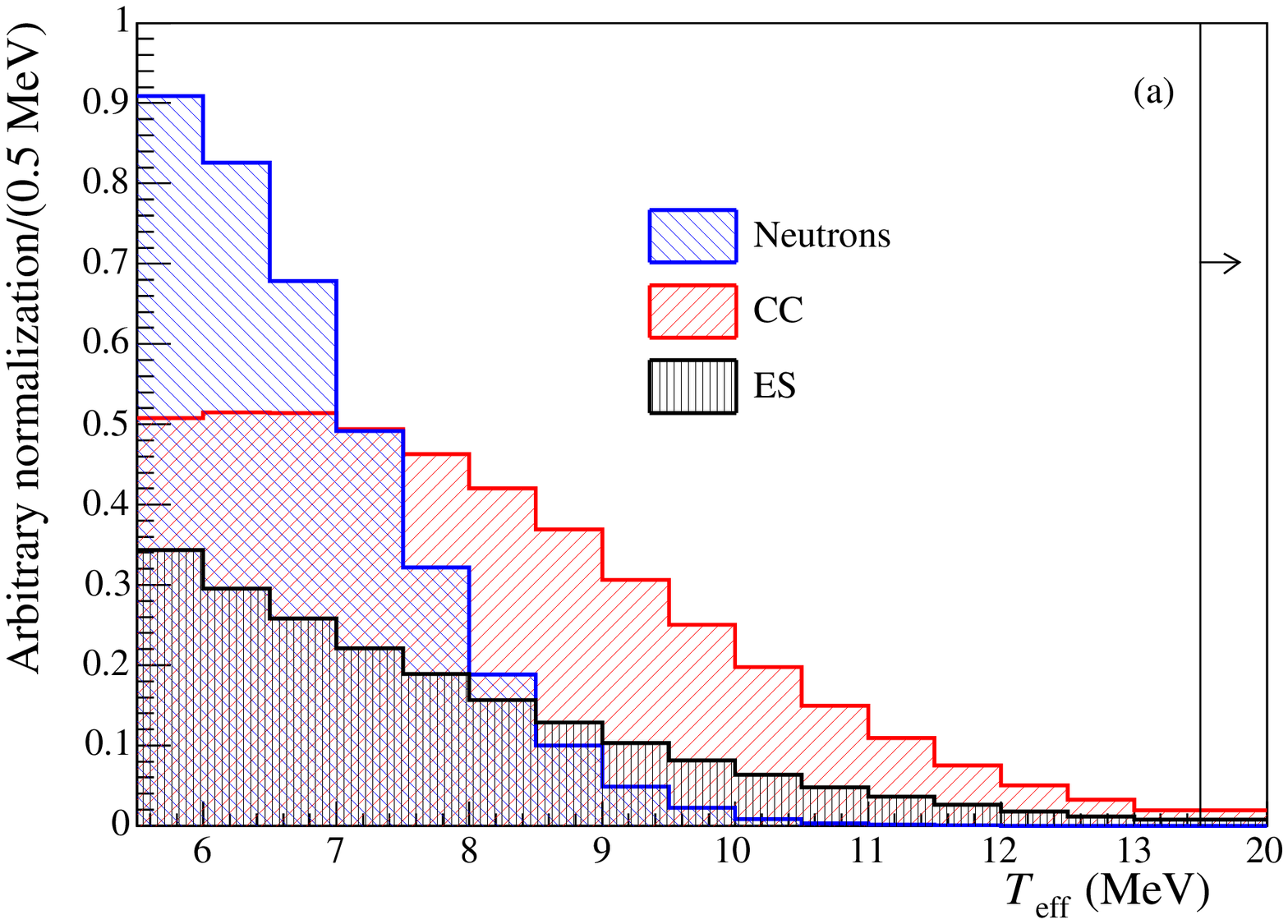}
\includegraphics[width=3.6in]{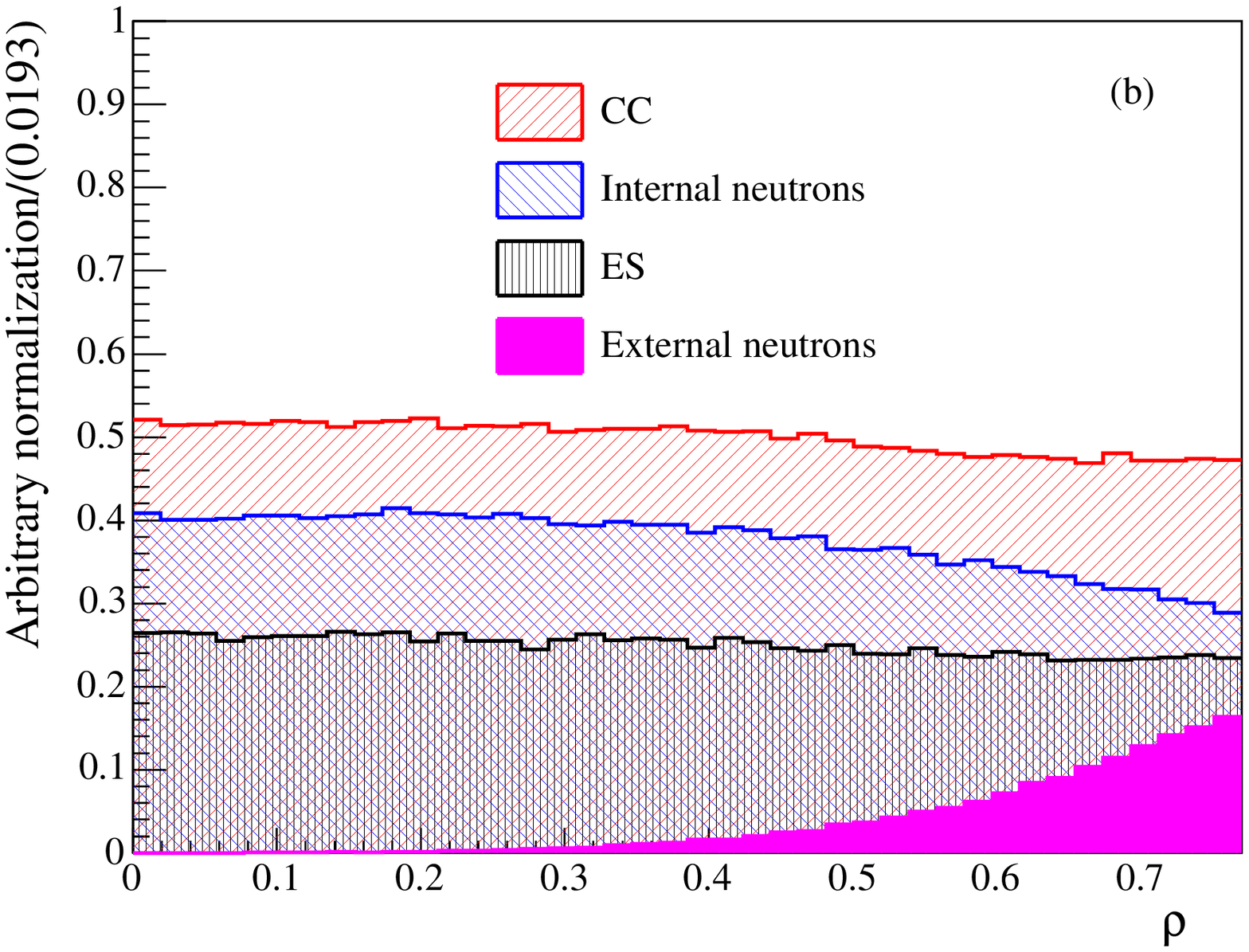}
\caption{\label{fig:salt_pdfs} (a) $T_{\text{eff}}$ and (b) $\rho$
  distributions for CC, ES, NC and external neutron events.  Where
  internal and external neutron distributions are identical the
  distribution is simply labeled neutrons.  Note that the distribution
  normalizations are arbitrary and chosen to allow the shape
  differences to be seen clearly.  The CC energy spectrum shape
  corresponds to an undistorted $^8$B model.}
\end{center}
\end{figure}

\begin{figure}
\begin{center}
\includegraphics[width=3.6in]{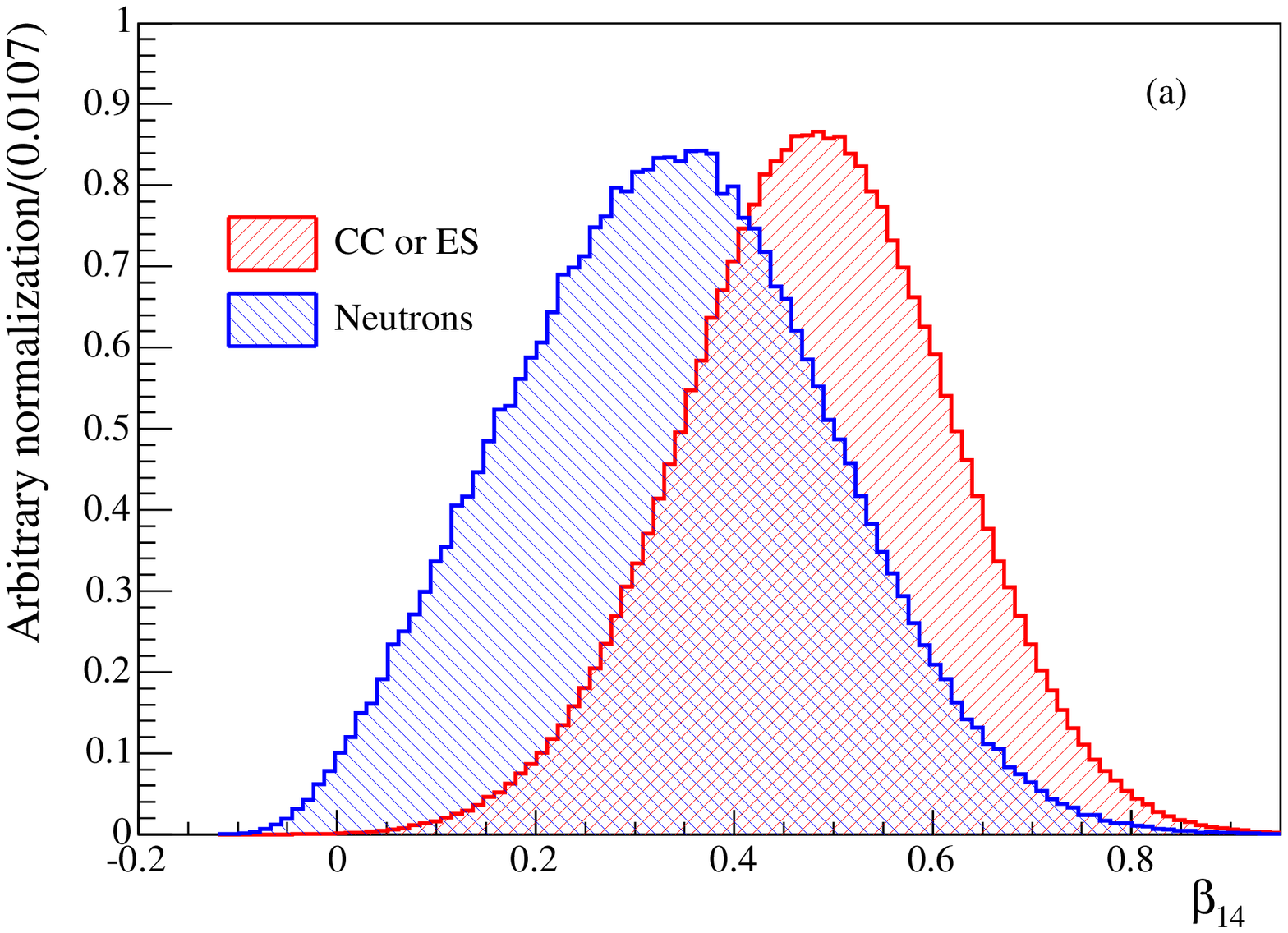}
\includegraphics[width=3.6in]{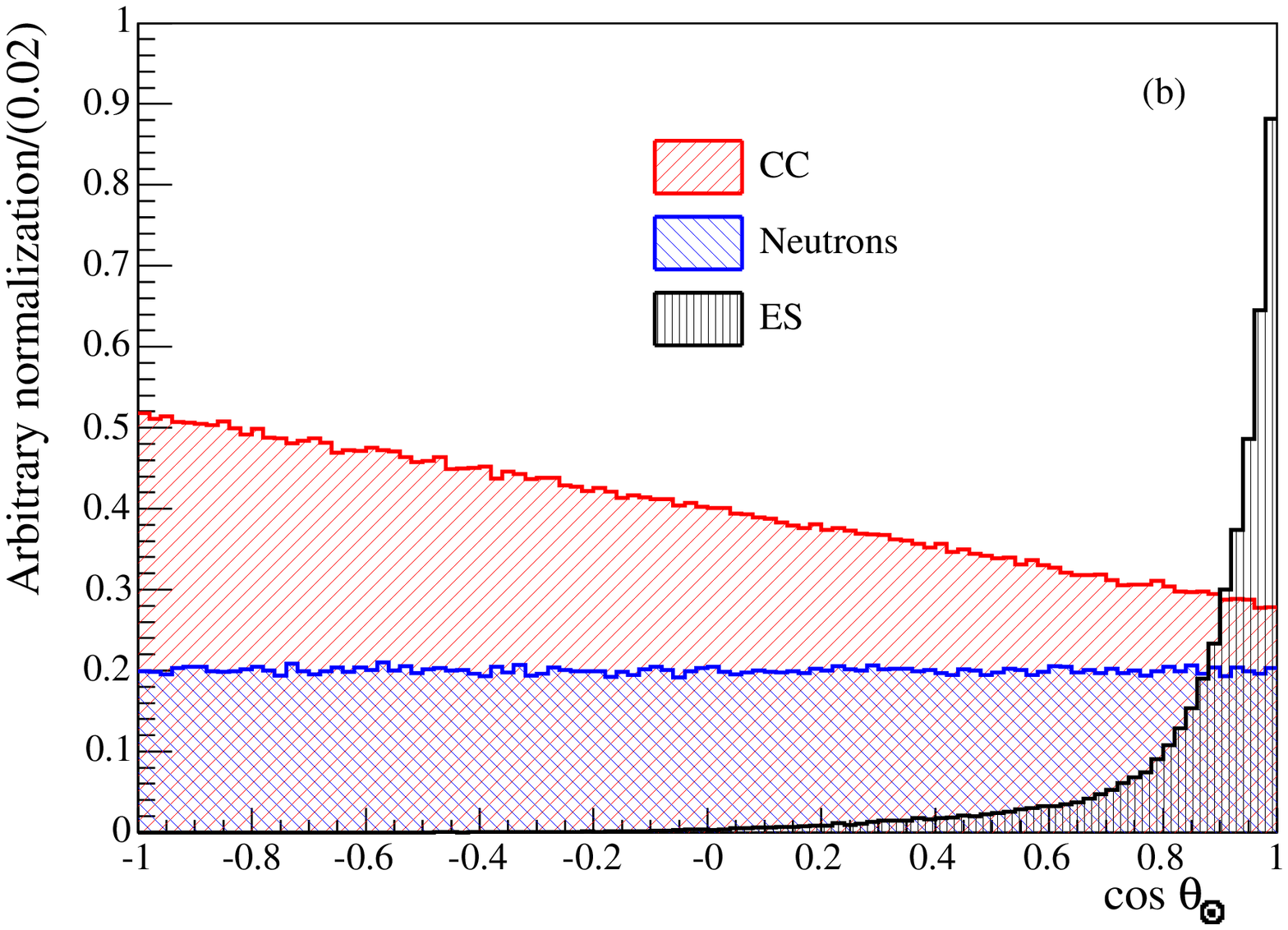}
\caption{\label{fig:salt_pdfsb} (a) $\beta_{14}$ and (b)
  $\cos{\theta_{\odot}}$ distributions for CC, ES, NC and external
  neutron events.  Where internal and external neutron distributions
  are identical the distribution is simply labeled neutrons.  Note
  that the distribution normalizations are arbitrary and chosen to
  allow the shape differences to be seen clearly.}
\end{center}
\end{figure}

Extraction of the electron energy spectrum, flux and day-night
asymmetry of the \be solar neutrinos is carried out via an extended
maximum likelihood fit of the event variables in the 4722-event data
set by MC generated PDFs.  The data variables utilized are $\teff$,
$\rho$, \costs, and \bof, and their distributions are shown in
Figs.~\ref{fig:salt_pdfs} and~\ref{fig:salt_pdfsb}.

In order to obtain the electron energy spectra of CC and ES
interactions, PDFs were created for $\teff$ intervals which spanned
the range from $5.5$~MeV to $13.5$~MeV in $0.5$~MeV steps.  For
$\teff$ values between $13.5$ and $20$~MeV, a single bin was used.
Minor adjustments were applied to the PDFs to take into account signal
loss due to instrumental cuts not modeled by the simulation
(Fig.~\ref{fig:cc_cut_eff}).  PDF normalizations for CC and ES
components were separately allowed to vary in each $\teff$ bin to
obtain model-independent spectra. Only the overall
normalizations of NC and external neutron components were allowed to
vary since their $\teff$ spectra are simply determined by the energy
release following neutron capture on $^{35}$Cl or $^2$H.

The parameter \bof is correlated with $\teff$ and, to a lesser
extent, with $\rho$. Similarly,
\costs is weakly correlated with $\teff$ and $\rho$. These dependencies
were taken into account through the use of a multi-dimensional PDF
$P(\teff,\beta_{14},\rho, \cos\theta_\odot)$ factorized as follows:
\begin{equation}
 P(\teff,\beta_{14},\rho, \cos\theta_\odot) = P(\teff,\beta_{14},\rho) \times
P(\cos\theta_\odot | \teff,\rho),
\label{eq:threed}
\end{equation}
 where the first factor is just the 3-dimensional PDF for the variables $\teff$,
\bof, and $\rho$, while the second factor is the conditional PDF for \costs given
$\teff$ and $\rho$.  This approach explicitly preserves all
correlations between the four relevant parameters with the exception of the
correlation between \costs and \bof, which is assumed to be linked only
through $\teff$ and $\rho$.

To confirm our understanding of these correlations and to verify the
results with an independent approach, extraction of solar neutrino
results has also been performed with the 3-dimensional PDF in
Eq.~\ref{eq:threed} further reduced to the following formulation:
\begin{equation}
P(\teff,\beta_{14},\rho,\cos\theta_\odot) = P(\teff,\beta_{14}) \times
P(\cos\theta_\odot) \times P(\rho).
\label{eq:twod}
\end{equation}
The signal extraction procedure was applied to 100 simulated data sets
each generated to simulate the expected characteristics of the data.
It was found that the parameterization in Eq.~(\ref{eq:twod}), which
ignored the correlations with $\rho$ and \costs, resulted in a small
bias compared with the approach in Eq.~(\ref{eq:threed}).  As a
further cross-check, both approaches were applied to the data and
yielded results that were consistent with this interpretation. After
applying corrections for the expected bias, the results were found to
be nearly identical.  The average difference in the extracted CC
signal between the two factorization approaches, after corrections, is
less than 1\%.

In the flux measurement, a ``constrained" fit, in which the CC energy
spectral shape is fixed as an undistorted \be spectrum
\cite{bib:ortiz}, and an ``unconstrained" fit, in which this
constraint is removed, are carried out.  The constrained analysis is
useful in testing the null hypothesis of neutrino flavor
transformation under the assumption of an undistorted $^8$B solar
neutrino spectrum.  The unconstrained analysis has the
advantage that this model assumption is removed.

Compared to the pure \dto phase, the addition of salt increases the
sensitivity at large radius to neutron capture (see
Fig.~\ref{fig:cf_eff}) making it possible to detect background
neutrons originating at or near the AV.  As shown in
Fig.~\ref{fig:salt_pdfs}(b), the differing radial profiles allow
external neutrons to be separated from neutrons generated within the
\dto volume.  In the pure \dto phase analysis~\cite{bib:snonc}, the
amplitudes of the photodisintegration neutron backgrounds from
internal radioactivity in the AV and the \hto were fixed in neutrino
signal decomposition.  In the current analysis, the $\rho$
distribution provides discrimination between the external neutron and
the signal flux contributions. The external neutron component is measured
simultaneously with the flux signals from the likelihood fit.

In the salt phase analysis, the amplitude of the $\rho$ PDF of the
external source neutrons is allowed to vary in the maximum likelihood
fit.  If the production rates of these external neutron backgrounds
were constant, they would contribute an amount smaller than the
systematic uncertainties reported in the earlier pure \dto phase
analysis.

Results from the measurements of the radioactive backgrounds
(Sec.~\ref{sec:backgrounds}) were incorporated in the extraction of
solar neutrino signals.  The internal photodisintegration neutron
backgrounds are subtracted from the fitted NC event total.  A separate
PDF for internal $\gamma$-rays from atmospheric neutrino interactions
was included with a normalization fixed to 3.2 events (in accordance
with Table~\ref{tab:letable}) and with a shape based on events
generated by the \ns calibration source because of their similarity.
Other backgrounds, listed in Table~\ref{tab:other_backgrounds}, were
treated as systematic uncertainties applied to the appropriate $\teff$
intervals for CC and NC components following signal extraction. The
distinctive dependence of ES events on \costs was assumed to reduce
the effect of these backgrounds on the ES results to negligible
levels.
\begingroup
\squeezetable
\begin{table}
\caption{\label{tab:other_backgrounds} Additional background contributions
treated as systematic uncertainties applied to appropriate $\teff$
intervals in the CC and NC spectra.  Note that the $\teff$ range
for the first bin is 5.5-6.0 MeV.}
\begin{center}
\begin{tabular}{lc}\hline\hline
Systematic parameter     & $1\sigma$ limits \\
\hline  
Internal $\gamma$ bkgd.    & $\pm$ 2.2 events
\\ \hline
Internal \Ckv bkgd. - Tl & $\le$ 0.53 events (1st CC bin)    \\ 
Internal \Ckv bkgd. - Bi & $\le$ 2.29 events (1st CC bin)     \\ 
Internal \Ckv bkgd. - Na & $\le$ 0.90 events (1st CC bin)     \\ 
External \Ckv bkgd.- PMT  & $\le$ 11.0 events (1st CC bin)     \\ 
External \Ckv bkgd.- H$_2$O & $\le$ 3.0 events  (1st CC bin)    \\
\hline
External \Ckv bkgd.- AV & $\le$ 7.0 events (1st CC bin)     \\ 
AV events                       & $\le \frac{6.55}{2}$ events  (1st CC bin)
\\
                                 & $\le \frac{6.55}{2}$ events  ($\teff>$6.5~MeV) \\ 
Instrumental bkgd.        & $\le$ 3 events (across $\teff$ range)   \\
\hline\hline

\end{tabular}
\end{center}
\end{table}
\endgroup

Day-night asymmetries are determined for both the constrained and
unconstrained fit cases.  As described below, asymmetries
are calculated both allowing all signal fluxes to vary and with the NC 
asymmetry constrained to zero.

The maximum likelihood fit returns the estimated numbers of CC, NC,
ES, and external neutron events, with statistical uncertainties.
Final fluxes are determined by normalizing to the solar-model
prediction calculated by MC simulation with several correction
factors, summarized in Table \ref{tab:corfac}, applied.  Additional
corrections are applied to adjust the predicted event rates as a
result of CC interactions on $^{17}$O, $^{18}$O, $^{23}$Na, and
$^{35,37}$Cl, which are not modeled in the MC simulation.  The MC
model uses the Effective Field Theory (EFT) calculation in
Ref.~\cite{bib:butler} to obtain the neutrino-deuteron cross sections.
This EFT calculation is normalized to the standard potential-model
calculation~\cite{bib:nsgk,bib:npa} by fixing the two-body axial
exchange-current counter term $L_{\rm 1,A}$.  Corrections are applied
to the flux results to account for small differences in the choices of
the axial coupling constant $g_A$~\cite{bib:beacom} that were used in
the theoretical calculations, and for normalization to the improved
potential-model calculation~\cite{bib:npa} (the EFT calculation was
done with $L_{\rm 1,A} = 5.6$ fm$^3$).  The values of the fundamental
constants~\cite{bib:PDG} used in the present calculations of the cross
sections, together with correction factors derived to bring the cross
section calculations in accordance with those constants, are listed in
Table \ref{tab:corfac}.  Radiative corrections included in
\cite{bib:npa} were taken out, and the corrections of Kurylov {\em et
al.}  \cite{bib:kmv}, parametrized as follows, were applied to the CC,
NC and ES cross sections:
\begin{eqnarray}\label{eq:cross}
   \omega_{\rm CC} &=& 1.0318-7.45\times 10^{-4} E_e + 4.72\times
10^{-6} E_e^2 \\ \omega_{\rm NC} &=& 1.0154 \label{eq:cross2}\\
\nonumber \omega_{\rm ES} &=& 0.9764 -7.81\times 10^{-4}\teff
\label{eq:cross3}\\ & & \hspace{0.2in} -1.31\times 10^{-4}\teff^2 +
3.64\times 10^{-6}\teff^3,
\end{eqnarray}
where $E_e$ is the true total energy of the electron.  Table
\ref{tab:corfac} also includes the corrections to the CC, ES, and NC
fluxes from CC interactions on oxygen, sodium, and chlorine, by which,
together with the radiative corrections, the EFT and ES cross sections
were multiplied before comparing with the data to extract the fluxes.

The flux measurements are presented in terms of the CC, NC, and ES
signals and as the flux of electron type ($\phi_e$) and non-electron
type ($\phi_{\mu\tau}$) active neutrinos.
\begin{table}
\caption{\label{tab:corfac}{Fundamental constants used in the MC and
post-fit correction factors to the neutrino flux in different reaction
channels.  The radiative correction functions are discussed in the
text.}}
\begin{center}
\begin{tabular}{lccc} \hline
\hline 
Constants used & & & \\
\hline
$g_A$ & \multicolumn{3}{l}{1.2670(30)} \\
$G_F$ &  \multicolumn{3}{l}{1.16639(1) $10^{-5}$ GeV$^{-2}$ } \\
$L_{1,A}$ & \multicolumn{3}{l}{4.0~fm$^3$} \\
$\sin^2 \theta_W(\overline{\rm MS})$ & 0.23113(15) & & \\
\hline
Correction & CC & ES & NC \\ \hline 
CC on O, Na and Cl & 1.0081 & 1.0 & 1.0\\ 
Radiative corrections & $\omega_{\rm CC}$ & $\omega_{\rm ES}$ & $\omega_{\rm NC}$\\ 
\hline\hline
\end{tabular}
\end{center}
\end{table} 
In general, the effects of systematic uncertainties are evaluated by
re-fitting the data after perturbing the model PDFs by the 1$\sigma$
uncertainties determined from calibration and background measurements.
The differences between the nominal flux fit values and those obtained
with the systematically shifted PDFs are quoted as 68\%
C.L. uncertainties.

\begin{figure}
\begin{center}
\includegraphics[width=3.6in]{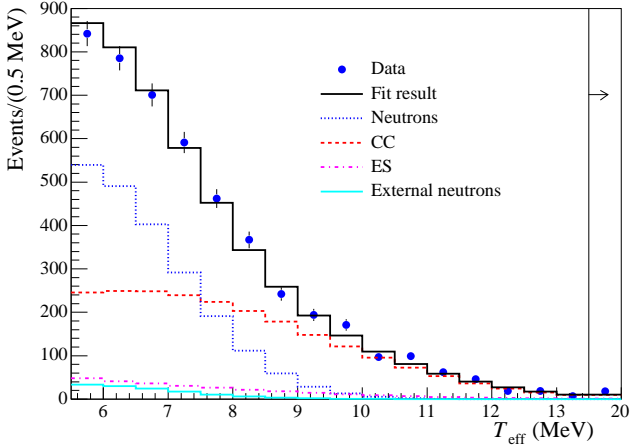}
\caption{\label{fig:fit_results}{Data $\teff$ spectrum with
statistical uncertainties. Included are MC spectra for neutron, CC,
ES, and external neutron distributions.  Note that an undistorted
$^8$B spectral shape has been assumed and each MC contribution has
been normalized to the number of corresponding fit events measured
by the energy-constrained signal extraction.}}
\end{center}
\end{figure}

Figure~\ref{fig:fit_results} shows the energy spectrum, with
statistical uncertainties, of the data that passes all selection cuts.
Included in the figure are MC generated spectra for CC, ES, internal
neutron, external neutron components, and their sum.  Note that the MC
generated distributions correspond to an undistorted $^8$B neutrino
spectrum and each spectrum has been normalized to correspond to the
total number of fit events for the given component as extracted by the
energy-constrained fit.

\section{\label{sec:spectrum}Spectrum}

\begin{figure}
\begin{center}
\includegraphics[width=3.6in]{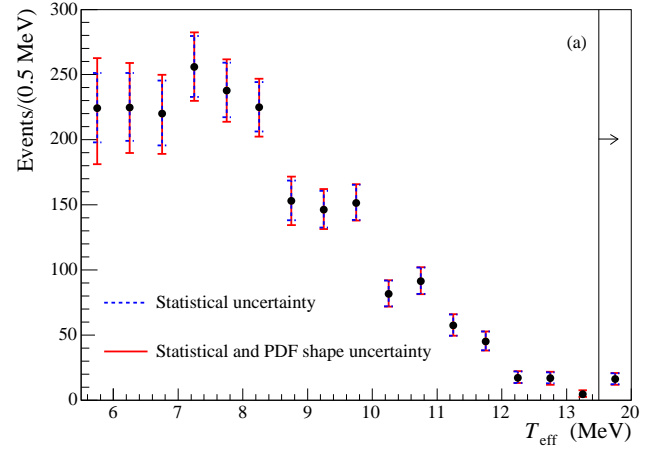}
\includegraphics[width=3.6in]{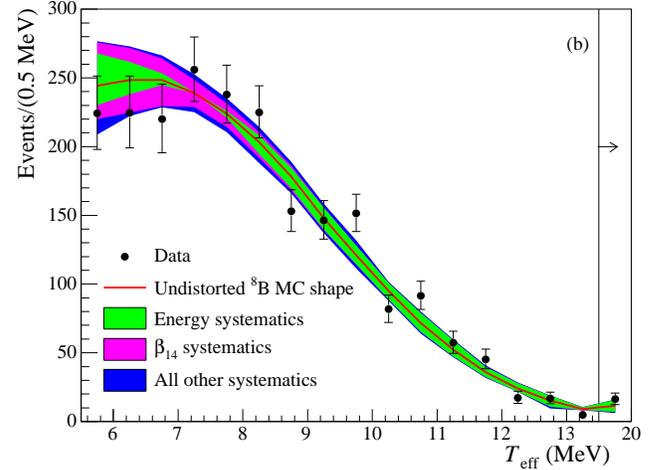}
\caption{\label{fig:cc_diff_fit}{(a) Extracted CC $\teff$ spectrum.
Systematic uncertainties have been combined in quadrature and include
only the effect of PDF shape change.  (b) Extracted CC $\teff$
spectrum with statistical error bars compared to predictions for an
undistorted \be shape with combined systematic uncertainties,
including both shape and acceptance components.  The systematic error
bands represent the fraction of the total uncertainty attributable to
the given quantity.}}
\end{center}
\end{figure}

\begin{figure}
\begin{center}
\includegraphics[width=3.6in]{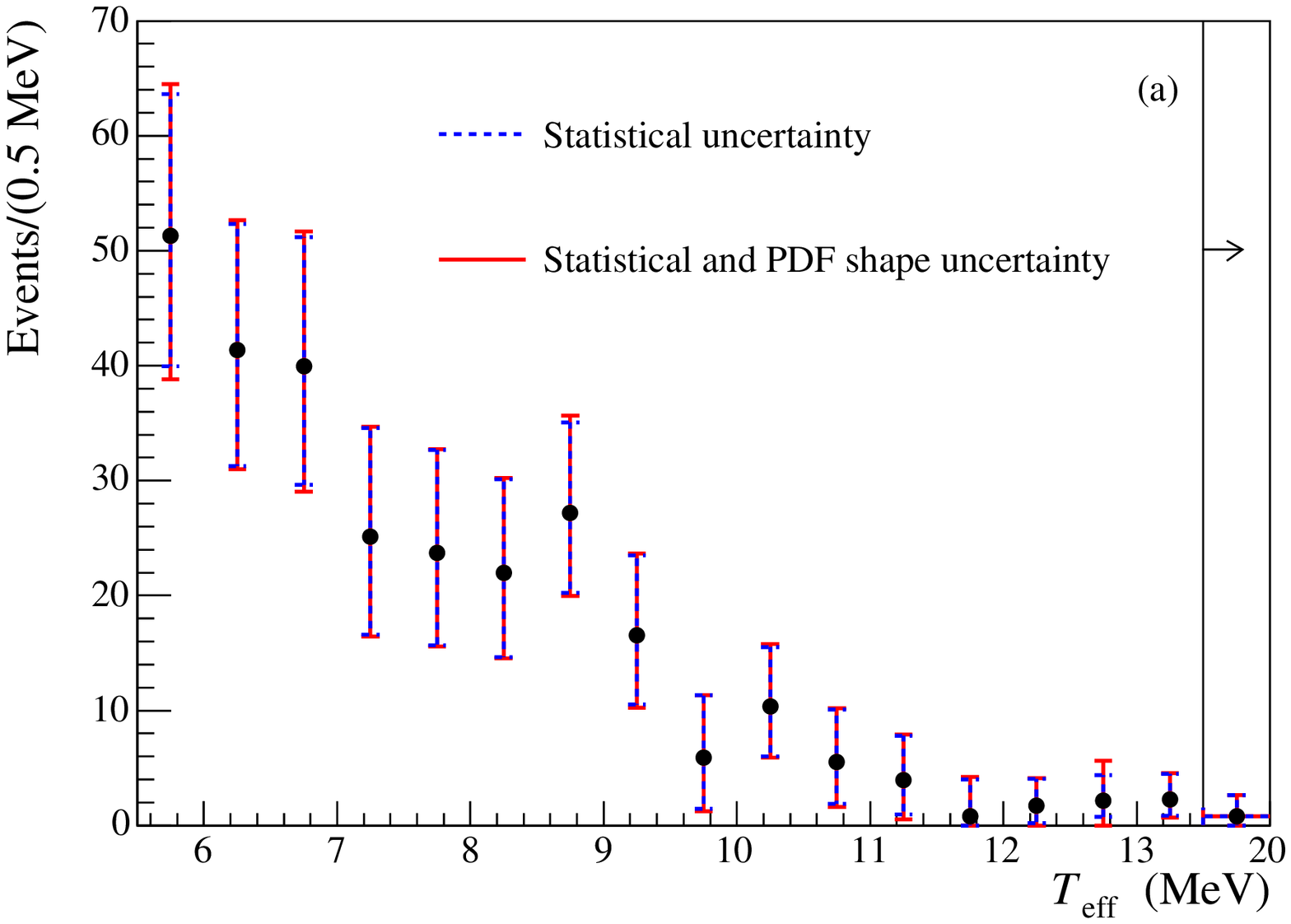}
\includegraphics[width=3.6in]{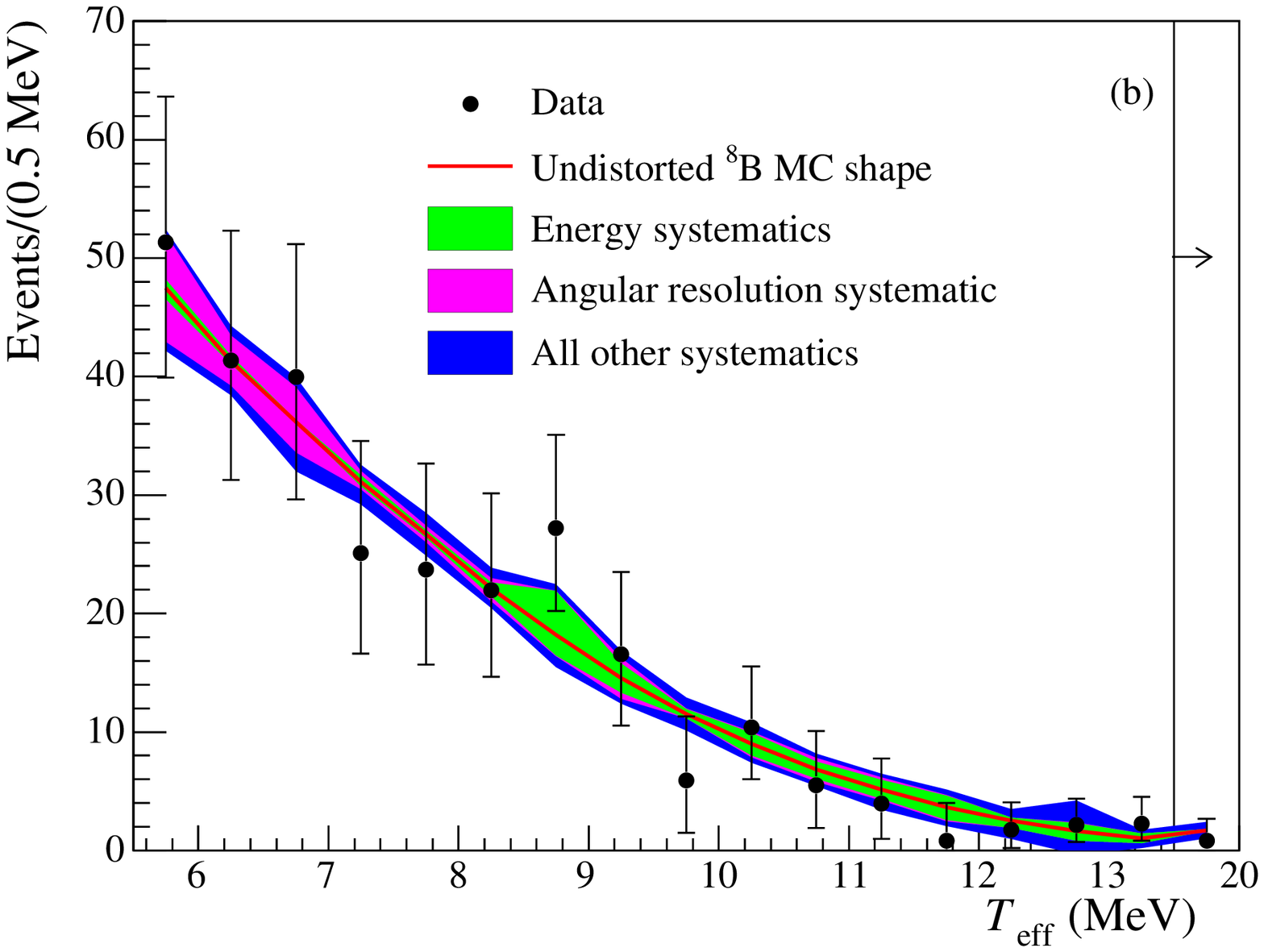}
\caption{\label{fig:es_diff_fit}{(a) Extracted ES $\teff$ spectrum.
Systematic uncertainties have been combined in quadrature and
include only the effect of PDF shape change.  (b)
Extracted ES $\teff$ spectrum with statistical error bars compared to
predictions for an undistorted \be shape with combined systematic
uncertainties, including both shape and acceptance components.  The
systematic error bands represent the fraction of the total uncertainty
attributable to the given quantity.}}
\end{center}
\end{figure}

Figure~\ref{fig:cc_diff_fit}(a) shows the CC energy spectrum extracted
from the energy-unconstrained fit. The PDF shape change component (see
below) of the systematic uncertainties is added in quadrature with the
statistical error to provide a combined error for each bin.  The
analogous ES spectrum is presented in Fig.~\ref{fig:es_diff_fit}(a).

Systematic uncertainties on the extracted CC energy spectrum are
calculated by separately varying the PDFs according to the estimated
1$\sigma$ uncertainties on the detector parameters. Signal extraction
is then repeated and the differences between the nominal fit values
and shifted PDF fit values taken as the spectrum systematic
uncertainties.
                                                                                
Systematic uncertainties on the spectrum are divided into those caused
by changes to the PDF shapes, and those resulting from uncertainties
in overall acceptance.  The former lead to errors in the fitted number
of charged current events in each spectral bin, while the latter lead
to errors in the translation of differential event counts into
differential neutrino fluxes.  For example, shifting the energy scale
alters both the PDF shapes and the acceptance in each bin, while a
change in the radial scale (and hence the effective fiducial volume)
primarily affects the acceptance.  For a particular neutrino spectrum
model, the fitting uncertainties from the PDFs shapes and the
uncertainties on the acceptance have correlations with each other
which must be taken into account when calculating bin-by-bin neutrino
fluxes.  A list of these uncertainties is given in
Table~\ref{tab:spect_sys} for the $\teff$ intervals $5.5-6.0$ MeV,
$7.0-7.5$ MeV and $9.0-9.5$ MeV.  Note that an undistorted $^8$B shape
has been assumed for generation of these uncertainties.  The complete
table of uncertainties for all energy bins is included in the
Appendix.

Figure~\ref{fig:flux_energysys} shows the CC spectrum PDF shape
systematic effects and combined systematics for each of the four
energy-related systematic uncertainties versus $\teff$ under the
assumption of an undistorted $^8$B shape.
Figures~\ref{fig:cc_diff_fit}(b) and \ref{fig:es_diff_fit}(b) show the
extracted CC and ES spectra as a functions of $\teff$, with
statistical error bars on the data, and combined systematic
uncertainty error bands centered on the prediction for an undistorted
$^8$B shape (normalized to the data).  Note that the error bands
correspond to the fraction of the total error, calculated as the ratio
of the square of the given uncertainty to the square of the total,
attributed to the indicated source (e.g., \bof systematic errors).
\begingroup
\squeezetable
\begin{table}
\caption{\label{tab:spect_sys} CC spectrum combined shape and
  acceptance systematic uncertainties for three sample energy
  ranges.  An undistorted $^8$B shape has been assumed.
Note that the full table is given in the Appendix.}
\renewcommand{\arraystretch}{1.1}
\begin{center}
\begin{tabular}{l c c c }\hline\hline
Source & \multicolumn{3}{c}{CC differential uncertainty (\%)}  \\
        & (5.5--6.0\,MeV) & (7.0--7.5\,MeV) &  (9.0--9.5\,MeV)    \\ \hline
           Energy scale (const.) &-8.0,\;\; 9.5  & 0.6,\;\; 0.0  & 4.7,\;\; -5.0\\
           Energy scale (E dep.) &-1.5,\;\; 1.6  &-0.4,\;\; 0.6  & 1.3,\;\; -0.4\\
              Energy radial bias &-5.7,\;\; 6.1  & 0.1,\;\;-0.1  & 3.8,\;\; -2.9\\
               Energy resolution & 4.9,\;\;-4.9  &-0.1,\;\; 0.1  &-2.6,\;\;  2.6\\
              $\beta_{14}$ scale & 7.1,\;\;-8.4  & 4.4,\;\;-4.6  & 1.1,\;\; -1.7\\
              $\beta_{14}$ width &-0.9,\;\;-0.1  & 0.0,\;\;-0.2  & 0.3,\;\; -0.2\\
         Radial scale (const.)   &-2.6,\;\; 2.5  &-2.4,\;\; 2.6  &-2.5,\;\;  2.8\\
           Radial scale (E dep.) & 0.2,\;\;-0.2  &-1.2,\;\; 1.2  &-2.1,\;\;  2.1\\
                      Vertex $x$ &-0.3,\;\; 0.0  &-0.5,\;\; 0.0  & 0.1,\;\; -0.1\\
                      Vertex $y$ &-0.2,\;\;-0.3  &-0.2,\;\; 0.0  & 0.1,\;\;  0.4\\
                      Vertex $z$ &-0.4,\;\;-0.2  & 0.2,\;\;-0.6  &-0.3,\;\; -0.1\\
               Vertex resolution &-0.4,\;\; 0.4  &-0.3,\;\; 0.3  & 0.4,\;\; -0.4\\
              Angular resolution &-0.2,\;\; 0.2  & 0.1,\;\;-0.1  &-0.6,\;\;  0.6\\
               Internal $\gamma$ & 0.4,\;\;-0.6  & 0.1,\;\;-0.1  & 0.0,\;\;  0.0\\
            Selection efficiency &-0.2,\;\; 0.2  &-0.1,\;\; 0.2  &-0.1,\;\;  0.2\\
                     Backgrounds &-8.5,\;\; 0.0  &-0.1,\;\; 0.0  & -0.1,\;\; +0.0\\
\hline\hline
\end{tabular}
\renewcommand{\arraystretch}{1.0}
\end{center}
\end{table}
\endgroup
\begin{figure}
\begin{center}
\includegraphics[width=3.6in]{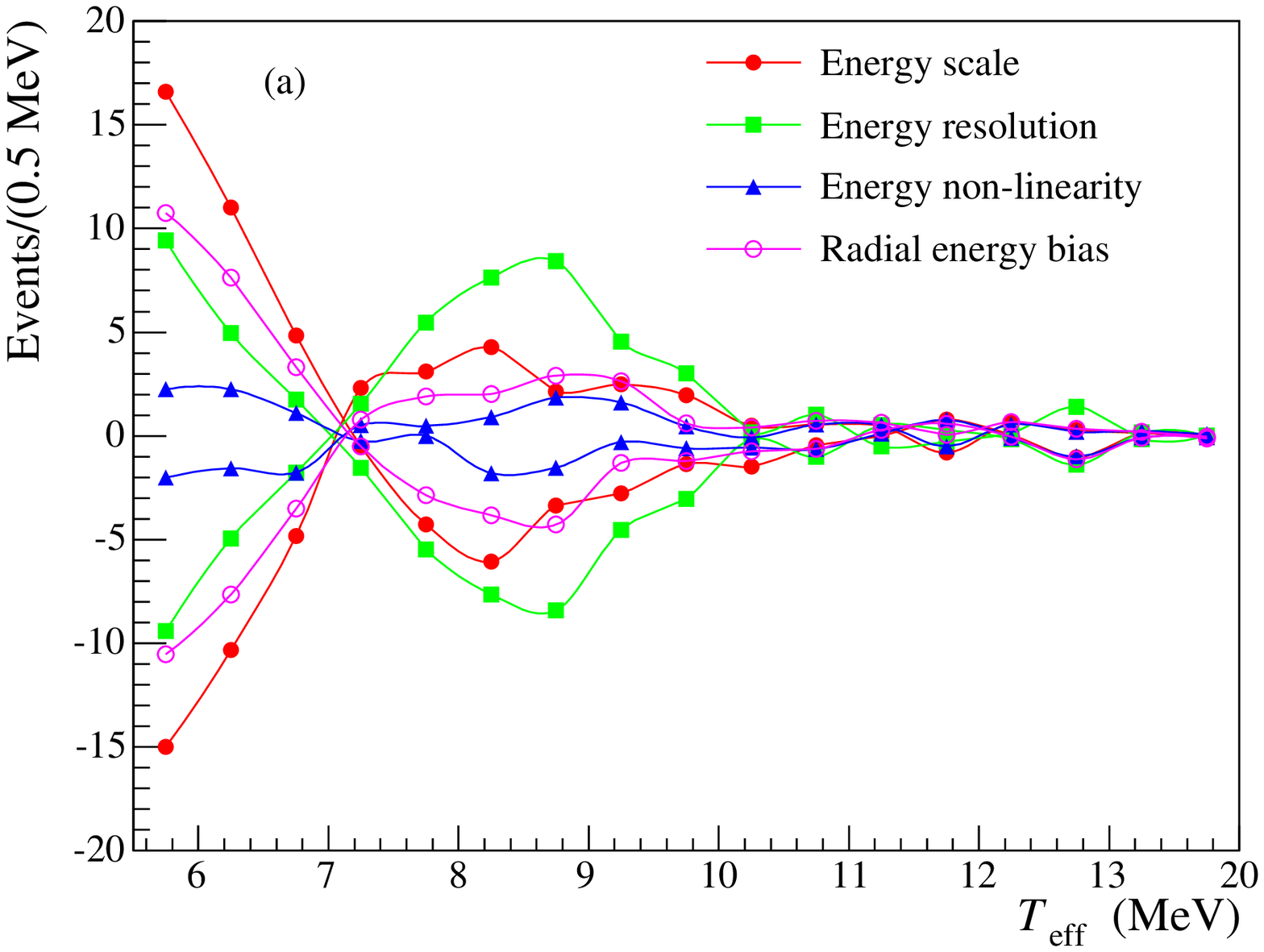}
\includegraphics[width=3.6in]{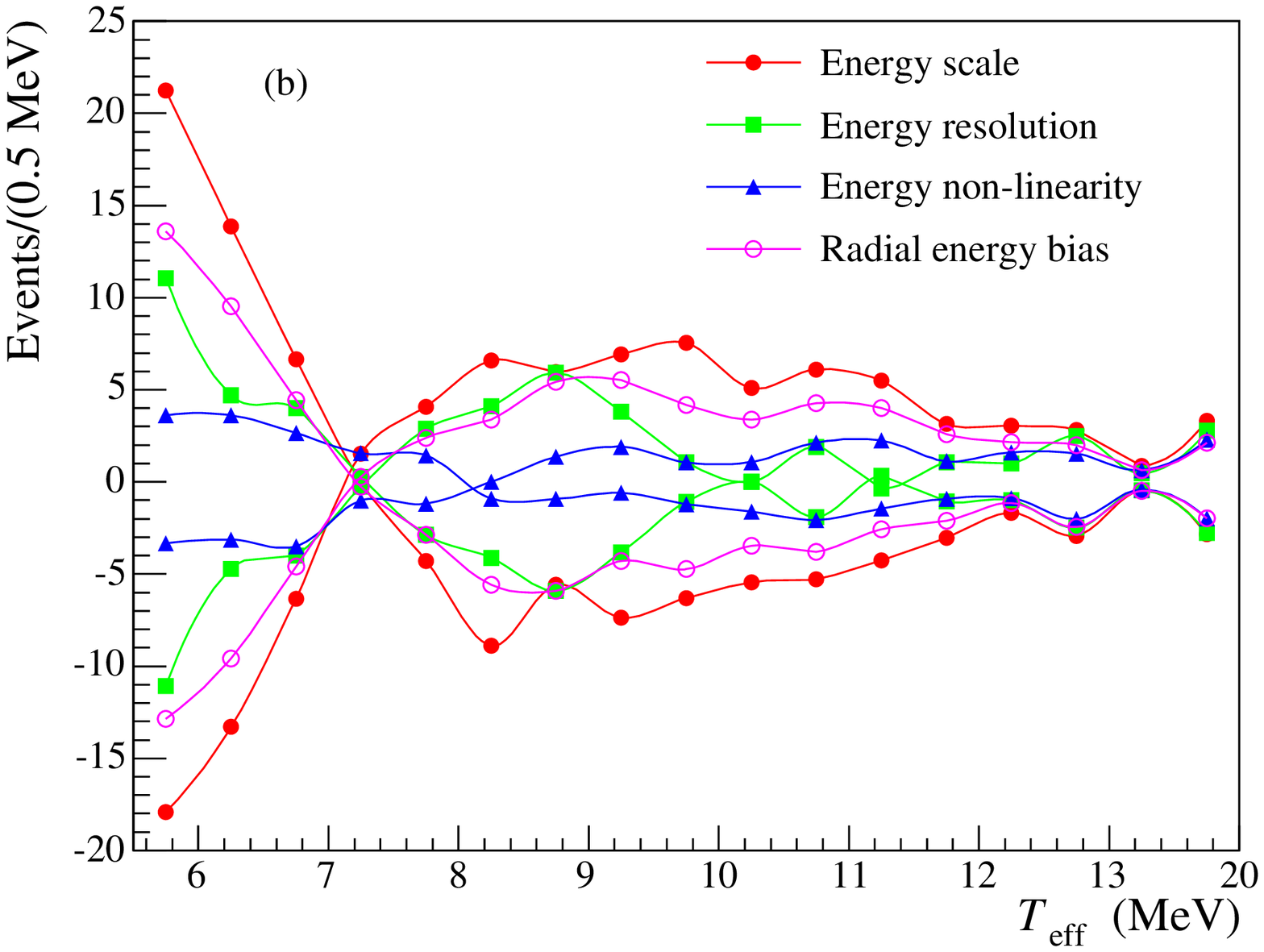}
\caption{\label{fig:flux_energysys}{(a) PDF shape change contributions
    and (b) total contributions from energy-related
systematic energy uncertainties to the extracted 
CC spectrum assuming an underlying, undistorted \be shape. }}
\end{center}
\end{figure}

\section{\label{sec:intflux}Solar Neutrino Flux Results}

\begin{figure}
\begin{center}
\psfrag{COSSUN}{~~~~~~$\cos \theta_{\odot}$}
\includegraphics[width=3.73in]{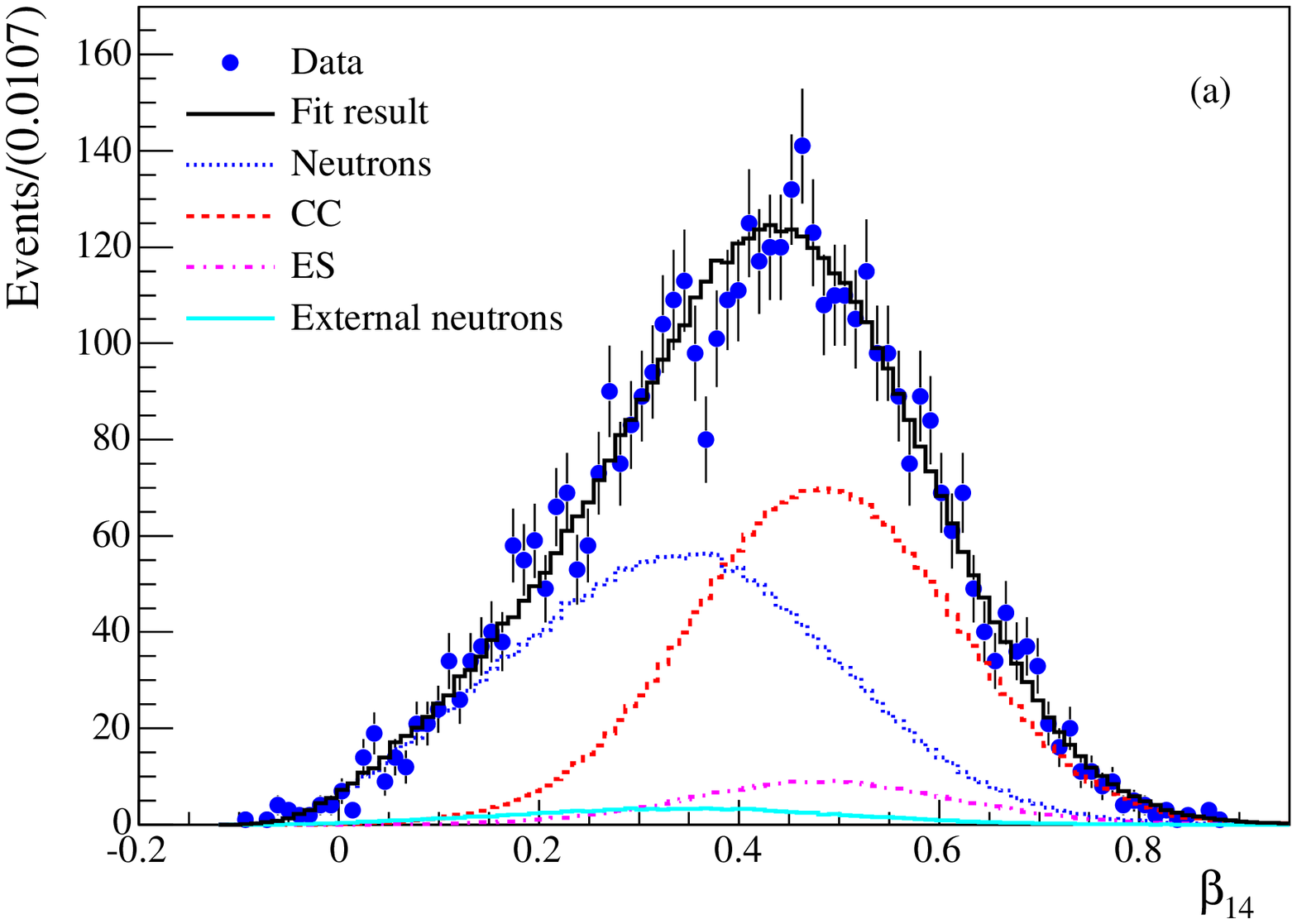}
\includegraphics[width=3.73in]{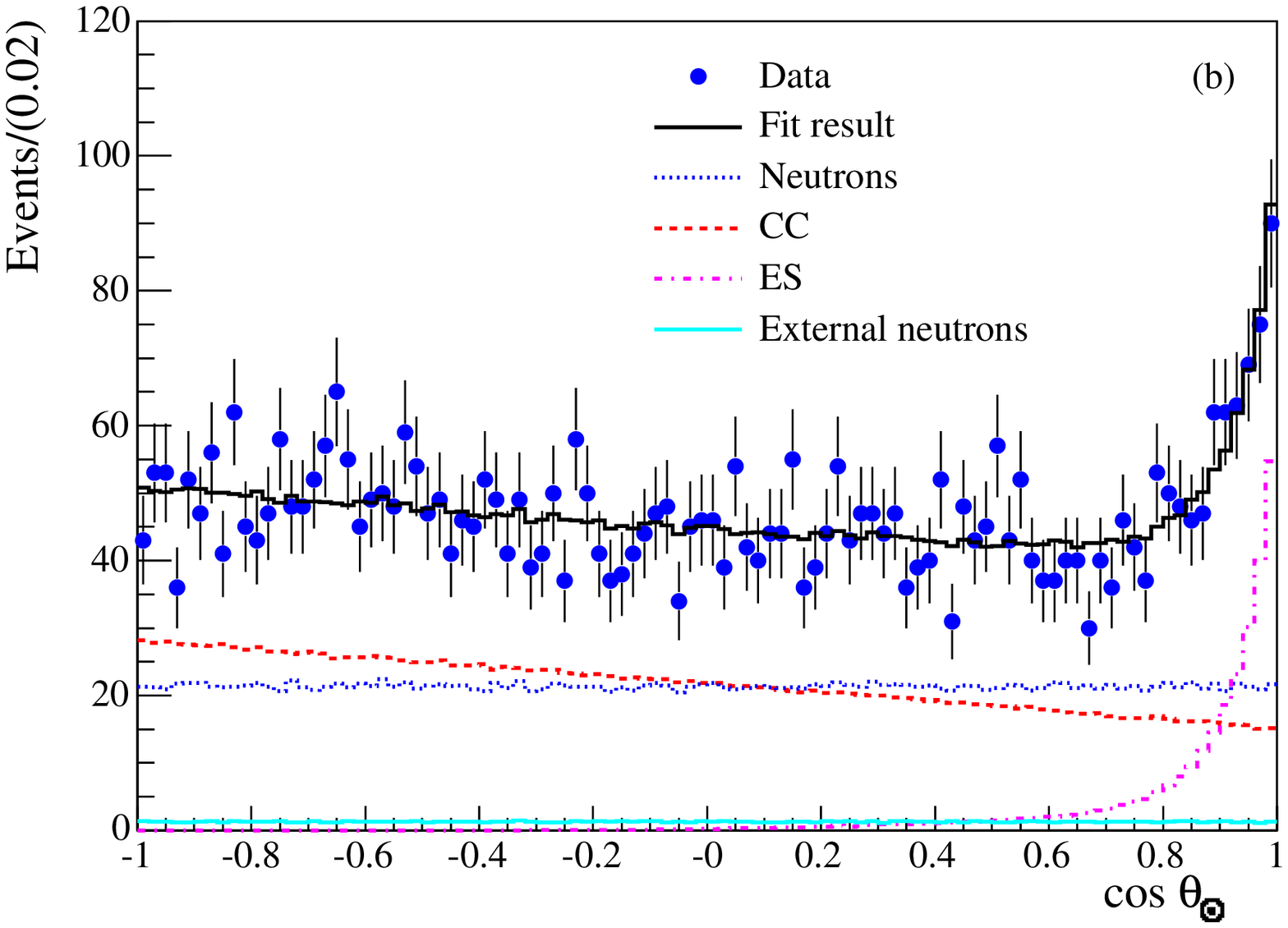}
\includegraphics[width=3.73in]{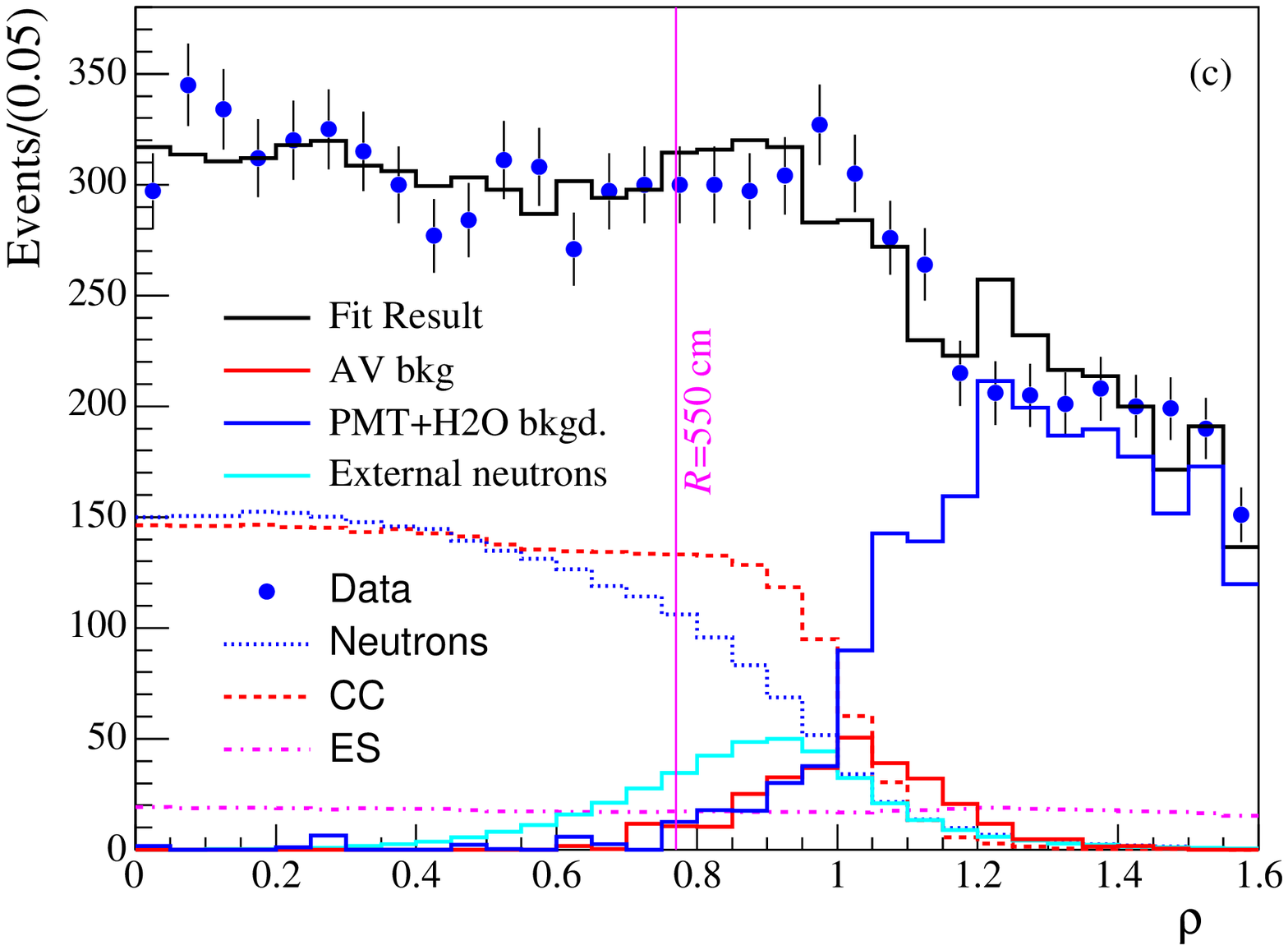}
\caption{\label{fig:salt_data}Distribution of (a) $\beta_{14}$, (b)
 $\cos \theta_{\odot}$ and (c) volume-weighted radius $\rho$.  Points
 with error bars represent data while the MC predictions for CC, ES,
 NC + internal and external-source neutron events, all scaled to the
 energy-unconstrained fit results, are as indicated in the legend.
 The dark solid lines represent the summed components.  The (a) and
 (b) distributions are for events with $T_{\rm eff}\geq 5.5$~MeV and
 R$_{\rm fit}\leq 550$~cm, and are averaged assuming an undistorted
 \be spectrum.  The same energy cut has been applied for (c) but
 events are shown out to $\rho<1.6$, where $\rho=1.0$ is the edge of
 the heavy water volume.  The dashed vertical line represents the
 550~cm fiducial volume cut.}
\end{center}
\end{figure}

As described in a previous section, a statistical analysis to separate
the neutrino candidate events into CC, NC, ES, and external-source
neutrons was performed using an extended maximum likelihood technique
using the \bof, \costs and $\rho$ distributions.  The extended maximum
likelihood analysis yielded $\nccfit$ CC, $\nesfit$ ES, $\nncfit$ NC,
and $\nncbefit$ external-source neutron events (recall that 125.4
internal neutrons and 3.5 internal $\gamma$-ray events have been
subtracted off).  Note that a fixed \textit{hep} contribution of $9.3
\times 10^{3}$ cm$^{-2}$ s$^{-1}$ has been assumed.  The systematic
uncertainties on the derived fluxes for this ``energy-unconstrained''
fit are shown in Table~\ref{tab:flux_uncert}.  The \bof,
$\cos\theta_{\odot}$, and $\rho$ distributions for the selected events
are presented in Fig.~\ref{fig:salt_data} with only statistical
uncertainties shown.
\begingroup
\squeezetable
\begin{table}
\caption{\label{tab:flux_uncert}Systematic uncertainties (\%) on fluxes for
  the energy-unconstrained analysis of the salt data
  set.  Note that ``const.'' denotes an energy-independent systematic component
and ``E dep'' an energy-dependent part. }
\begin{center}
\begin{ruledtabular}
\begin{tabular}{llllll}
  Source       & NC uncert. (\%) & CC uncert. (\%) & ES uncert. (\%) \\ 
         \hline 
              Energy scale (const.)   &-3.3,\;\; +3.8  &-0.9,\;\; +1.0  &-1.6,\;\; +1.9  \\
               Energy scale (E dep.)  &-0.1,\;\; +0.1  &-0.1,\;\; +0.1  &-0.1,\;\; +0.1  \\
                  Energy radial bias  &-2.0,\;\; +2.1  &-0.6,\;\; +0.7  &-1.1,\;\; +1.2  \\
                    Energy resolution &-0.8,\;\; +0.8  &-0.2,\;\; +0.2  &-0.7,\;\; +0.7   \\
        $\beta_{14}$ mean (const.)   &-3.6,\;\; +4.5  &-4.0,\;\; +3.7  &-1.2,\;\; +1.3  \\
         $\beta_{14}$ mean (E dep.)  &-0.1,\;\; +0.2  &-0.2,\;\; +0.0  &-0.0,\;\; +0.1  \\
                 $\beta_{14}$ width   &-0.0,\;\; +0.0  &-0.2,\;\; +0.2  &-0.2,\;\; +0.2  \\
              Radial scale (const.)   &-3.0,\;\; +3.3  &-2.6,\;\; +2.5  &-2.6,\;\; +3.0  \\
               Radial scale (E dep.)  &-0.6,\;\; +0.5  &-0.9,\;\; +0.8  &-0.7,\;\; +0.8  \\
                          Vertex $x$  &-0.0,\;\; +0.0  &-0.0,\;\; +0.0  &-0.1,\;\; +0.1  \\
                          Vertex $y$  &-0.1,\;\; +0.0  &-0.0,\;\; +0.0  &-0.1,\;\; +0.1  \\
                          Vertex $z$  &-0.2,\;\; +0.2  &-0.1,\;\; +0.1  &-0.0,\;\; +0.0  \\
                    Vertex resolution &-0.1,\;\; +0.1  &-0.1,\;\; +0.1  &-0.1,\;\; +0.1  \\
                  Angular resolution  &-0.2,\;\; +0.2  &-0.4,\;\; +0.4  &-5.1,\;\; +5.1  \\
                Internal neutron bkgd.&-1.9,\;\; +1.6  &-0.0,\;\; +0.0  &-0.0,\;\; +0.0  \\
              Internal $\gamma$ bkgd. &-0.1,\;\; +0.1  &-0.1,\;\; +0.1  &-0.0,\;\; +0.0  \\
                  Internal \Ckv bkgd. &-0.9,\;\; +0.0  &-0.9,\;\; +0.0  &-0.0,\;\; +0.0  \\
                  External \Ckv bkgd. &-0.2,\;\; +0.0  &-0.2,\;\; +0.0  &-0.0,\;\; +0.0  \\
                   Instrumental bkgd. &-0.4,\;\; +0.0  &-0.3,\;\; +0.0  &-0.0,\;\; +0.0  \\
                 Neutron capture eff. &-2.3,\;\; +2.1  &-0.0,\;\; +0.0  &-0.0,\;\; +0.0  \\
\hline
Total systematic                      &-6.9,\;\; +7.6  &-5.1,\;\; +4.7  &-6.2,\;\; +6.5  \\
\hline
Cross section~\cite{bib:crosssection} & $\pm 1.1$  & $\pm 1.2 $& $\pm
  0.5$ \\
\hline
Total statistical    &$\pm4.2$  &$\pm3.7$  &$\pm9.3$\\
\end{tabular}
\end{ruledtabular}
\end{center}
\end{table}
\endgroup
\begingroup
\squeezetable
\begin{table}
\caption{\label{tab:flux_uncert_con}Systematic uncertainties (\%) on fluxes for
  the energy-constrained analysis of the salt data
  set.  Note that ``const.'' denotes an energy-independent systematic component
and ``E dep'' an energy-dependent part.  }
\begin{center}
\begin{ruledtabular}
\begin{tabular}{lllllll}
  Source       & NC uncert. (\%) & CC uncert. (\%) & ES uncert. (\%) \\ 
         \hline 
       Energy scale (const.)    &-0.3,\;\; +0.7  &-3.7,\;\; +3.9  &-1.8,\;\; +1.6  \\
        Energy scale (E dep.)   &-0.9,\;\; +1.0  &-1.0,\;\; +1.0  &-0.2,\;\; +0.2  \\
            Energy radial bias  &-0.1,\;\; +0.1  &-2.5,\;\; +2.6  &-1.0,\;\; +0.9  \\
             Energy resolution  &-2.1,\;\; +2.1  &-1.1,\;\; +1.1  &-0.6,\;\; +0.6 \\
 $\beta_{14}$ mean (const.)    &-2.2,\;\; +3.0  &-2.4,\;\; +2.0  &-0.5,\;\; +2.3  \\
  $\beta_{14}$ mean (E dep.)   &-0.2,\;\; +0.2  &-0.2,\;\; +0.2  &-0.7,\;\; +0.7  \\
        $\beta_{14}$ width      &-0.0,\;\; +0.0  &-0.1,\;\; +0.1  &-0.8,\;\; +0.8  \\
        Radial scale (const.)   &-3.0,\;\; +3.3  &-2.7,\;\; +2.6  &-1.9,\;\; +2.9  \\
         Radial scale (E dep.)  &-0.2,\;\; +0.2  &-1.3,\;\; +1.2  &-0.8,\;\; +0.8  \\
                    Vertex $x$  &-0.0,\;\; +0.1  &-0.0,\;\; +0.0  &-0.1,\;\; +0.1  \\
                    Vertex $y$  &-0.1,\;\; +0.0  &-0.0,\;\; +0.0  &-0.2,\;\; +0.2  \\
                    Vertex $z$  &-0.1,\;\; +0.1  &-0.1,\;\; +0.0  &-0.0,\;\; +0.0  \\
              Vertex resolution &-0.1,\;\; +0.1  &-0.2,\;\; +0.2  &-0.7,\;\; +0.7  \\
            Angular resolution  &-0.2,\;\; +0.2  &-0.4,\;\; +0.4  &-4.9,\;\; +4.9  \\
         Internal neutron bkgd. &-1.9,\;\; +1.6  &-0.0,\;\; +0.0  &-0.0,\;\; +0.0  \\
         Internal $\gamma$ bkgd.&-0.2,\;\; +0.1  &-0.1,\;\; +0.0  &-0.0,\;\; +0.1  \\
            Internal \Ckv bkgd. &-0.9,\;\; +0.0  &-0.8,\;\; +0.0  &-0.0,\;\; +0.0  \\
            External \Ckv bkgd. &-0.2,\;\; +0.0  &-0.2,\;\; +0.0  &-0.0,\;\; +0.0  \\
             Instrumental bkgd. &-0.4,\;\; +0.0  &-0.3,\;\; +0.0  &-0.0,\;\; +0.0  \\
           Neutron capture eff. &-2.3,\;\; +2.1  &-0.0,\;\; +0.0  &-0.0,\;\; +0.0  \\
\hline
 Total systematic              &-5.4,\;\; +5.7  &-6.2,\;\; +6.0  &-5.9,\;\; +6.6  \\ 
\hline
Cross section~\cite{bib:crosssection} & $\pm 1.1$  & $\pm 1.2 $& $\pm
         0.5$ \\
\hline
Total Statistical &$\pm3.9$ &$\pm3.1$ &$\pm9.8$\\
\end{tabular}
\end{ruledtabular}
\end{center}
\end{table}
\endgroup

For the unconstrained analysis, the quoted CC and ES fluxes are the
equivalent fluxes of $^8$B electron-neutrinos, assuming an undistorted
$^8$B energy spectral shape, that would produce the same CC and ES
event rates above the analysis threshold of $\teff = 5.5$ MeV.  For
the NC case, the quoted flux is the flux of all active neutrino types
that would produce the same NC rate above the reaction threshold of
2.2 MeV.  The fitted numbers of events give the equivalent $^8$B
fluxes~\cite{bib:hep_footnote,bib:es_crosssection}~(in units of
$10^6~{\rm cm}^{-2} {\rm s}^{-1}$)
\begin{eqnarray*}
\phi^{\text{uncon}}_{\text{CC}} & = & \snoccfluxunc \\
\phi^{\text{uncon}}_{\text{ES}} & = & \snoesfluxunc \\
\phi^{\text{uncon}}_{\text{NC}} & = & \snoncfluxunc~\mbox{,}
\end{eqnarray*} and the ratios of the CC flux to NC and ES
respectively are
\begin{eqnarray*}
\frac{\phi^{\text{uncon}}_{\text{CC}}}{\phi^{\text{uncon}}_{\text{NC}}}
= \snoccncratiounc \\
\frac{\phi^{\text{uncon}}_{\text{CC}}}{\phi^{\text{uncon}}_{\text{ES}}}
= \snoccesratiounc. \\
\end{eqnarray*}
Note that the uncertainties on the ratios are not normally
distributed.

\begin{figure}
\begin{center}
\includegraphics[width=3.73in]{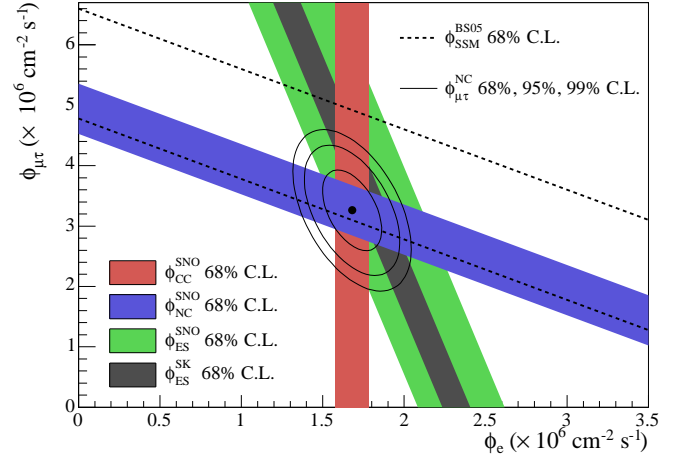}
\caption{\label{fig:salt_flavor}Flux of $\mu+\tau$ neutrinos versus
 flux of electron neutrinos.  CC, NC and ES flux measurements are
 indicated by the filled bands.  The total \be solar neutrino flux
 predicted by the Standard Solar Model~\cite{bib:bs05} is shown as
 dashed lines, and that measured with the NC channel is shown as the
 solid band parallel to the model prediction.  The narrow band
 parallel to the SNO ES result correponds to the Super-Kamiokande
 result in \cite{bib:superk}.  The intercepts of these bands with the
 axes represent the $\pm 1\sigma$ uncertainties.  The non-zero value
 of $\phi_{\mu\tau}$ provides strong evidence for neutrino flavor
 transformation.  The point represents $\phi_e$ from the CC flux and
 $\phi_{\mu\tau}$ from the NC-CC difference with 68\%, 95\%, and 99\%
 C.L. contours included.}
\end{center}
\end{figure}

The non-$\nu_{\text{e}}$ active neutrino component ($\phi_{\mu\tau}$)
of the $^8$B flux can be determined by subtracting the $\phi_e$
component, as measured by the CC flux, from the NC and ES fluxes.
Whereas the NC measurement is equally sensitive to all active
neutrinos, the ES measurement has reduced sensitivity to non-electron
neutrinos in the form $\phi_{\text{ES}}=\phi_e +
0.1553\phi_{\mu\tau}$.  The resulting $\phi_{\mu\tau}$ fluxes, in
units of $10^6$ cm$^{-2}$ s$^{-1}$, are
\begin{eqnarray*}
\phi^{\text{NC,uncon}}_{\mu\tau} = \snoncmccunc \\
\phi^{\text{ES,uncon}}_{\mu\tau} = \snoesmccunc.\\
\end{eqnarray*}
Figure~\ref{fig:salt_flavor} shows the flux of non-electron flavor
active neutrinos ($\phi_{\mu\tau}$) versus the flux of electron
neutrinos ($\phi_e$).  The error ellipses shown are the 68\%, 95\%
and 99\% joint probability contours for $\phi_{\mu\tau}$ and
$\phi_e$. 

Adding the constraint of an undistorted ${}^{8}$B energy spectrum to
the signal extraction yields, for comparison with earlier results~(in units of
$10^6~{\rm cm}^{-2} {\rm s}^{-1}$):
\begin{eqnarray*}
\phi^{\text{con}}_{\text{CC}} & = & \snoccfluxconc \\
\phi^{\text{con}}_{\text{ES}} & = & \snoesfluxconc \\
\phi^{\text{con}}_{\text{NC}} & = & \snoncfluxconc,
\end{eqnarray*}
with corresponding ratios
\begin{eqnarray*}
\frac{\phi^{\text{con}}_{\text{CC}}}{\phi^{\text{con}}_{\text{NC}}}&=&\snoccncratiocon\\
\frac{\phi^{\text{con}}_{\text{CC}}}{\phi^{\text{con}}_{\text{ES}}}&=&\snoccesratiocon,
\end{eqnarray*}
and $\phi_{\mu\tau}$ values,  in
units of $10^6$ cm$^{-2}$ s$^{-1}$, 
\begin{eqnarray*}
\phi^{\text{NC,con}}_{\mu\tau} = \snoncmcccon \\
\phi^{\text{ES,con}}_{\mu\tau} = \snoesmcccon.\\
\end{eqnarray*}
The ``energy-constrained'' fit is sensitive to somewhat different
systematic uncertainties than the unconstrained fit as shown in
Table \ref{tab:flux_uncert_con}.  The correlation matrix  for the
constrained fit, including
correlations with the external-neutron component, is given in Table
\ref{tab:concor}.
\begin{table}
\caption{\label{tab:concor}Correlation matrix for the
  constrained fit.  External neutrons component is labeled EN.}
\begin{center}
\begin{tabular}{ccccc} 
\hline
\hline 
&NC & CC & ES & EN \\
\hline
NC &   1.000 & & &\\
CC &  -0.400 & 1.000 & &\\
ES &  -0.073 &-0.168 & 1.000 &\\
EN &  -0.472 &-0.039 &-0.012 & 1.000\\
\hline\hline
\end{tabular}
\end{center}
\end{table} 

Compared to the initial salt phase results \cite{bib:saltprl}, some
systematic uncertainties have slightly increased.  In particular, more
detailed analyis of calibration source data during the full salt
data set has generated larger systematic uncertainty estimates on mean
\bof and its energy dependence, and on angular resolution.  The
combined systematic uncertainties for the CC and NC fluxes have not
increased.  The ES flux systematic uncertainties have increased, but
the ES measurement is still dominated by statistical uncertainty which
has decreased with the increased statistics.  For the $\phi_{CC}$/$\phi_{NC}$ ratios,
however, the effects of \bof systematic uncertainties are highly
anti-correlated, and consequently the $\phi_{CC}$/$\phi_{NC}$ systematic uncertainties are
larger than that reported in \cite{bib:saltprl}.

\begin{figure}
\begin{center}
\includegraphics[width=3.73in]{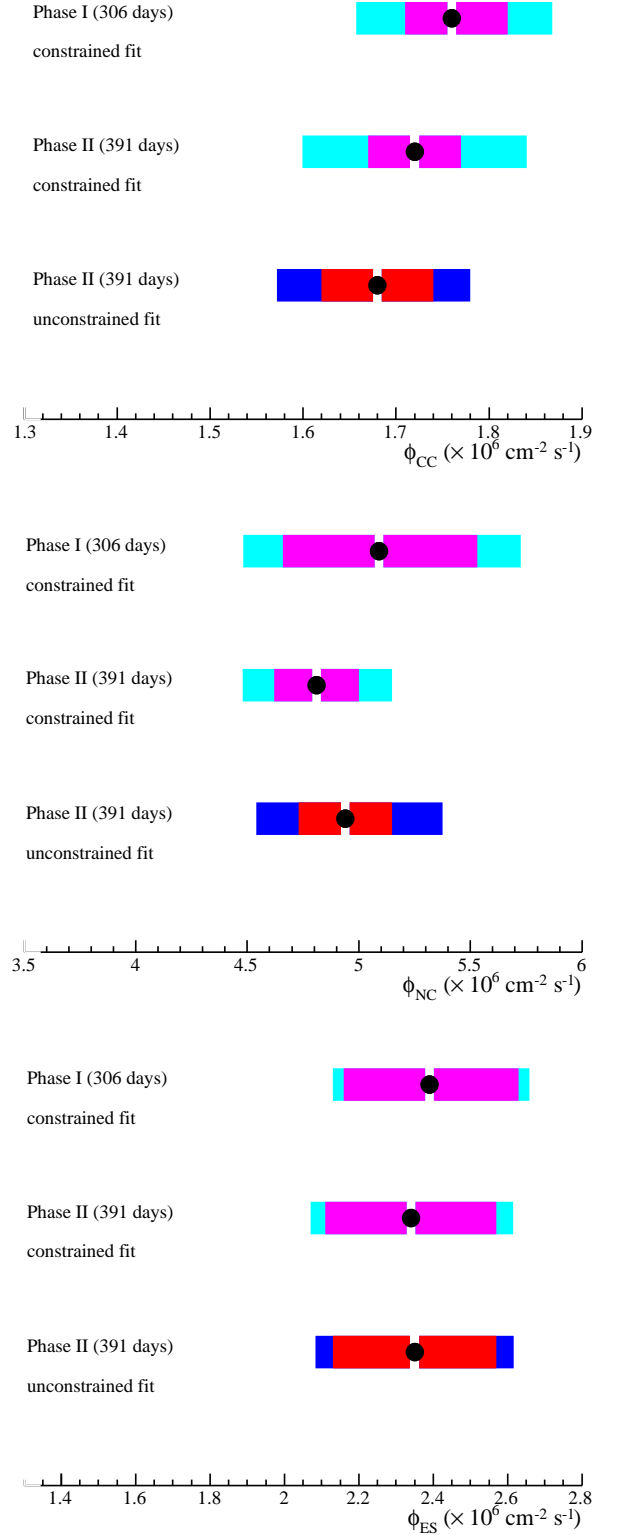}
\caption{\label{fig:d2osalt}Comparison of phase I (306 days)
  and phase II (391 days) flux results.  For each case the inner error bar
  represents the statistical uncertainty while the full error bar
  represents the combined statistical and systematic uncertainty. }
\end{center}
\end{figure}

As shown in Fig.~\ref{fig:d2osalt} and Table~\ref{tab:allresults},
these results are consistent with the pure \dto phase
results~\cite{bib:snocc,bib:snonc}.  Comparisons with the 254-day salt
data measured fluxes ~\cite{bib:saltprl}, also given in
Table~\ref{tab:allresults}, show some small differences for the
unconstrained fit case.  The differences are consistent with
$\sim1\sigma$ statistical fluctuations, including covariances, between
the 254-day data set of Ref.~\cite{bib:saltprl} and the additional
137-day data set included here. Part of the difference may also be
attributed to uncorrelated components of the systematic uncertainties
between the 254-day and 137-day data sets.
\begin{table}
\caption{\label{tab:allresults} Constrained and unconstrained flux
  results from phase I and phase II SNO data sets in units of
$10^6~{\rm cm}^{-2} {\rm s}^{-1}$.  Note that the phase
  I $\teff$ threshold was lower than the phase II threshold.}
\begin{center}
\begin{tabular}{lccc} \hline\hline
 \multicolumn{4}{c}{Constrained fit} \\
\hline   
Data Set& 
$\phi^{\text{con}}_{\text{CC}}$ & 
$\phi^{\text{con}}_{\text{NC}}$ & 
$\phi^{\text{con}}_{\text{ES}}$ \vspace{0.1cm}\\
\hline
\vspace{0.2cm}Phase I (306 days)\cite{bib:snonc} & 
$1.76^{+0.06+0.09}_{-0.05-0.09}$ &
$5.09^{+0.44+0.46}_{-0.43-0.43}$ &
$2.39^{+0.24+0.12}_{-0.23-0.12}$ \\ \vspace{0.1cm}
Phase II (254 days)\cite{bib:saltprl} & 
$1.70^{+0.07+0.09}_{-0.07-0.10}$ &
$4.90^{+0.24+0.29}_{-0.24-0.27}$ &
$2.13^{+0.29+0.15}_{-0.28-0.08}$ \\\vspace{0.1cm}
Phase II (391 days) & 
$1.72^{+0.05+0.11}_{-0.05-0.11}$ &
$4.81^{+0.19+0.28}_{-0.19-0.27}$ &
$2.34^{+0.23+0.15}_{-0.23-0.14}$ \\
\hline
 \multicolumn{4}{c}{Unconstrained Fit}\\
\hline
Data Set &$\phi^{\text{uncon}}_{\text{CC}}$ & 
$\phi^{\text{uncon}}_{\text{NC}}$ & 
$\phi^{\text{uncon}}_{\text{ES}}$ \vspace{0.1cm}\\
\hline
Phase I (306 days)\cite{bib:snonc}&& $6.42^{+1.57+0.55}_{-1.57-0.58}$&\\\vspace{0.1cm}
Phase II (254 days)\cite{bib:saltprl}&
$1.59^{+0.08+0.06}_{-0.07-0.08}$ &
$5.21^{+0.27+0.38}_{-0.27-0.38}$ &
$2.21^{+0.31+0.10}_{-0.26-0.10}$ \\\vspace{0.1cm}
Phase II (391 days)&
$1.68^{+0.06+0.08}_{-0.06-0.09}$ &
$4.94^{+0.21+0.38}_{-0.21-0.34}$ &
$2.35^{+0.22+0.15}_{-0.22-0.15}$ \\
\hline\hline
\end{tabular}
\end{center}
\end{table}

\section{Day-Night Asymmetries of Solar Neutrino Flux}
\label{sec:dnresults}

For certain ranges of mixing parameters, matter-enhanced neutrino
oscillations predict $\nu_e$ regeneration inside the
Earth~\cite{bib:theo1,bib:theo2,bib:theo3,
bib:theo4,bib:theo5,bib:theo6,bib:theo7}.  This effect would manifest
itself as a day-night asymmetry in the charged current and elastic
scattering rates in SNO.  For standard neutrino oscillations between
active flavors, the neutral current rate should not vary between day
and night.  An observation of a day-night asymmetry in the neutral
current rate could be evidence for an admixture of sterile neutrinos,
or for unexpected matter interactions inside the earth.  To search for
day-night asymmetries in solar neutrino reaction rates, the day-night
asymmetry ratio $\asymmA = 2(\phi_{N}-\phi_{D})/(\phi_{N}+\phi_{D})$
is constructed using the day and night fluxes $\phi_{D}$ and
$\phi_{N}$ for each reaction.  The day and night neutrino fluxes at
SNO can be measured either allowing for a neutral current day-night
asymmetry, or constraining $\asymmA_{\rm NC} = 0$ as predicted by
standard neutrino oscillations with only active flavors.

SNO's data set was divided into day and night portions, defined by the
Sun being above or below the horizon respectively.  Separate day and
night signal probability distributions were built for CC, ES, and NC
events from Monte Carlo simulations that properly included the live
time exposure of the data set.  The day-night analysis used the same
event selection, analysis cuts, and background estimates as the
integral flux and energy spectral analyses.  Each background was
divided between the day and night data sets according to its measured
diurnal rate asymmetry.  Fits of the signal and background PDFs to the
data sets determined the night and day neutrino fluxes for each type
of interaction.  In addition to day and night live time
corrections, the fluxes were corrected for seasonal variations in
the neutrino rate due to the eccentricity of the Earth's orbit.  The
eccentricity corrections were determined from Monte Carlo simulations.

Because most systematics cancel when forming a day-night ratio, the
dominant uncertainties on day-night ratios are statistical.  To avoid
introducing statistical biases into the analysis, the entire data set
was divided into a 20\% ``open'' portion and an 80\% ``closed''
portion.  The 20\% open fraction was sampled uniformly from each run.
All analysis cuts and procedures were developed and tested based on
measurement of day-night ratios on only the 20\% open data set.  Then the
analysis procedures were frozen, and day-night asymmetries were
calculated for the 391-day data set.  Day-night results for the 20\%
open data set and for the full data set are statistically consistent.
Only results for the total data set are reported here.

Systematic uncertainties on the day-night asymmetries are described in
Sec.~\ref{sec:dnspecific}.  The effects of day-night differences in
detector response have been determined by perturbing the day and night
signal PDFs by the detector response uncertainty for each systematic.
This perturbation technique is described in Sec.~\ref{sec:sigex}.
Uncertainties in backgrounds have been propagated by varying the
amplitude and the day-night rate asymmetry of each background in the
flux fits.

\subsection{Total Event Rate}

The simplest day-night analysis is to compare the total event rates
(signals + backgrounds) between day and night.
Table~\ref{tab:dn_rawrates} shows these results.  The day and night
rates are statistically consistent.  Because the external neutron
background is determined from fits to the data itself, backgrounds
cannot be subtracted from the raw event rates without doing a full
signal extraction fit.
\begin{table}
\caption{\label{tab:dn_rawrates}Event totals and rates for the day and night data sets.}
\begin{center}
\begin{tabular}{lcc} \hline\hline
  & Events & Rate (day$^{-1}$) \\
\hline
Day & 2134 & $12.09 \pm 0.26$ \\
Night & 2588 & $12.04 \pm 0.24$ \\ \hline\hline
\end{tabular}
\end{center}
\end{table}

\subsection{Model-Independent Day-Night Asymmetries}
\label{sec:dn_modelindependent}

The most general day-night analysis is to fit for the day and night
neutrino fluxes separately, placing no constraint on
$\asymmA_{\rm NC}$ and making no assumption about the energy
dependence of the $\nu_e$ oscillation probability.  The results
include day and night NC fluxes, and separate day and night CC energy
spectra. 
\begingroup
\squeezetable
\begin{table}
\caption{Day-night integral fluxes and asymmetries from a
shape-unconstrained signal extraction.  Fluxes are in units of $10^6$
neutrinos cm$^{-2}$ sec$^{-1}$.  The 
systematic uncertainties on the day and
night fluxes include large correlated systematics that cancel in the
day-night asymmetry ratio $A$.
\label{tab:dn_fluxes_unconstrained} }
\begin{center}
\begin{tabular}{lccc} \hline\hline
Signal  & Day flux & Night flux & $A$ \\
\hline
CC & $   1.73\pm 0.09 \pm 0.10$ &$ 1.64\pm 0.08 \pm 0.09$ & $ -0.056\pm
  0.074\pm  0.053$ \\
NC & $ 4.81\pm  0.31\pm 0.39$ & $ 5.02 \pm 0.29  \pm 0.41$ & $0.042 \pm
   0.086\pm 0.072$ \\
ES & $2.17 \pm 0.34 \pm 0.14$ & $2.52\pm 0.32  \pm 0.16$ & $ 0.146 \pm
0.198 \pm 0.033 $ \\ \hline\hline
\end{tabular}
\end{center}
\end{table}
\endgroup

Table~\ref{tab:dn_fluxes_unconstrained} presents the day and night
integral fluxes from the shape-unconstrained analysis.  Each pair of day-night
fluxes shares some large common systematics, as calculated for the
integral flux analysis in Sec.~\ref{sec:intflux}.  The day and
night fluxes are statistically independent, however.  The asymmetry
ratio $A$ for each flux includes a statistical uncertainty,
and a systematic uncertainty due to day-night specific effects as
described in Sec.~\ref{sec:dnspecific}.  All asymmetries are
consistent with zero.

\begin{figure}
\begin{center}
\includegraphics[width=3.6in]{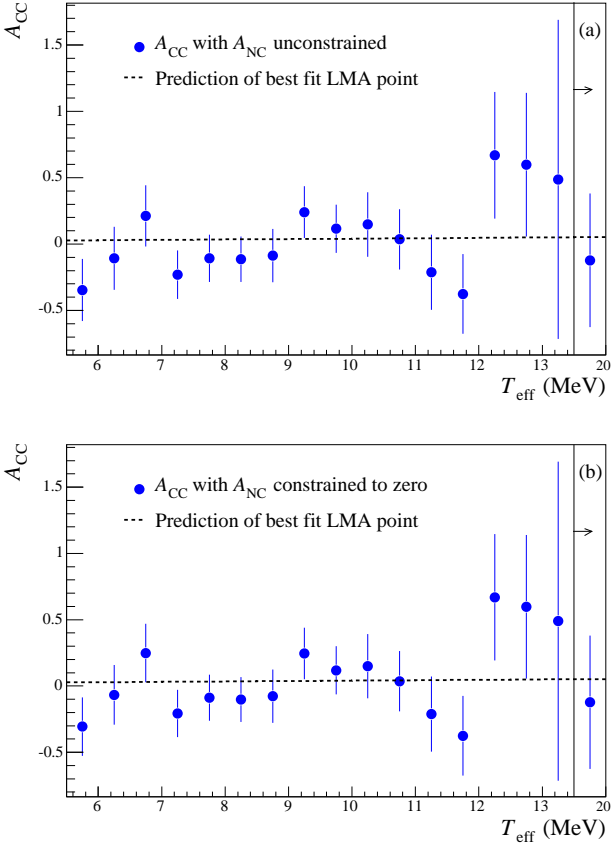}
\caption{\label{fig:cc_asymmetry_spectrum} Day-night asymmetries on each 
CC energy bin as a function of electron energy.  Panel (a) shows the case in 
which no constraint is made on $\asymmA_{\rm NC}$.  Panel (b) shows the case 
in which $\asymmA_{\rm NC}$ is constrained to zero.  Uncertainties are 
statistical only.  The vertical lines in each figure show the expectation for 
$\Delta m^2 = 7 \times 10^{-5}$~eV$^2$ and $\tan^2 \theta = 0.40$.}
\end{center}
\end{figure}

Figure~\ref{fig:cc_asymmetry_spectrum}(a) shows the value of
$\asymmA_{\rm CC}$ in each energy bin $\asymmA_{{\rm CC},i}$.
Overlaid is the expectation for the previous best-fit mixing
parameters $\Delta m^2 = 7 \times 10^{-5}$~eV$^2$ and $\tan^2 \theta =
0.40$~\cite{bib:saltprl}.  The dependence of $\asymmA_{\rm CC}$ on CC electron energy is
consistent with this expectation, but is also consistent with no
day-night asymmetries.

\begin{figure}
\begin{center}
\includegraphics[width=3.6in]{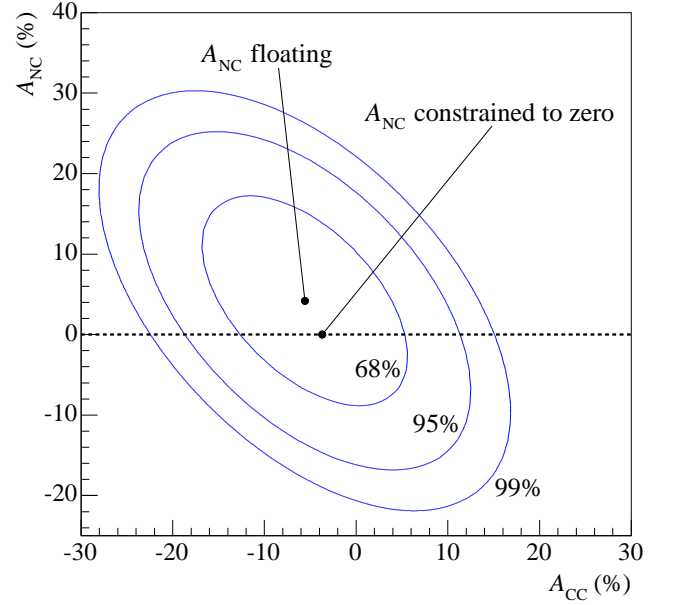}
\caption{\label{fig:acc_vs_anc} Joint probability contours for
$\asymmA_{\rm NC}$ versus $\asymmA_{\rm CC}$ (as \%), statistical uncertainties
only.  The points indicate the
results when $\asymmA_{\rm NC}$ is allowed to float and when it is
constrained to zero.  }
\end{center}
\end{figure}

Although the day and night fluxes are statistically independent, the
CC, ES, and NC fluxes for either day or night are statistically
correlated since they are produced from a common fit.  As a result,
$\asymmA_{\rm CC}$, $\asymmA_{\rm NC}$, and $\asymmA_{\rm ES}$ are
statistically correlated, with correlation coefficients given by:
\begin{eqnarray}
\nonumber
\rho(\mbox{CC, NC}) &=& -0.532 \\ \nonumber
\rho(\mbox{CC, ES}) &=& -0.147 \\ \nonumber
\rho(\mbox{ES, NC}) &=& -0.064.
\end{eqnarray}

Similarly, the $\asymmA_{{\rm CC},i}$ are modestly correlated between
different energy bins at lower energies due to their common covariance
with $\asymmA_{\rm NC}$.  Figure~\ref{fig:acc_vs_anc} is a contour
plot of $\asymmA_{\rm NC}$ versus $\asymmA_{\rm CC}$, illustrating the
covariances.  

\begingroup
\squeezetable
\begin{table}
\caption{Systematic uncertainties on day-night asymmetries for the
shape-unconstrained signal extraction. 
For presentation, uncertainties have been symmetrized and rounded.
\label{tab:dn_syserr_unconstrained}
}
\begin{center}
\begin{tabular}{lccc} \hline\hline
Systematic & $\asymmA_{\rm CC}$ uncert. & $\asymmA_{\rm NC} $ uncert. &
$\asymmA_{\rm ES}$ uncert. \\
\hline
Diurnal energy scale               & 0.004   & 0.015   & 0.007 \\
Directional energy scale           & 0.001   & 0.000   & 0.014 \\
Long-term energy scale variation   & 0.002   & 0.010   & 0.001 \\
Diurnal energy resolution          & 0.003   & 0.006   & 0.004 \\
Directional energy resolution      & 0.001   & 0.001   & 0.003 \\
Diurnal vertex shift               & 0.008   & 0.012   & 0.007 \\
Directional vertex shift           & 0.000   & 0.000   & 0.003 \\
Diurnal vertex resolution          & 0.002   & 0.006   & 0.002 \\
Directional vertex resolution      & 0.000   & 0.000   & 0.001 \\
Diurnal isotropy                    & 0.050  & 0.064   & 0.017 \\
Directional isotropy               & 0.002   & 0.002   & 0.004 \\
Long-term isotropy variation       & 0.014   & 0.015   & 0.006 \\
Directional angular resolution     & 0.001   & 0.001   & 0.020 \\
Live time                          & 0.000   & 0.000   & 0.000 \\
Cut acceptance                     & 0.003   & 0.004   & 0.003 \\
External \Ckv tail bkgd. amplitude  & 0.002   & 0.003   & 0.000 \\
External \Ckv tail bkgd. asymmetry  & 0.003   & 0.004   & 0.000 \\
Internal \Ckv tail bkgd. amplitude  & 0.001   & 0.000   & 0.000 \\
Internal \Ckv tail bkgd. asymmetry  & 0.001   & 0.001   & 0.000 \\
Internal neutron bkgd. amplitude    & 0.000   & 0.003   & 0.000 \\
Internal neutron bkgd. asymmetry    & 0.000   & 0.015   & 0.000 \\
Internal $\gamma$ bkgd. amplitude   & 0.000   & 0.000   & 0.000 \\
Internal $\gamma$ bkgd. asymmetry   & 0.000   & 0.000   & 0.000 \\
Isotropic AV bkgd. amplitude        & 0.001   & 0.001   & 0.000 \\
Isotropic AV bkgd. asymmetry        & 0.002   & 0.002   & 0.000 \\
Instrumental bkgd. amplitude        & 0.000   & 0.001   & 0.000 \\
Instrumental bkgd. asymmetry        & 0.001   & 0.002   & 0.000 \\
\hline
Total                               & 0.053  & 0.072  & 0.033  \\
\hline\hline
\end{tabular}
\end{center}
\end{table}
\endgroup

Table~\ref{tab:dn_syserr_unconstrained} lists each systematic
uncertainty on the integral flux asymmetries from the
shape-unconstrained day-night analysis.  The largest systematics are
uncertainties on diurnal variations of the isotropy parameter
$\beta_{14}$, and diurnal variations in energy scale and vertex
shift.  For the ES rate, directional systematics are significant.
However, the overall uncertainties on all asymmetries are ultimately
limited by statistics. 

\subsection{Shape-Constrained Day-Night Asymmetries}
\label{sec:dn_shapeconstrained}

A variant of the preceding analysis is to constrain the day and night
$\nu_e$ energy spectra to follow an undistorted $^8$B shape. This
corresponds to an energy-independent oscillation probability that
varies between night and day.  The NC rate was again allowed to vary
in the fit.  It should be noted that standard neutrino oscillations
with mixing parameters in the LMA region do not predict
energy-independent day-night asymmetries.  The derived asymmetries
under the assumption that the CC and ES energy spectra are
undistorted, but allowing them to have different normalizations
between night and day, are
\begin{eqnarray}
\asymmA_{\rm CC} &=&  -0.021 \pm 0.063\pm 0.035\nonumber\\
\asymmA_{\rm NC} &=&  0.018 \pm  0.079\pm 0.052\nonumber\\
\asymmA_{\rm ES} &=&  0.066\pm 0.198 \pm 0.057.
\end{eqnarray}

\subsection{Day-Night Asymmetries with the Constraint
${\bm{\asymmA_{\rm NC}\equiv 0}}$} 

In the previous two subsections the NC flux was allowed to vary in the
fit between the day and night data sets.  Under the standard picture of
matter-enhanced neutrino oscillations, $\asymmA_{\rm NC}$ should be
zero.  This prediction has been confirmed by the results of the
previous two subsections.  When determining the best estimate of the
day-night asymmetry on the electron neutrino flux, {\em assuming
standard neutrino oscillations}, it is appropriate to constrain
$\asymmA_{\rm NC}$.  This constraint has been applied by
simultaneously fitting the day and night data sets, not allowing
$\phi_{\rm NC}$ to vary between night and day.

The additional constraint of $\asymmA_{\rm NC} \equiv 0$ reduces the
statistical uncertainties on $\asymmA_{\rm CC}$ and
$\asymmA_{\rm ES}$.  It also produces a modest covariance between the
day and night fluxes, due to their common covariance with
$\phi_{\rm NC}$.  In contrast, without a constraint on $\asymmA_{\rm NC}$
the day and night neutrino fits are statistically independent.
\begingroup
\squeezetable
\begin{table}
\caption{Day-night integral fluxes from a shape-unconstrained signal
extraction, with the constraint $\asymmA_{\rm NC}\equiv 0$.  Fluxes are in units of $10^6$
neutrinos cm$^{-2}$ sec$^{-1}$.
\label{tab:dn_fluxes_unconstrained_anc0}
}
\begin{center}
\begin{tabular}{lccc} \hline\hline
Signal  & Day flux & Night flux & Asymmetry \\
\hline
CC &  $  1.71\pm 0.08 \pm 0.09$&$1.65 \pm  0.08 \pm 0.09$& $-0.037\pm
0.063\pm 0.032$ \\
ES &$  2.18\pm 0.34\pm 0.14$ & $ 2.53\pm   0.32\pm 0.16$ & $ 0.153\pm
0.198 \pm 0.030 $\\
\hline
NC & \multicolumn{2}{c}{ $ 4.93 \pm 0.21 \pm 0.36$ } &
${\asymmA}_{\rm NC}\equiv{0}$ \\ \hline\hline
\end{tabular}
\end{center}
\end{table}
\endgroup

The day and night neutrino fluxes were fit in a shape-unconstrained
analysis, requiring $\asymmA_{\rm NC}\equiv 0$.
Table~\ref{tab:dn_fluxes_unconstrained_anc0} gives the day and night
integral fluxes from this fit, and the NC flux.  No statistically
significant asymmetries are observed.  Forcing $\asymmA_{\rm NC}\equiv 0$ results
in some reduction in $|\asymmA_{\rm CC}|$, as expected from the
anti-correlation of CC and NC event totals in the signal extraction.  
\begin{figure}
\begin{center}
\includegraphics[width=3.6in]{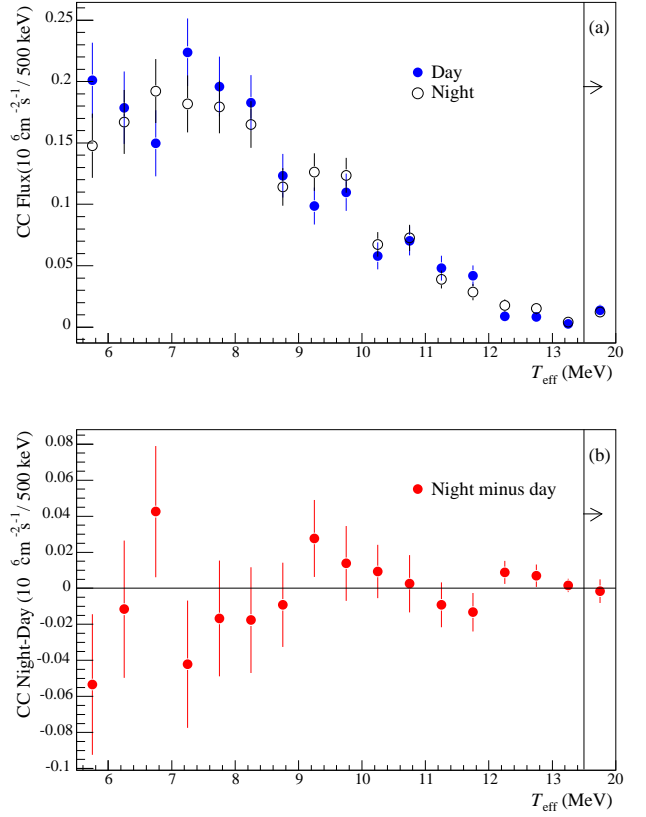}
\caption{\label{fig:dn_spectra_anc0} 
(a) Day and night extracted CC energy
spectra with statistical uncertainties only.  Bin values are expressed in
units of equivalent $^8$B flux, normalized such that the sum of the flux bin
values above 5.5 MeV equals the total integral $^8$B neutrino flux above 0
MeV, as determined for the day and night integral fluxes quoted in section
XII(D) (see Appendix A).  (b) Difference \emph{night - day}
between the spectra.  In both figures, the final bin extends to $20$~MeV.}
\end{center}
\end{figure}

Figures~\ref{fig:cc_asymmetry_spectrum}(b) and
\ref{fig:dn_spectra_anc0} show the CC asymmetry as a function of
electron energy and the day and night CC energy spectra, binned by
electron energies.  The additional constraint does not significantly
change the results.  Table~\ref{tab:dn_syserr_unconstrained_anc0}
presents the systematic uncertainties on $\asymmA_{\rm CC}$ and
$\asymmA_{\rm ES}$ for this analysis.
\begingroup
\squeezetable
\begin{table}
\caption{Systematic uncertainties on day-night asymmetries for the
shape-unconstrained signal extraction, with the constraint 
$\asymmA_{\rm NC}\equiv0$.
For presentation, uncertainties have been symmetrized and rounded.
\label{tab:dn_syserr_unconstrained_anc0}
}
\begin{center}
\begin{tabular}{lcc} \hline\hline
Systematic  & $\asymmA_{\rm CC}$ uncert. & $\asymmA_{\rm ES}$ uncert. \\
\hline
Diurnal energy scale               &   0.009  & 0.009   \\
Directional energy scale           &   0.001  & 0.014   \\
Long-term energy scale variation   &   0.006  & 0.002   \\
Diurnal energy resolution          &   0.002  & 0.004   \\
Directional energy resolution      &   0.001  & 0.002   \\
Diurnal vertex shift               &   0.013  & 0.009   \\
Directional vertex shift           &   0.000  & 0.003   \\
Diurnal vertex resolution          &   0.001  & 0.002   \\
Directional vertex resolution      &   0.000  & 0.001   \\
Diurnal isotropy                   &   0.022  & 0.009   \\
Directional isotropy               &   0.001  & 0.005   \\
Long-term isotropy variation       &   0.013  & 0.003   \\
Directional angular resolution     &   0.002  & 0.019   \\
Live time                          &   0.000  & 0.000   \\
Cut acceptance                     &   0.003  & 0.003   \\
External \Ckv tail bkgd. amplitude  &   0.002  & 0.000   \\
External \Ckv tail bkgd. asymmetry  &   0.003  & 0.000   \\
Internal \Ckv tail bkgd. amplitude  &   0.001  & 0.000   \\
Internal \Ckv tail bkgd. asymmetry  &   0.001  & 0.000   \\
Internal neutron bkgd. amplitude    &   0.002  & 0.001   \\
Internal neutron bkgd. asymmetry    &   0.007  & 0.002   \\
Internal $\gamma$ bkgd. amplitude   &   0.000  & 0.000   \\
Internal $\gamma$ bkgd. asymmetry   &   0.000  & 0.000   \\
Isotropic AV bkgd. amplitude        &   0.001  & 0.000   \\
Isotropic AV bkgd. asymmetry        &   0.002  & 0.000   \\
Instrumental bkgd. amplitude        &   0.001  & 0.000   \\
Instrumental bkgd. asymmetry        &   0.001  & 0.000   \\
\hline
Total                         & 0.032  & 0.030   \\ \hline \hline
\end{tabular}
\end{center}
\end{table}
\endgroup

\subsection{Shape-Constrained Day-Night Asymmetries with the
Constraint ${\bm{\asymmA_{\rm NC} \equiv 0}}$}
\label{sec:dn_allconstrained}

For the sake of completeness the analysis of
Sec.~\ref{sec:dn_shapeconstrained} has been repeated with the
additional constraint that $\asymmA_{\rm NC} \equiv 0$.  The results
are
\begin{eqnarray}
\asymmA_{\rm CC} &=&  -0.015\pm  0.058 \pm 0.027 \nonumber \\
\asymmA_{\rm ES} &=& 0.070\pm 0.197  \pm 0.054.
\end{eqnarray}

\subsection{Combined Day-Night Asymmetries from the Salt and D$_2$O Data
Sets}

During the first phase of the SNO experiment, the asymmetry on the
electron neutrino flux was measured to be $\asymmA_e = 0.070 \pm 0.049
^{+0.013}_{-0.012}$, assuming a standard $^8$B shape and constraining
$\asymmA_{\rm NC} = 0$.  Although an asymmetry ratio formed from two
normally-distributed variables is not necessarily normally
distributed, for the case of SNO's day and night fluxes a normal
distribution is an excellent approximation for the true distribution
we calculate for $\asymmA_{\rm CC}$ or $\asymmA_e$.  The asymmetry
results from SNO's first phase are statistically independent of the
results from the salt data set and statistical uncertainties dominate
over systematics for the asymmetries.  Hence, combining values and
uncertainties for $\asymmA_{\rm CC}$ from
Sec.~\ref{sec:dn_allconstrained} and $\asymmA_e$ from above can be
done trivially to produce a combined, albeit model-dependent,
day-night asymmetry of $\asymmA_{e,{\rm combined}} = 0.037 \pm 0.040$.
A future SNO analysis will address the issue of doing a joint
shape-unconstrained fit to the D$_2$O and salt data sets.
                                                                                
The Super-Kamiokande collaboration has measured the day-night rate
asymmetry of ES interactions above 5~MeV to be $\asymmA_{\rm ES} =
0.021 \pm 0.020 ^{+0.012}_{-0.013}$~\cite{bib:sk_dn}.  Because ES
interactions can be initiated by either $\nu_e$ or $\nu_{\mu\tau}$,
the day-night asymmetry for ES events is diluted by a factor of
$(\phi_e + 0.1576(\phi_{tot}-\phi_e))/((1-0.1576)\phi_e) = 1.55$.
Assuming an energy-independent conversion mechanism and only active
neutrinos, the Super-Kamiokande result scales to a $\phi_e$ asymmetry
of $\asymmA_{e,{\rm SK}} = 0.033 \pm 0.031 ^{+0.019}_{-0.020}$.
Combining the SNO D$_2$O and SNO salt values for $\asymmA_e$ with the
equivalent Super-Kamiokande value $\asymmA_{e,{\rm SK}}$ gives
$\asymmA_{e,{\rm combined}} = 0.035 \pm 0.027$.  This result
explicitly assumes a shape-constrained $^8$B spectrum for SNO and
ignores energy dependence of the oscillation probability over the
energy range in question.

\section{\label{sec:physint}MSW Interpretation of Results}

The observation of a substantially suppressed $\nu_e$ flux with the CC 
reaction in SNO compared to the total active flux measured by the NC 
reaction in SNO provides clear evidence for neutrino flavor change that 
can be analyzed in terms of neutrino oscillations.  Constraints on neutrino 
mixing parameters can be derived by comparing neutrino oscillation model 
predictions with experimental data, as has been done in, for example, 
\cite{bib:theo6,bib:bahcall_global,bib:balantekin_global,bib:fogli_global,
bib:smirnov_global,bib:aliani_global} and 
in previous SNO analyses \cite{bib:snodn,bib:saltprl}.

A two-flavor, active neutrino oscillation model has two parameters: 
$\Delta m^2$, the difference between the square of the masses of the 
relevant eigenstates of propagation for the neutrinos, and $\tan^2 \theta$ 
that quantifies the strength of the mixing between flavor and mass 
eigenstates.  Note that the three-flavor mixing matrix element $U_{e2}$ 
can be written as $\cos \theta_{13} \sin \theta_{12}$ \cite{bib:mnsp}, 
which is approximately equal to $\sin \theta$ for two-flavor solar neutrino 
oscillations when $\theta_{13}$ is small and when 
$\Delta m^2_{\rm sol} \ll \Delta m^2_{\rm atm}$.  The MSW effect 
\cite{bib:msw} can result in neutrinos above a few MeV emerging from the 
Sun essentially as a pure $\nu_2$ state (e.g.\ for oscillation parameters 
in the Large Mixing Angle region).  To the degree to which this statement 
is true, SNO's $\phi_{CC}$/$\phi_{NC}$  ratio, a direct measure of the $\nu_e$ survival 
probability, is also a direct measure of $\left| U_{e2} \right|^2$ and 
thus should be approximately equal to $\sin^2 \theta$. For the sake of 
comparison with other past and present oscillation analyses, this present 
work still employs $\tan^2 \theta$ to quantify the mixing angle for solar 
neutrino oscillations.

For each pair of parameters, the oscillation model predicts the expected 
rates in the Cl \cite{bib:homestake} and Ga experiments 
\cite{bib:sage,bib:gno}, Super-Kamiokande zenith spectra \cite{bib:superk}, 
and SNO rates and spectra.  The model prediction accounts for MSW propagation 
of neutrino states through dense matter in the Sun and the Earth and so allows 
for the regeneration of $\nu_e$ flavor for neutrinos passing through the 
Earth at night.  A global $\chi^2$ calculation can be performed; best-fit 
parameters can be determined and allowed parameter regions can be identified 
using $\Delta \chi^2$ confidence levels for two degrees of freedom.  The 
same neutrino oscillation model can also predict rates and spectra for 
the KamLAND experiment \cite{bib:kamland,bib:kl2004}, assuming CPT 
invariance (since KamLAND detects $\bar{\nu}_e$).  The likelihood values from 
a KamLAND oscillation analysis can be easily combined with that from the global 
solar neutrino analysis to further restrict the allowed oscillation 
parameter space.

For the analysis presented in this paper, 
earlier data from SNO-I (pure D$_2$O phase)
have been included.  SNO-I day and night spectra
have been interpreted in a similar manner as before 
\cite{bib:snodn,bib:pured2o_howto,bib:saltprl}.  Summed 
spectra (CC+ES+NC+background) predictions were compared to the number 
of counts in each spectral bin from the SNO-I data, for both day and 
night. A minor improvement in the analysis of this
earlier data is that this part of the calculation now 
includes energy-dependent $\nu$-$d$ radiative corrections for the 
CC reaction.  Previously the CC radiative correction was included
as an energy-independent factor.

In \cite{bib:saltprl}, salt phase ``fluxes'' (i.e., CC, NC and ES fluxes 
inferred from rates) were added to the global $\chi^2$ analysis. The 
present work has a new oscillation analysis using data from the 391-day 
data set of the salt phase which have been analyzed and extracted as CC 
spectra and NC and ES integrated fluxes, separately for day and night.  
This information is included in the global $\chi^2$ analysis in lieu of just 
salt phase fluxes inferred from rates.  This allows CC spectral shape 
information and day-night rate asymmetry information from the salt phase 
to be included in the global oscillation analysis.  CC-NC separation is 
preserved in this analysis since the SNO unconstrained signal extraction 
utilized information from event isotropy $\beta_{14}$ and angular 
correlation $\cos \theta_\odot$ distributions for separating the salt 
NC and ES fluxes from the CC spectra.

SNO's unconstrained signal extraction produced two 19$\times$19 
statistical covariance matrices (one for day and one for night) for 
17 spectral bins of the CC spectrum, starting from 5.5 MeV kinetic 
energy up to 13.5 MeV, in 0.5 MeV steps, with one extra bin 
integrating from 13.5--20.0 MeV, plus the NC and ES fluxes.  These 
statistical covariance matrices are required in the calculation of 
$\chi^2$ and are available in Appendix A\@. 
Day and night data are statistically independent from 
each other and the results with no constraint on $\asymmA_{\rm NC}$ were used.

Systematic uncertainties also have bin-to-bin correlations and unlike 
the statistical correlations from SNO's signal extraction may also 
include correlations that extend across day and night spectra.
Experimental spectral shape systematic uncertainties were described
in Section~\ref{sec:spectrum} and were included in this oscillation analysis.  
The uncertainty in the shape of the $^8$B neutrino
spectrum has also been included in this $\chi^2$ analysis.  The $^8$B spectrum
used in our model is the one from \cite{bib:ortiz}; however, the more generous
uncertainties from \cite{bib:jnb_b8} were employed in the systematics 
calculation in our $\chi^2$ analysis. 

Day-night systematics, though small, were also included in the global 
$\chi^2$ analysis.  The significant day-night systematics are 
diurnal energy scale, long-term energy scale variation, diurnal 
vertex shift, diurnal isotropy variation, long-term isotropy variation, and 
internal neutron background asymmetry.  Other day-night systematics 
discussed in this paper are smaller 
in magnitude and were averaged together in the $\chi^2$ calculation.
Note that some directional systematics have a non-negligible effect 
on the day-night asymmetry of ES events; however, the impact of the 
day-night asymmetry of ES events in SNO on the oscillation analysis 
is not that significant so combining these systematics is also
reasonable.  The technique for including systematic uncertainties 
and bin-to-bin correlations in the $\chi^2$ analysis is the conventional one, 
as in \cite{bib:stand_chi}.  Thus $\chi^2_{\rm SNO-II}$ from SNO's 
391-day data set is defined as: 
\begin{equation}
\chi^2_{\rm SNO-II} = \sum_{i,j=1}^{38} (Y_{i}^{\rm data} - Y_{i}^{\rm model})
       [\sigma_{ij}^2(\mbox{tot})]^{-1} (Y_{j}^{\rm data} - Y_{j}^{\rm model}),
\shwlabel{eqnchi2}
\end{equation}
where $Y_{i}^{\rm data}$ is the SNO experimental value in one of the 17 CC 
spectral bins, or the NC or ES flux, day or night, 
and $Y_{i}^{\rm model}$ is the model predicted
value for bin $i$ based on the neutrino oscillation hypothesis and the
set of parameters being evaluated.

The error matrix for the calculation $\sigma^2_{ij}(\mbox{tot})$, is 
composed of  statistical and systematic components: 
\begin{equation}
\sigma^2_{ij}(\mbox{tot}) = \sigma^2_{ij}(\mbox{stat}) + 
                            \sigma^2_{ij}(\mbox{syst}), 
\end{equation}
with $\sigma^2_{ij}(\mbox{stat})$ containing the elements from the 
statistical covariance matrices from SNO's unconstrained signal 
extraction and $\sigma^2_{ij}(\mbox{syst})$ containing contributions 
from systematic uncertainties.  The spectral systematics error matrix 
is formed 
from the partial derivatives that relate the rate $Y^{\rm model}$ 
in the $i$th bin to the uncertainty in each one of the $K$ spectral 
systematics $S_k$:
\begin{equation}
  \sigma_{ij}^2\mbox{(syst)} 
      = \sum_{k=1}^{K}
        \frac{\partial Y_{i}}{\partial S_k}
        \frac{\partial Y_{j}}{\partial S_k}
        (\Delta S_k)^2,
\end{equation}
where $\Delta S_{k}$ is the uncertainty estimated for
spectral systematic $S_k$.  Note that all systematic uncertainties 
have an effect on the extracted CC spectra, and possibly an 
energy-dependent effect; thus, all systematics are spectral 
systematics. In this standard $\chi^2$ treatment, bin-to-bin 
correlations are included for the systematics; however, possible 
correlations among the various systematic uncertainties were
neglected.

\begin{figure}[ht]
\begin{center}
\includegraphics[width=3.6in]{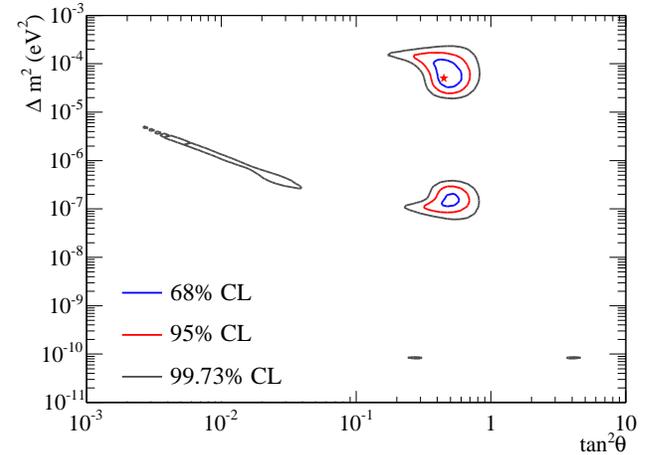}
\caption[f_snoonly]{SNO-only neutrino oscillation analysis, including 
pure D$_2$O phase day and night spectra, and salt 
extracted CC spectra, NC and ES fluxes, day and night.  
The $^8$B flux was free in the fit; $hep$ solar  neutrinos were fixed 
at $9.3 \times 10^{3}$ cm$^{-2}$ s$^{-1}$.  The 
star is plotted at the best-fit parameters
from the $\chi^2$ analysis, listed in
Table~\ref{tab:mswparms}.}
\label{fig:snoonly} 
\end{center}
\end{figure}

Figure~\ref{fig:snoonly} shows the allowed regions for neutrino 
oscillation parameters when only SNO data (SNO-I and SNO-II)
are analyzed.  The inclusion of CC spectral data, 
improved measurement of the NC flux from the larger data set, 
and the addition of separate day and night results 
compared with \cite{bib:saltprl} produce slightly smaller
allowed ranges of parameters. The best-fit 
parameters from a SNO-only analysis are: 
$\Delta m^2 = 5.0 \times 10^{-5}$ eV$^2$, 
$\tan^2 \theta = 0.45$, $f_B = 5.11 \times 10^{6}$ cm$^{-2}$ s$^{-1}$,
which is the total active $^8$B solar neutrino flux, a free parameter
during $\chi^2$ minimization.
The best-fit $\chi^2$ is 68.9 for 69 degrees of freedom in the SNO-only
oscillation analysis.

\begin{figure}[ht]
\begin{center}
\includegraphics[width=3.6in]{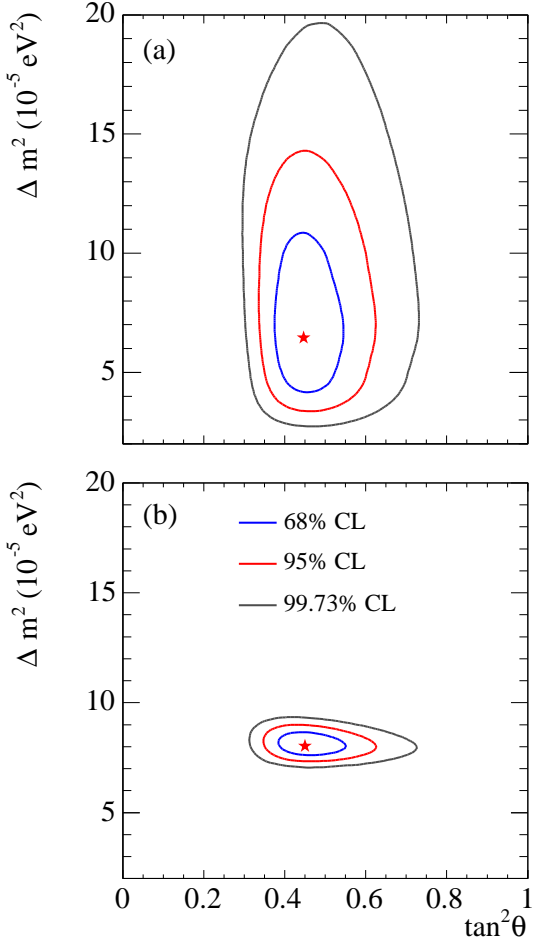}
\caption[f_global]{(a) Global neutrino oscillation analysis using only
solar neutrino data, and (b) including KamLAND 766 ton-year data.
The solar neutrino data included
SNO's pure D$_2$O phase day and night spectra, SNO's salt phase
extracted day and night CC spectra and ES and NC fluxes, the
rate measurements from the 
Cl, SAGE, Gallex/GNO, and SK-I zenith spectra.  
The $^8$B flux was free in the fit; $hep$ solar neutrinos
were fixed at $9.3 \times 10^{3}$ cm$^{-2}$ s$^{-1}$.  
The stars are plotted at the best-fit parameters
from the $\chi^2$ analysis, listed in
Table~\ref{tab:mswparms}.}
\label{fig:global} 
\end{center}
\end{figure}

The top panel in Fig.~\ref{fig:global} shows the allowed region for
a global oscillation analysis that included data from all solar
neutrino experiments.  The best-fit oscillation parameters, with
1$\sigma$ uncertainties on the 2-dimensional parameter region given, are
$\Delta m^2 = 6.5^{+4.4}_{-2.3} \times 10^{-5}$ eV$^2$, $\tan^2 \theta
= 0.45^{+0.09}_{-0.08}$, with a best-fit $\chi^2 = 113.1$ for 116
degrees of freedom in the global solar $\chi^2$ analysis.  The lower
panel shows the results of the analysis when the 766 ton-year data
from KamLAND \cite{bib:kl2004} were also included.  The best-fit
parameters from the global solar plus KamLAND analysis are: $\Delta
m^2 = 8.0^{+0.6}_{-0.4} \times 10^{-5}$ eV$^2$, $\theta =
33.9^{+2.4}_{-2.2}$ degrees, $f_B = 4.93 \times 10^{6}$ cm$^{-2}$
s$^{-1}$, where the 1$\sigma$ uncertainties on the 2-dimensional
parameter region are given.  The inclusion of KamLAND data shifts the
best-fit $\Delta m^2$ value but this shift is perfectly consistent
with the global solar neutrino constraints and gives a $\chi^2 =
113.6$ for the solar neutrino part of the calculation.  A summary of
the best-fit oscillation parameters and their ranges within the
allowed LMA regions appears in Table~\ref{tab:mswparms}.  SNO data are
providing strong constraints on the mixing angle.

\begin{table}[h]
\caption
  {Best-fit neutrino oscillation parameters.  Uncertainties listed 
   are $\pm 1 \sigma$ for the 2-D parameter regions (and only
   within the LMA region for the SNO-only analysis).
  \label{tab:mswparms}
  }
\begin{center}
\begin{tabular}{lcc}
\hline
\hline
Oscillation analysis & $\Delta m^2 \; (10^{-5}$ eV$^2$) & $\tan^2\theta$ \\
\hline
SNO-only             & $5.0^{+6.2}_{-1.8}$ & $0.45^{+0.11}_{-0.10}$  \\
Global solar         & $6.5^{+4.4}_{-2.3}$ & $0.45^{+0.09}_{-0.08}$  \\
Solar plus KamLAND   & $8.0^{+0.6}_{-0.4}$ & $0.45^{+0.09}_{-0.07}$  \\
\hline
\hline
\end{tabular}
\end{center}
\end{table}

Compared to \cite{bib:saltprl} the inclusion of the 391-day salt 
data set (with spectral and day-night information) 
in the oscillation analysis moves the allowed oscillation region to 
slightly larger mixing angles.  This is due to the larger central 
value of the $\phi_{CC}$/$\phi_{NC}$ ratio found in the present analysis.  
The 2004 KamLAND data
\cite{bib:kl2004} have already tightly constrained
the parameter $\Delta m^2$.  In terms of individual uncertainties 
the results become $\Delta m^2 = 8.0^{+0.4}_{-0.3} \times 10^{-5}$ eV$^2$ and 
$\theta = 33.9^{+1.6}_{-1.6}$ degrees, where the uncertainties were
obtained as 1-dimensional projections of the respective parameter while
marginalizing the uncertainties in the other.

\begingroup
\squeezetable
\begin{table}[h]
\caption
  {Comparison of SNO total active $^8$B solar neutrino flux
   measurements and solar model predictions.
  \label{tab:8bflux}
  }
\begin{center}
\begin{tabular}{ll}
\hline
\hline
Source                      & Total $^8$B Flux ($10^{6}$ cm$^{-2}$ s$^{-1}$) \\
\hline
SNO pure D$_2$O phase NC    &
${5.09^{+0.44}_{-0.43}\mbox{(stat.)}^{+0.46}_{-0.43}\mbox{(syst.)}}$ \\
above, energy unconstrained & 
${6.42 \pm 1.57\mbox{(stat.)}^{+0.55}_{-0.58}\mbox{(syst.)}}$       \\
SNO salt phase NC           &  
${4.94 \pm 0.21\mbox{(stat.)}^{+0.34}_{-0.38}\mbox{(syst.)}}$       \\
$\;$$\;$$\;$SNO salt day NC             &
${4.81 \pm 0.31\mbox{(stat.)}\pm0.39\mbox{(syst.)}}$       \\
$\;$$\;$$\;$SNO salt night NC           & 
${5.02 \pm 0.29\mbox{(stat.)}\pm0.41\mbox{(syst.)}}$       \\
\hline
SNO-only oscillation fit & 5.11 \\
global solar fit         & 5.06 \\
solar plus KamLAND fit   & 4.93 \\
\hline
BS05(OP) \cite{bib:bs05}         & $5.69 \pm 0.91$ \\
BS05(AGS,OP) \cite{bib:bs05}     & $4.51 \pm 0.72$ \\
BP04 \cite{bib:bp2004}             & $5.79 \pm 1.33$ \\
BP2000 \cite{bib:bp2000}           & $5.05^{+1.01}_{-0.81}$ \\
TC04 tac A \cite{bib:tc04}       & $4.25$  \\
TC04 seismic \cite{bib:tc04}     & $5.31 \pm 0.6$  \\
\hline
\hline
\end{tabular}
\end{center}
\end{table}
\endgroup

The total active $^8$B solar neutrino flux, measured by the NC reaction,
has been presented in several ways in SNO analyses.  
Table~\ref{tab:8bflux} lists SNO measured (or fit) values and 
fluxes predicted by solar models.  In the first row,
the SNO NC flux was extracted assuming an undistorted $^8$B spectrum (for the
null hypothesis test).  All subsequent values in the table are free from
that assumption.  The salt phase NC value (this work) is the most precise and
appropriate one to compare with solar models.  The agreement between
solar models and this measurement is good.

\begin{figure}[ht]
\begin{center}
\includegraphics[width=3.6in]{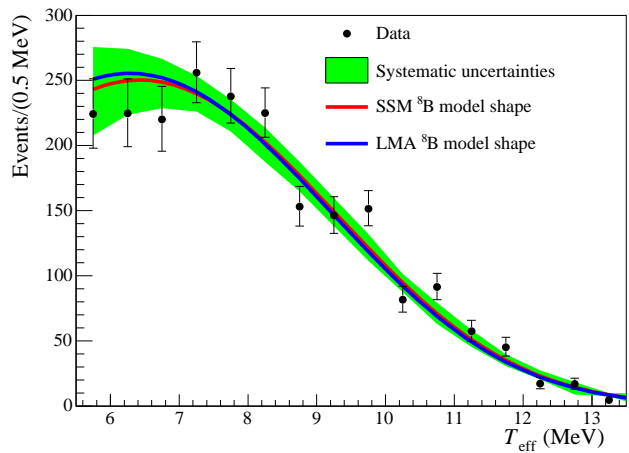}
\caption{Extracted CC $\teff$ spectrum compared to that predicted with 
the best-fit LMA parameters.  Only statistical uncertainties are shown 
in the data spectrum.  
The band on the undistorted $^8$B model shape represents the 1$\sigma$ 
uncertainty determined from detector systematic uncertainties.  
The predicted spectrum is normalized to the same number of counts as 
the data spectrum.  Note that the data points, especially the first three
points, are statistically correlated as well as having correlated
systematics as indicated by the error band.}
\label{fig:lmaspect} 
\end{center}
\end{figure}

Based on the best-fit parameters from the global solar plus KamLAND 
analysis, the predicted CC electron energy spectrum is determined.  
In Fig.~\ref{fig:lmaspect}, this prediction is compared to the 
measured CC spectrum.  The $\chi^2$ between the extracted spectrum
and the expected shape for the best-fit LMA parameters, calculated
with all statistical correlations and systematic uncertainties as
described above for the global oscillation analysis, is 27.2 for
16 degrees of freedom (17 spectral bins minus a floating normalization
factor).  The probability of observing a $\chi^2 > 27.2$ under the
assumption that the data are drawn from the expected LMA spectrum
is 3.9\%.

\section{\label{sec:summary}Summary}

An extensive analysis of the data from the full running period with
salt added to the heavy water in SNO has been presented. The salt
additive enables a statistical separation of NC events from CC and ES
events by measuring event isotropy. In addition to new results for
integral fluxes, energy spectral information from the CC
reaction is presented, with complete statistical and systematic
uncertainties. Separate day and night spectra and day-night
integrated-flux asymmetries are also presented. The flux measurements
are in agreement with, and slightly more precise than, previous
measurements \cite{bib:saltprl}. The energy spectrum derived
from the CC reaction is consistent with the expected spectrum assuming an
undistorted $^8$B shape and also with the predicted spectrum
corresponding to the best-fit LMA parameters for a global oscillation
analysis with solar neutrino and KamLAND reactor neutrino data included.
Within uncertainties, no significant day-night asymmetries are
observed as expected for the best-fit LMA solution.  Detailed MSW fits
find a single allowed region in oscillation parameter space and
tightly constrained values for $\Delta m^2$ and the mixing angle
$\theta$.  These data provide further confirmation of flavor change
for solar neutrinos and for the oscillation of massive neutrinos as
the dominant flavor change mechanism. The total flux of all active
neutrino types for $^8$B solar neutrinos is in agreement with the most
recent solar model calculations \cite{bib:bs05,bib:tc04}.

\begin{acknowledgments}
This research was supported by: Canada: Natural Sciences and
Engineering Research Council, Industry Canada, National Research
Council, Northern Ontario Heritage Fund, Atomic Energy of
Canada, Ltd., Ontario Power Generation, High Performance Computing
Virtual Laboratory, Canada Foundation for Innovation; US: Dept.\ of
Energy, National Energy Research Scientific Computing Center; UK:
Particle Physics and Astronomy Research Council. We thank the SNO
technical staff for their strong contributions.
We thank Inco for hosting this project.
\end{acknowledgments}

\bibliography{sno_nsp}

\begin{thebibliography}{25}
\bibitem[{mgcl()}]{bib:mgcl_footnote}
\bibinfo{note}{In Ref.~\cite{bib:sno_nim}, the proposal of adding MgCl$_2$, instead of \nacl, to the \dto target to enhance the neutrino detection efficiency is discussed.}
\bibitem[{es_()}]{bib:es_crosssection}
\bibinfo{note}{Electron neutrino cross sections are used to calculate all
  fluxes.}
\end{thebibliography}

\pagebreak

\appendix

\section{SNO Data in an Oscillation Analysis}

The following tables contain a subset of SNO's salt phase results 
and some supporting information that may be needed in a neutrino oscillation 
analysis.  
The values in these tables include outputs from
the signal extraction (described earlier in this paper) and results
from systematic uncertainty studies.  Day and night 
values were extracted separately without any constraints on the NC rate
asymmetry or energy spectrum shapes.

Table~\ref{tab:dnfluxvalues} contains day and night CC spectra, expressed
as fluxes.  The meaning of these fluxes (for example the day flux in the 
6.0--6.5 MeV bin of $0.182 \times 10^6$ cm$^{-2}$ s$^{-1}$) is that the 
number of events SNO observed in the
salt day data set attributed to CC interactions by the signal extraction, 
with an electron kinetic energy between 6.0--6.5 MeV, 
is equal to the number of all CC events that would be observed above
kinetic energy 5.5 MeV, if the integral flux (from zero to
endpoint) of $\nu_e$ had the value of $0.182 \times 10^6$ cm$^{-2}$ s$^{-1}$
and had an undistorted $^8$B spectral shape \cite{bib:ortiz}.
The $^8$B spectral shape aspect of this definition
is only for normalization.  There is no assumption of
any spectral shape when extracting the actual number of
events in each bin for SNO's salt phase.  This normalization was chosen so 
that the sum of the values for all bins equals the integral day and night
$^8$B fluxes quoted in Section~\ref{sec:dn_modelindependent}.

When calculating the theoretical CC flux and spectra
for a set of oscillation parameters,
for comparison with SNO data, one should be aware of the above definition.
Thus, a model prediction for the value in the above mentioned example bin 
would be:
\begin{widetext}
\begin{equation}
\int_0^\infty \phi(E_\nu) \, d E_\nu \;
\frac{\int_0^\infty \int_0^\infty \int_{6.0}^{6.5} \phi(E_\nu) P_{ee}(E_\nu) \frac{d \sigma}{d T_e}(E_\nu,T_e) R(T_e,\teff) \, d E_\nu \, d T_e \, d \teff}
{\int_0^\infty \int_0^\infty \int_{5.5}^\infty \phi(E_\nu) \frac{d \sigma}{d T_e}(E_\nu,T_e) R(T_e,\teff) \, d E_\nu \, d T_e \, d \teff}
\end{equation}
\end{widetext}
where $P_{ee}$ is the survival probability for a $\nu_e$ produced
in the Sun to be detected as a $\nu_e$, $\phi(E_\nu)$ is the 
flux of $^8$B solar neutrinos as a function of neutrino energy,
$\frac{d \sigma}{d T_e}$ is the
differential cross section for the CC reaction
and $R(T_e,\teff)$ is the energy response function,
\begin{equation}
 R(T_e,\teff) = \frac{1}{\sqrt{2\pi} \sigma_T}
            \exp \left[
                -\frac{(T_e - \teff)^2}{2\sigma_T^2}
                 \right],
\end{equation}
and $T_e$ is the true recoil electron kinetic energy and $\teff$ is the
observed electron kinetic energy, with resolution,
\begin{equation}
\sigma_T(T_e) = -0.131 + 0.383\sqrt{T_e} + 0.03731 \, T_e
\end{equation}
in units of MeV\@.

Tables~\ref{tab:daycorr} and \ref{tab:nightcorr} contain the statistical
correlation coefficients from SNO signal extraction, for salt phase
day and night data respectively.  The numbering of the CC spectral bins 
listed in these tables follows the ordering of energy bins as listed in 
Table~\ref{tab:dnfluxvalues}.  These correlation coefficients are necessary 
to include in an oscillation analysis that includes SNO CC spectral
information.

The systematic uncertainties for the SNO extracted CC spectrum are listed
in Table~\ref{tab:ccsyserr}.  These uncertainties (in percent) can be 
used as the partial derivatives in a
bin-to-bin correlated systematics part of a $\chi^2$ calculation, 
as described in Section~\ref{sec:physint}.  The spectral shape of these 
uncertainties can also be used as the shape of the related
day-night asymmetry systematic.  The magnitude of a day-night spectral 
systematic can be estimated by taking the ratio of the size of the 
day-night flux asymmetry uncertainty and the total flux uncertainty, 
for each relevant systematic, and using this ratio in each bin to scale 
that spectral systematic.  Note that NC-only systematics (internal neutron
background and neutron capture efficiency uncertainties) are not listed in
Table~\ref{tab:ccsyserr} since they do not affect the CC spectrum,
but are necessary to include in an oscillation analysis and can be 
found in Table~\ref{tab:flux_uncert}.

Table~\ref{tab:zenlive} contains the live time distribution for SNO's 391-day
salt data set as a function of cosine of the zenith angle of the Sun.
For detailed calculations of neutrino survival probabilities
including propagation through the Earth, the live times at different
zenith angles can be used as weighting factors.

\begin{table}[h]
\caption
  {Charged-current reaction recoil electron kinetic energy spectra 
   from the 391-day SNO salt phase, expressed in units of
   equivalent $^8$B fluxes.  The normalization is such that the sum
   over all bins equals the day or night integral 
   $^8$B solar neutrino flux above 0 MeV, as determined and quoted in
   Section~\ref{sec:dn_modelindependent}.
   Day and night extracted values are listed with their statistical 
   uncertainties from SNO's signal extraction.
    \label{tab:dnfluxvalues}
  }
\begin{center}
\begin{tabular}{lll}
\hline
\hline
CC electron kinetic & salt phase day  & salt phase night  \\
energy bin (MeV)    & ($10^6$ cm$^{-2}$ s$^{-1})$ & ($10^6$ cm$^{-2}$ s$^{-1})$ \\
\hline
5.5--6.0   &  $0.205 \pm 0.032$   &  $0.145 \pm 0.027    $ \\
6.0--6.5   &  $0.182 \pm 0.030$   &  $0.164 \pm 0.027	 $ \\
6.5--7.0   &  $0.153 \pm 0.028$   &  $0.190 \pm 0.026	 $ \\
7.0--7.5   &  $0.226 \pm 0.028$   &  $0.180 \pm 0.024	 $ \\
7.5--8.0   &  $0.198 \pm 0.025$   &  $0.178 \pm 0.022	 $ \\
8.0--8.5   &  $0.184 \pm 0.023$   &  $0.164 \pm 0.019	 $ \\
8.5--9.0   &  $0.124 \pm 0.018$   &  $0.114 \pm 0.015	 $ \\
9.0--9.5   &  $0.099 \pm 0.015$   &  $0.126 \pm 0.015	 $ \\
9.5--10.0  &  $0.110 \pm 0.015$   &  $0.124 \pm 0.014	 $ \\
10.0--10.5 &  $0.058 \pm 0.011$   &  $0.067 \pm 0.010	 $ \\
10.5--11.0 &  $0.070 \pm 0.012$   &  $0.073 \pm 0.011	 $ \\
11.0--11.5 &  $0.048 \pm 0.010$   &  $0.039 \pm 0.007	 $ \\
11.5--12.0 &  $0.042 \pm 0.008$   &  $0.029 \pm 0.007	 $ \\
12.0--12.5 &  $0.0088 \pm 0.0038$ &  $0.018 \pm 0.005	 $ \\
12.5--13.0 &  $0.0082 \pm 0.0040$ &  $0.015 \pm 0.005	 $ \\
13.0--13.5 &  $0.0025 \pm 0.0028$ &  $0.0042 \pm 0.0025  $ \\
13.5--20.0 &  $0.014 \pm 0.005$   &  $0.012 \pm 0.004$ \\
\hline
\hline
\end{tabular}
\end{center}
\end{table}

\begin{table}[h]
\caption
  {Zenith-live time distribution for the 391-day SNO salt phase data set.
   There are sixty equally-spaced bins in $\cos \theta_{z}$ with
   $\theta_{z}$ being the zenith angle of the Sun
   (e.g., $\cos \theta_z = -1$ would be the value if the Sun was directly 
   below the SNO detector).  The amount of live time in each bin is 
   listed in seconds.
   The bins start from
   $\cos \theta_z = -1$ to -0.9667 at the top left of the table and increase
   going down the column.  Continuing at the top of the second column,
   the top bin has $\cos \theta_z = -0.5$ to -0.4667.  
   The first two columns contain
   night live times while the final two columns are day live times.
    \label{tab:zenlive}
  }
\begin{center}
\begin{tabular}{rrrr}
\hline
\hline
\multicolumn{4}{c}{$\cos \theta_z$} \\
$-1.0$ to $-0.5$ & $-0.5$ to $0.0$ & $0.0$ to $0.5$ & $0.5$ to $1.0$ \\
\hline
0      &  705975 & 606806 & 493795 \\
0      &  743962 & 608735 & 471616 \\
432202 &  761182 & 606873 & 447927 \\
545975 &  814881 & 607682 & 445073 \\
592399 & 1107970 & 616573 & 473089 \\
566033 &  887601 & 617365 & 491700 \\
570422 &  777241 & 625955 & 453238 \\
539112 &  728515 & 651235 & 458976 \\
527208 &  698101 & 682766 & 466092 \\
571086 &  672856 & 752869 & 449265 \\
594019 &  652494 & 844742 & 433395 \\
613845 &  639256 & 633145 & 388327 \\
622530 &  624127 & 576052 & 255268 \\
672151 &  615141 & 570776 & 0     \\
684213 &  608656 & 521223 & 0     \\
\hline
\hline
\end{tabular}
\end{center}
\end{table}

\begin{turnpage}
\begin{table}[h]
\caption
  {Statistical correlation coefficients from SNO's signal extraction for the 
  salt phase day data.  The numbering of the CC spectral bins follows the 
  ordering of energy bins in Table~\ref{tab:dnfluxvalues}.
    \label{tab:daycorr}
  }
\begin{center}
\begin{tabular}{rrrrrrrrrrrrrrrrrrrr}
\hline
\hline
   & NC & CC1 & CC2 & CC3 & CC4 & CC5 & CC6 & CC7 & CC8 & CC9 & CC10 & CC11 & CC12 &
CC13 & CC14 & CC15 & CC16 & CC17 & ES \\
\hline
NC   &	  1.0000  &  -0.3478  &  -0.3320  &  -0.3281  &  -0.2488  &  -0.2000  &  -0.1555  &  -0.0877  &  -0.0618  &  -0.0329  &  -0.0152  &  -0.0063  &  -0.0025  &   0.0003  &  -0.0000  &  -0.0000  &  -0.0000  &  -0.0000  &  -0.0610  \\
CC1  &   -0.3478  &   1.0000  &   0.1528  &   0.1440  &   0.1124  &   0.0865  &   0.0676  &   0.0425  &   0.0272  &   0.0135  &   0.0061  &   0.0030  &   0.0011  &   0.0001  &   0.0000  &  -0.0000  &   0.0000  &   0.0000  &  -0.0695  \\
CC2  &   -0.3320  &   0.1528  &   1.0000  &   0.1459  &   0.1147  &   0.0873  &   0.0684  &   0.0440  &   0.0275  &   0.0135  &   0.0061  &   0.0031  &   0.0011  &   0.0001  &   0.0000  &  -0.0000  &   0.0000  &   0.0000  &  -0.0507  \\
CC3  &   -0.3281  &   0.1440  &   0.1459  &   1.0000  &   0.1073  &   0.0824  &   0.0644  &   0.0407  &   0.0259  &   0.0129  &   0.0058  &   0.0029  &   0.0010  &   0.0001  &   0.0000  &  -0.0000  &   0.0000  &   0.0000  &  -0.0525  \\
CC4  &   -0.2488  &   0.1124  &   0.1147  &   0.1073  &   1.0000  &   0.0643  &   0.0503  &   0.0322  &   0.0202  &   0.0100  &   0.0045  &   0.0023  &   0.0008  &   0.0001  &  -0.0000  &  -0.0000  &   0.0000  &  -0.0000  &  -0.0556  \\
CC5  &   -0.2000  &   0.0865  &   0.0873  &   0.0824  &   0.0643  &   1.0000  &   0.0387  &   0.0243  &   0.0155  &   0.0078  &   0.0035  &   0.0017  &   0.0006  &   0.0000  &  -0.0000  &   0.0000  &   0.0000  &   0.0000  &  -0.0637  \\
CC6  &   -0.1555  &   0.0676  &   0.0684  &   0.0644  &   0.0503  &   0.0387  &   1.0000  &   0.0190  &   0.0122  &   0.0061  &   0.0027  &   0.0014  &   0.0005  &   0.0000  &   0.0000  &   0.0000  &   0.0000  &   0.0000  &  -0.0648  \\
CC7  &   -0.0877  &   0.0425  &   0.0440  &   0.0407  &   0.0322  &   0.0243  &   0.0190  &   1.0000  &   0.0077  &   0.0037  &   0.0017  &   0.0009  &   0.0003  &   0.0000  &   0.0000  &   0.0000  &   0.0000  &   0.0000  &  -0.0595  \\
CC8  &   -0.0618  &   0.0272  &   0.0275  &   0.0259  &   0.0202  &   0.0155  &   0.0122  &   0.0077  &   1.0000  &   0.0024  &   0.0011  &   0.0005  &   0.0002  &   0.0000  &   0.0000  &   0.0000  &   0.0000  &   0.0000  &  -0.0430  \\
CC9  &   -0.0329  &   0.0135  &   0.0135  &   0.0129  &   0.0100  &   0.0078  &   0.0061  &   0.0037  &   0.0024  &   1.0000  &   0.0006  &   0.0003  &   0.0001  &   0.0000  &   0.0000  &   0.0000  &   0.0000  &  -0.0000  &  -0.0495  \\
CC10 &   -0.0152  &   0.0061  &   0.0061  &   0.0058  &   0.0045  &   0.0035  &   0.0027  &   0.0017  &   0.0011  &   0.0006  &   1.0000  &   0.0001  &   0.0000  &   0.0000  &   0.0000  &  -0.0000  &  -0.0000  &   0.0000  &  -0.0466  \\
CC11 &   -0.0063  &   0.0030  &   0.0031  &   0.0029  &   0.0023  &   0.0017  &   0.0014  &   0.0009  &   0.0005  &   0.0003  &   0.0001  &   1.0000  &   0.0000  &   0.0000  &   0.0000  &   0.0000  &   0.0000  &   0.0000  &  -0.0487  \\
CC12 &   -0.0025  &   0.0011  &   0.0011  &   0.0010  &   0.0008  &   0.0006  &   0.0005  &   0.0003  &   0.0002  &   0.0001  &   0.0000  &   0.0000  &   1.0000  &   0.0000  &   0.0000  &   0.0000  &  -0.0000  &   0.0000  &  -0.0570  \\
CC13 &    0.0003  &   0.0001  &   0.0001  &   0.0001  &   0.0001  &   0.0000  &   0.0000  &   0.0000  &   0.0000  &   0.0000  &   0.0000  &   0.0000  &   0.0000  &   1.0000  &   0.0000  &  -0.0000  &  -0.0000  &  -0.0000  &   0.0000  \\
CC14 &   -0.0000  &   0.0000  &   0.0000  &   0.0000  &  -0.0000  &  -0.0000  &   0.0000  &   0.0000  &   0.0000  &   0.0000  &   0.0000  &   0.0000  &   0.0000  &   0.0000  &   1.0000  &  -0.0000  &  -0.0000  &  -0.0000  &  -0.0000  \\
CC15 &   -0.0000  &  -0.0000  &  -0.0000  &  -0.0000  &  -0.0000  &   0.0000  &   0.0000  &   0.0000  &   0.0000  &   0.0000  &  -0.0000  &   0.0000  &   0.0000  &  -0.0000  &  -0.0000  &   1.0000  &   0.0000  &   0.0000  &  -0.0163  \\
CC16 &   -0.0000  &   0.0000  &   0.0000  &   0.0000  &   0.0000  &   0.0000  &   0.0000  &   0.0000  &   0.0000  &   0.0000  &  -0.0000  &   0.0000  &  -0.0000  &  -0.0000  &  -0.0000  &   0.0000  &   1.0000  &  -0.0000  &  -0.0379  \\
CC17 &	 -0.0000  &   0.0000  &   0.0000  &   0.0000  &  -0.0000  &   0.0000  &   0.0000  &   0.0000  &   0.0000  &  -0.0000  &   0.0000  &   0.0000  &   0.0000  &  -0.0000  &  -0.0000  &   0.0000  &  -0.0000  &   1.0000  &   0.0000  \\
ES   &   -0.0610  &  -0.0695  &  -0.0507  &  -0.0525  &  -0.0556  &  -0.0637  &  -0.0648  &  -0.0595  &  -0.0430  &  -0.0495  &  -0.0466  &  -0.0487  &  -0.0570  &   0.0000  &  -0.0000  &  -0.0163  &  -0.0379  &   0.0000  &   1.0000  \\
\hline
\hline
\end{tabular}
\end{center}
\end{table}
\end{turnpage}

\begin{turnpage}
\begin{table}[h]
\caption
  {Statistical correlation coefficients from SNO's signal extraction for the 
  salt phase night data.  The numbering of the CC spectral bins follows the 
  ordering of energy bins in Table~\ref{tab:dnfluxvalues}.
    \label{tab:nightcorr}
  }
\begin{center}
\begin{tabular}{rrrrrrrrrrrrrrrrrrrr}
\hline
\hline
   & NC & CC1 & CC2 & CC3 & CC4 & CC5 & CC6 & CC7 & CC8 & CC9 & CC10 & CC11 & CC12 &
CC13 & CC14 & CC15 & CC16 & CC17 & ES \\
\hline
NC   &	  1.0000  &  -0.3485  &  -0.3437  &  -0.3060  &  -0.2669  &  -0.2089  &  -0.1571  &  -0.1034  &  -0.0545  &  -0.0317  &  -0.0138  &  -0.0067  &  -0.0010  &  -0.0008  &   0.0001  &   0.0000  &   0.0000  &   0.0000  &  -0.0693  \\
CC1  &   -0.3485  &   1.0000  &   0.1580  &   0.1400  &   0.1201  &   0.0967  &   0.0671  &   0.0477  &   0.0237  &   0.0143  &   0.0085  &   0.0027  &   0.0017  &   0.0003  &   0.0000  &  -0.0000  &  -0.0000  &  -0.0000  &  -0.0684  \\
CC2  &   -0.3437  &   0.1580  &   1.0000  &   0.1379  &   0.1183  &   0.0952  &   0.0661  &   0.0469  &   0.0234  &   0.0141  &   0.0084  &   0.0027  &   0.0016  &   0.0003  &   0.0000  &  -0.0000  &  -0.0000  &  -0.0000  &  -0.0463  \\
CC3  &   -0.3060  &   0.1400  &   0.1379  &   1.0000  &   0.1048  &   0.0843  &   0.0586  &   0.0416  &   0.0207  &   0.0125  &   0.0074  &   0.0024  &   0.0014  &   0.0002  &   0.0000  &  -0.0000  &  -0.0000  &  -0.0000  &  -0.0743  \\
CC4  &   -0.2669  &   0.1201  &   0.1183  &   0.1048  &   1.0000  &   0.0723  &   0.0505  &   0.0357  &   0.0178  &   0.0107  &   0.0062  &   0.0020  &   0.0012  &   0.0002  &   0.0000  &  -0.0000  &  -0.0000  &  -0.0000  &  -0.0527  \\
CC5  &   -0.2089  &   0.0967  &   0.0952  &   0.0843  &   0.0723  &   1.0000  &   0.0404  &   0.0287  &   0.0143  &   0.0086  &   0.0051  &   0.0016  &   0.0010  &   0.0002  &   0.0000  &  -0.0000  &  -0.0000  &  -0.0000  &  -0.0483  \\
CC6  &   -0.1571  &   0.0671  &   0.0661  &   0.0586  &   0.0505  &   0.0404  &   1.0000  &   0.0199  &   0.0101  &   0.0060  &   0.0033  &   0.0012  &   0.0006  &   0.0001  &  -0.0000  &   0.0000  &  -0.0000  &   0.0000  &  -0.0489  \\
CC7  &   -0.1034  &   0.0477  &   0.0469  &   0.0416  &   0.0357  &   0.0287  &   0.0199  &   1.0000  &   0.0070  &   0.0042  &   0.0025  &   0.0008  &   0.0005  &   0.0001  &   0.0000  &   0.0000  &   0.0000  &   0.0000  &  -0.0417  \\
CC8  &   -0.0545  &   0.0237  &   0.0234  &   0.0207  &   0.0178  &   0.0143  &   0.0101  &   0.0070  &   1.0000  &   0.0021  &   0.0012  &   0.0004  &   0.0002  &   0.0000  &   0.0000  &   0.0000  &  -0.0000  &   0.0000  &  -0.0692  \\
CC9  &   -0.0317  &   0.0143  &   0.0141  &   0.0125  &   0.0107  &   0.0086  &   0.0060  &   0.0042  &   0.0021  &   1.0000  &   0.0007  &   0.0002  &   0.0001  &   0.0000  &   0.0000  &   0.0000  &  -0.0000  &   0.0000  &  -0.0410  \\
CC10 &   -0.0138  &   0.0085  &   0.0084  &   0.0074  &   0.0062  &   0.0051  &   0.0033  &   0.0025  &   0.0012  &   0.0007  &   1.0000  &   0.0001  &   0.0001  &   0.0000  &   0.0000  &  -0.0000  &  -0.0000  &  -0.0000  &   0.0016  \\
CC11 &   -0.0067  &   0.0027  &   0.0027  &   0.0024  &   0.0020  &   0.0016  &   0.0012  &   0.0008  &   0.0004  &   0.0002  &   0.0001  &   1.0000  &   0.0000  &   0.0000  &  -0.0000  &  -0.0000  &  -0.0000  &  -0.0000  &  -0.0359  \\
CC12 &   -0.0010  &   0.0017  &   0.0016  &   0.0014  &   0.0012  &   0.0010  &   0.0006  &   0.0005  &   0.0002  &   0.0001  &   0.0001  &   0.0000  &   1.0000  &   0.0000  &   0.0000  &  -0.0000  &  -0.0000  &  -0.0000  &   0.0003  \\
CC13 &   -0.0008  &   0.0003  &   0.0003  &   0.0002  &   0.0002  &   0.0002  &   0.0001  &   0.0001  &   0.0000  &   0.0000  &   0.0000  &   0.0000  &   0.0000  &   1.0000  &   0.0000  &   0.0000  &  -0.0000  &   0.0000  &  -0.0229  \\
CC14 &    0.0001  &   0.0000  &   0.0000  &   0.0000  &   0.0000  &   0.0000  &  -0.0000  &   0.0000  &   0.0000  &   0.0000  &   0.0000  &  -0.0000  &   0.0000  &   0.0000  &   1.0000  &  -0.0000  &  -0.0000  &  -0.0000  &  -0.0139  \\
CC15 &    0.0000  &  -0.0000  &  -0.0000  &  -0.0000  &  -0.0000  &  -0.0000  &   0.0000  &   0.0000  &   0.0000  &   0.0000  &  -0.0000  &  -0.0000  &  -0.0000  &   0.0000  &  -0.0000  &   1.0000  &  -0.0000  &  -0.0000  &  -0.0188  \\
CC16 &    0.0000  &  -0.0000  &  -0.0000  &  -0.0000  &  -0.0000  &  -0.0000  &  -0.0000  &   0.0000  &  -0.0000  &  -0.0000  &  -0.0000  &  -0.0000  &  -0.0000  &  -0.0000  &  -0.0000  &  -0.0000  &   1.0000  &   0.0000  &  -0.0085  \\
CC17 &	  0.0000  &  -0.0000  &  -0.0000  &  -0.0000  &  -0.0000  &  -0.0000  &   0.0000  &   0.0000  &   0.0000  &   0.0000  &  -0.0000  &  -0.0000  &  -0.0000  &   0.0000  &  -0.0000  &  -0.0000  &   0.0000  &   1.0000  &  -0.0183  \\
ES   &   -0.0693  &  -0.0684  &  -0.0463  &  -0.0743  &  -0.0527  &  -0.0483  &  -0.0489  &  -0.0417  &  -0.0692  &  -0.0410  &   0.0016  &  -0.0359  &   0.0003  &  -0.0229  &  -0.0139  &  -0.0188  &  -0.0085  &  -0.0183  &   1.0000  \\
\hline
\hline
\end{tabular}
\end{center}
\end{table}
\end{turnpage}

\begin{turnpage}
\begin{table}[h]
\caption
  {Complete CC spectrum systematic uncertainties (in percent) from SNO's
  unconstrained fit signal extraction of the 391-day salt phase data set.
  The numbering of the CC spectral bins follows the ordering of 
  energy bins in Table~\ref{tab:dnfluxvalues}.
    \label{tab:ccsyserr}
  }
\begin{center}
\begin{tabular}{rcccccccccc}
\hline
\hline
 Uncertainty  & NC & ES & CC1 & CC2 & CC3 & CC4 & CC5 & CC6 & CC7 & CC8  \\
\hline

                Energy scale (const.) & 3.8,\;\; -3.3 & 1.9,\;\; -1.6 & -8.0,\;\; 9.5 & -5.9,\;\; 6.2 & -2.9,\;\; 3.0 & 0.6,\;\; 0.0 & 1.7,\;\; -1.8 & 2.9,\;\; -3.9 & 3.9,\;\; -3.6 & 4.7,\;\; -5.0 \\

            Energy scale (E dep.) & 0.1,\;\; -0.1 & 0.1,\;\; -0.1 & -1.5,\;\; 1.6 & -1.4,\;\; 1.6 & -1.6,\;\; 1.2 & -0.4,\;\; 0.6 & -0.5,\;\; 0.6 & 0.0,\;\; -0.4 & 0.9,\;\; -0.6 & 1.3,\;\; -0.4 \\
              Energy radial bias & 2.1,\;\; -2.0 & 1.2,\;\; -1.1 & -5.7,\;\; 6.1 & -4.3,\;\; 4.2 & -2.1,\;\; 2.0 & 0.1,\;\; -0.1 & 1.0,\;\; -1.2 & 1.5,\;\; -2.5 & 3.6,\;\; -3.9 & 3.8,\;\; -2.9 \\
               Energy resolution & 0.8,\;\; -0.8 & 0.7,\;\; -0.7 & 4.9,\;\; -4.9 & 2.1,\;\; -2.1 & 1.8,\;\; -1.8 & 0.1,\;\; -0.1 & -1.2,\;\; 1.2 & -1.8,\;\; 1.8 & -3.9,\;\; 3.9 & -2.6,\;\; 2.6 \\
                       \bof mean & -3.6,\;\; 4.5 & 1.3,\;\; -1.2 & 7.1,\;\; -8.4 & 7.0,\;\; -8.3 & 6.3,\;\; -7.9 & 4.4,\;\; -4.6 & 3.3,\;\; -4.0 & 2.3,\;\; -2.9 & 2.6,\;\; -2.9 & 1.1,\;\; -1.7 \\
                      \bof width & 0.0,\;\; 0.0 & 0.2,\;\; -0.2 & -0.9,\;\; -0.1 & -0.7,\;\; 0.0 & -0.8,\;\; -0.3 & 0.0,\;\; -0.2 & -0.2,\;\; -0.2 & 0.0,\;\; -0.2 & -0.1,\;\; 0.3 & 0.3,\;\; -0.2 \\
                 Radial scale (const.) & -3.0,\;\; 3.3 & -2.6,\;\; 3.0 & -2.6,\;\; 2.5 & -3.4,\;\; 2.3 & -1.7,\;\; 2.4 & -2.4,\;\; 2.6 & -2.5,\;\; 2.3 & -2.8,\;\; 2.4 & -2.6,\;\; 2.9 & -2.5,\;\; 2.8 \\
                 Radial scale (E dep.) & -0.6,\;\; 0.5 & -0.7,\;\; 0.8 & 0.2,\;\; -0.2 & -0.4,\;\; 0.4 & -0.9,\;\; 0.9 & -1.2,\;\; 1.2 & -1.5,\;\; 1.5 & -1.7,\;\; 1.7 & -2.0,\;\; 2.0 & -2.1,\;\; 2.1 \\
                      Vertex $x$ & 0.0,\;\; -0.0 & 0.1,\;\; -0.1 & -0.3,\;\; 0.0 & -0.2,\;\; -0.5 & -0.3,\;\; 0.0 & -0.5,\;\; 0.0 & -0.1,\;\; 0.0 & -0.2,\;\; -0.1 & -0.1,\;\; -0.1 & 0.1,\;\; -0.1 \\
                      Vertex $y$ & 0.0,\;\; -0.1 & 0.1,\;\; -0.1 & -0.2,\;\; -0.3 & 0.0,\;\; -0.2 & 0.1,\;\; -0.1 & -0.2,\;\; 0.0 & 0.1,\;\; 0.0 & -0.3,\;\; 0.0 & 0.0,\;\; -0.3 & 0.1,\;\; 0.4 \\
                      Vertex $z$ & -0.2,\;\; 0.2 & 0.0,\;\; 0.0 & -0.4,\;\; -0.2 & -0.9,\;\; -0.6 & 0.4,\;\; 0.1 & 0.2,\;\; -0.6 & 0.0,\;\; -0.3 & 0.0,\;\; -0.1 & 0.5,\;\; -0.2 & -0.3,\;\; -0.1 \\
               Vertex resolution & 0.1,\;\; -0.1 & 0.1,\;\; -0.1 & -0.4,\;\; 0.4 & -0.2,\;\; 0.2 & 1.4,\;\; -1.4 & -0.3,\;\; 0.3 & 0.1,\;\; -0.1 & -0.6,\;\; 0.6 & 0.0,\;\; 0.0 & 0.4,\;\; -0.4 \\
              Angular resolution & -0.2,\;\; 0.2 & 5.1,\;\; -5.1 & -0.2,\;\; 0.2 & 0.4,\;\; -0.4 & 0.2,\;\; -0.2 & 0.1,\;\; -0.1 & -0.2,\;\; 0.2 & 0.0,\;\; 0.0 & 0.0,\;\; 0.0 & -0.6,\;\; 0.6 \\
                  Internal $\gamma$ & 0.0,\;\; -0.1 & 0.0,\;\; -0.0 & 0.4,\;\; -0.6 & 0.3,\;\; -0.4 & 0.2,\;\; -0.2 & 0.1,\;\; -0.1 & 0.0,\;\; 0.0 & 0.0,\;\; 0.0 & 0.0,\;\; 0.0 & 0.0,\;\; 0.0 \\
            Selection efficiency & -0.1,\;\; 0.2 & -0.2,\;\; 0.2 & -0.2,\;\; 0.2 & -0.1,\;\; 0.2 & -0.1,\;\; 0.2 & -0.1,\;\; 0.2 & -0.1,\;\; 0.2 & -0.1,\;\; 0.2 & -0.1,\;\; 0.2 & -0.1,\;\; 0.2 \\

                     Backgrounds & -1.0,\;\; 0.0 & -0.0,\;\; 0.0 & -8.5,\;\; 0.0 &
               -0.1,\;\; 0.0 & -0.1,\;\; 0.0 & -0.1,\;\; 0.0 &
               -0.1,\;\; 0.0 & -0.1,\;\; 0.0 & -0.1,\;\; 0.0 &
               -0.1,\;\; 0.0 \\  
\hline
& CC9 & CC10 & CC11 & CC12 &
CC13 & CC14 & CC15 & CC16 & CC17 & \\
\hline

                    Energy scale (const.) & 5.0,\;\; -4.2 & 6.2,\;\; -6.7 & 6.7,\;\; -5.8 & 9.6,\;\; -7.4 & 7.0,\;\; -6.7 & 17.6,\;\; -9.7 & 16.6,\;\; -17.4 & 18.2,\;\; -10.2 & 20.5,\;\; -17.5 \\  

            Energy scale (E dep.) & 0.7,\;\; -0.8 & 1.3,\;\; -2.0 & 2.3,\;\; -2.3 & 3.9,\;\; -2.5 & 2.5,\;\; -2.0 & 9.3,\;\; -5.1 & 9.1,\;\; -11.9 & 13.8,\;\; -8.8 & 14.1,\;\; -12.2 \\  
              Energy radial bias & 2.8,\;\; -3.1 & 4.1,\;\; -4.3 & 4.7,\;\; -4.2 & 7.0,\;\; -4.5 & 5.7,\;\; -4.7 & 12.4,\;\; -6.6 & 11.8,\;\; -14.6 & 13.5,\;\; -10.3 & 13.2,\;\; -12.3 \\  
               Energy resolution & -0.7,\;\; 0.7 & 0.0,\;\; 0.0 & -2.1,\;\; 2.1 & 0.6,\;\; -0.6 & -2.3,\;\; 2.3 & -5.8,\;\; 5.8 & -14.8,\;\; 14.8 & -9.9,\;\; 9.9 & -17.2,\;\; 17.2 \\  
                       \bof mean & 0.6,\;\; -0.4 & 0.2,\;\; 0.4 & -0.1,\;\; -0.1 & 1.4,\;\; -0.1 & 0.3,\;\; 0.1 & -1.0,\;\; 4.4 & 2.2,\;\; -1.4 & -5.9,\;\; 9.0 & 3.8,\;\; -2.1 \\  
                      \bof width & 0.0,\;\; -0.& -0.3,\;\; 0.6 & 0.0,\;\; 0.2 & 0.0,\;\; 0.2 & -0.4,\;\; 0.0 & -0.5,\;\; 1.9 & 0.5,\;\; -0.1 & -0.1,\;\; 6.5 & -0.2,\;\; -0.2 \\  
                 Radial scale (const.) & -2.8,\;\; 3.5 & -2.0,\;\; 2.0 & -2.2,\;\; 1.9 & -1.7,\;\; 2.4 & -2.4,\;\; 3.7 & -2.5,\;\; 5.6 & -12.6,\;\; 3.9 & -0.4,\;\; -1.4 & -2.8,\;\; 2.9 \\  
                 Radial scale (E dep.) & -2.2,\;\; 2.2 & -2.3,\;\; 2.5 & -2.5,\;\; 2.7 & -2.7,\;\; 2.9 & -2.9,\;\; 3.1 & -3.1,\;\; 3.3 & -3.3,\;\; 3.5 & -3.6,\;\; 3.8 & -3.9,\;\; 4.3 \\  
                      Vertex $x$ & -0.1,\;\; -0.1& 0.0,\;\; -0.5 & -0.1,\;\; 0.3 & -0.2,\;\; 0.0 & 0.7,\;\; -0.4 & 0.2,\;\; -0.2 & -1.2,\;\; 0.0 & -0.9,\;\; -0.6 & -0.2,\;\; 0.5 \\  
                      Vertex $y$ & 0.0,\;\; 0.1  & 0.5,\;\; -0.2 & 0.0,\;\; 0.0 & 0.5,\;\; 0.0 & 0.0,\;\; 0.3 & 0.5,\;\; -0.1 & -0.5,\;\; 0.2 & -1.3,\;\; -0.4 & -0.1,\;\; 0.1 \\  
                      Vertex $z$ & 0.7,\;\; 0.0  & -0.2,\;\; -0.3 & 0.1,\;\; -0.3 & 0.7,\;\; 0.1 & 0.8,\;\; -1.0 & 0.5,\;\; 1.3 & -1.2,\;\; 1.0 & -2.4,\;\; 1.0 & 0.1,\;\; -0.9 \\  
               Vertex resolution & 0.0,\;\; 0.0  & -0.2,\;\; 0.2 & 0.1,\;\; -0.1 & -0.1,\;\; 0.1 & 1.0,\;\; -1.0 & 3.3,\;\; -3.3 & -8.7,\;\; 8.7 & -4.1,\;\; 4.1 & 0.8,\;\; -0.8 \\  
              Angular resolution & 0.0,\;\; 0.0  & -0.6,\;\; 0.6 & -0.7,\;\; 0.7 & -0.8,\;\; 0.8 & -0.8,\;\; 0.8 & 0.3,\;\; -0.3 & 0.5,\;\; -0.5 & 0.4,\;\; -0.4 & 0.0,\;\; 0.0 \\  
                  Internal $\gamma$ & 0.0,\;\; 0.0  & 0.0,\;\; 0.0 & 0.0,\;\; 0.0 & 0.0,\;\; 0.0 & 0.0,\;\; 0.0 & 0.0,\;\; 0.0 & 0.0,\;\; 0.0 & 0.0,\;\; 0.0 & 0.0,\;\; 0.0 \\  
            Selection efficiency & -0.1,\;\; 0.2 & -0.2,\;\; 0.2 & -0.2,\;\; 0.2 & -0.2,\;\; 0.2 & -0.2,\;\; 0.2 & -0.2,\;\; 0.2 & -0.2,\;\; 0.2 & -0.2,\;\; 0.2 & -0.3,\;\; 0.3 \\  

                     Backgrounds & -0.1,\;\; 0.0 & -0.2,\;\; 0.0 & -0.2,\;\; 0.0 & -0.3,\;\; 0.0 & -0.3,\;\; 0.0 & -0.9,\;\; 0.0 & -0.9,\;\; 0.0 & -3.2,\;\; 0.0 & -12.3,\;\; 0.0 \\  

\hline
\hline
\end{tabular}
\end{center}
\end{table}
\end{turnpage}

\end{document}